\documentclass{ptephy}
\usepackage{bm}
\usepackage{slashbox}

\begin{document}

\title{A model of charmed baryon-nucleon potential and 2- and 3-body bound states with charmed baryon}

\author{\name{Saori Maeda}{1,\ast}, \name{Makoto Oka}{1,2}, \name{Akira Yokota}{1}, \name{Emiko Hiyama}{3}, 
and \name{Yan-Rui Liu}{4}}

\address{\affil{1}{Department of Physics, Tokyo Institute of Technology, O-okayama 2-12-1, Meguro, Tokyo 152-8551, Japan}
\affil{2}{Advanced Science Research Center, Japan Atomic Energy Agency, Tokai, Ibaraki 319-1195, Japan}
\affil{3}{Nishina Center for Accelerator-Based Science, The institute of Physical and Chemical Research(RIKEN), Wako 351-0198, Japan}
\affil{4}{Department of Physics, Shandong University, Jinan, Shandong 250100, China}
\email{s-maeda@th.phys.titech.ac.jp}}

\begin{abstract}

A potential model for the interaction between a charmed baryon ($\Lambda_c$, $\Sigma_c$, and $\Sigma_c^*$) and the nucleon ($N$) is constructed. 
The model contains a long-range meson ($\pi$ and $\sigma$) exchange part and a short-distance quark exchange part. 
The quark cluster model is used to evaluate the short-range repulsion and a monopole type form factor is introduced 
to the long-range potential to reflect the extended structure of hadrons. 
We determine the cutoff parameters in the form factors by fitting the $NN$ scattering data with the same approach 
and we obtain four sets of parameters (a -- d). 
The most attractive potential (d) leads to bound $\Lambda_c N$ states with $J^\pi= 0^+$ and $1^+$ 
once the channel couplings among $\Lambda_c$, $\Sigma_c$ and $\Sigma_c^*$ are taken into account. 
One can also investigate many-body problems with the model.
Here, we construct an effective $\Lambda_c N$ one-channel potential 
with the parameter set (d) and apply it to the 3-body $\Lambda_c NN$ system.
The bound states with $J=1/2$ and $3/2$ are predicted.
\end{abstract}


\maketitle

\section{Introduction}
\label{sec:intro}

Recent development of hadron spectroscopy revealed that there may exist various molecular bound states of hadrons (observed as hadron resonances) . 
In particular, the observation of the unexpected X, Y, and Z mesons and the follow by 
theoretical studies indicates that heavy quark molecules are more plausible.
This can be understood from the balance between the kinetic term 
and the potential in the Hamiltonian: a heavier system has a smaller kinetic energy \cite{ref-i,ref-j,ref-k}. 

The above naive expectation motivates us to explore possible bound states composed of charmed (or bottomed) 
baryons $Y_{c(b)}$ ($\Lambda_{c(b)}$, $\Sigma_{c(b)}$, $\ldots$ ) and the nucleon ($N$) or nucleus.
This is, of course, a natural extension of the hypernucleus, which is a nuclear bound state with one or more strange baryons $Y$ ( $\Lambda$, $\Sigma$, $\Xi$ $\ldots$ ).
Hypernuclear spectroscopy in the last decades played a key role in analysing structures of hypernuclei and extracting
information on the $YN$ and $YY$ interactions.
Because there is no two-body bound states in $\Lambda N$ or $\Sigma N$ systems and it is difficult to perform
direct scattering experiments for the hyperons, it is important to get information on their interactions
from the three-body or heavier nucleus with strange baryon(s). 

The idea of the charmed hypernucleus is, in fact, old \cite{ref-a,ref-b,ref-c,ref-d,ref-e,ref-1,ref-01,ref-02}.
It was pointed out \cite{ref-c} that the SU(4) symmetry, 
though it is badly broken, for the one boson exchange (OBE) models predicts a weaker attraction between $Y_c$ and $N$, 
because the $K$ exchange is replaced by the $D$ exchange, which is suppressed by the heavier mass of the $D$ meson.
Recent re-analysis with a model based on the heavy quark effective theories suggested the possibility of bound $Y_c N$ states.
However, the model has a difficulty in describing the short-range interaction and its prediction was very sensitive to the
cutoff parameters in the one-boson exchange interaction \cite{ref-1}.

In this paper, we construct a potential model consisting of the long-range one-pion and one-sigma exchange interaction and the quark exchange effect based on the quark cluster model for the baryon-baryon interaction \cite{ref-2,ref-3}. 
The latter provides a short range repulsion between $Y_c$ and $N$.
Here $Y_c$ represents $\Lambda_c$, $\Sigma_c$ and $\Sigma^*_c$. 
The model also contains the off-diagonal components, 
which couple $\Lambda_cN$, $\Sigma_cN$ and $\Sigma^*_c N$ channels of the same total quantum numbers.
By adjusting the meson exchange parameters to reproduce low-energy data of the NN system, 
we obtain four sets of the $Y_cN$ potential.

We solve the coupled channel Schr\"odinger equation using the Gaussian expansion method (GEM) \cite{ref-6} and obtain $\Lambda_c N$ bound states.
The potential is also applied to the three-body $\Lambda_c NN$ system. Bound state solutions are found 
and their properties are analysed.

The paper is organized as follows. 
In Sec.~2, we present our model of the $Y_c N$ potential and explain how to determine the model parameters.
In Sec.~3, the two-body $Y_cN$ systems are analysed with the obtained potentials and the results are examined.
In Sec.~4, we construct an effective one-channel $\Lambda_c N$ potential model and apply it to the three-body $\Lambda_cNN$
system. 
Conclusions are given in Sec.~5.

\section{Model of $Y_{c}N$ interactions}
\label{sec:form}

We first consider the two-baryon systems, with isospin $I=\frac{1}{2}$ and spin-parity $J^{\pi}=0^{+}$ or $1^{+}$.
Table \ref{tb:pot1} shows possible channels contributed by $\Lambda_{c}N$, $\Sigma_{c}N$, and $\Sigma_{c}^{*}N$ for each $J$.
The channels with the orbital angular momentum $L=0$ and $2$ will be coupled by tensor force, 
while those with the same $L$ are coupled mainly by central force.
We solve coupled channel Schr$\ddot{\rm{o}}$dinger equation in a hybrid potential model.
The model includes a pion exchange potential and a scalar meson exchange potential for long-range interactions \cite{ref-1} 
and a short-range repulsive potential coming from the quark exchanges between the baryons \cite{ref-2}.

\begin{table}[t]
\centering
\small
\begin{tabular}{|l|ccccccc|} \hline
Channels & 1 & 2 & 3 & 4 & 5 & 6 & 7 \\ \hline
$J^{\pi} = 0^{+}$ & $\Lambda_{c}N({}^{1}S_{0})$ & $\Sigma_{c}N({}^{1}S_{0})$ & $\Sigma_{c}^{*}N({}^{5}D_{0})$ &   &   &   &   \\
$J^{\pi} = 1^{+}$ & $\Lambda_{c}N({}^{3}S_{1})$ & $\Sigma_{c}N({}^{3}S_{1})$ & $\Sigma_{c}^{*}N({}^{3}S_{1})$ & $\Lambda_{c}N({}^{3}D_{1})$ & $\Sigma_{c}N({}^{3}D_{1})$ & $\Sigma_{c}^{*}N({}^{3}D_{1})$ & $\Sigma_{c}^{*}N({}^{5}D_{1})$ \\ \hline
\end{tabular}
\caption{The S-wave $\Lambda_{c}N$ states and the channels coupling to them \cite{ref-1}}
\label{tb:pot1}
\end{table}

\subsection{One boson exchange potential}
\label{sec:pot}

The meson exchange part contains four kinds of terms: spin-independent, spin-spin, spin-orbit, and tensor,  
\begin{eqnarray}
V_{\pi}(i,j) & = & C_{\pi}(i,j)\frac{m_{\pi}^{3}}{24\pi f_{\pi}^{2}}\left\{ \left< \bm{\mathcal{O}}_{spin} \right>_{ij} Y_{1}(m_{\pi}, \Lambda_{\pi} ,r) + \left< \bm{\mathcal{O}}_{ten} \right>_{ij} H_{3}(m_{\pi}, \Lambda_{\pi} ,r) \right\} \nonumber \\
V_{\sigma}(i,j) & = & C_{\sigma}(i,j)\frac{m_{\sigma}}{16\pi}\left\{ \left< \bm{1} \right>_{ij}4Y_{1}(m_{\sigma}, \Lambda_{\sigma} ,r) + \left< \bm{\mathcal{O}}_{LS} \right>_{ij} \left( \frac{m_{\sigma}}{M_{N}} \right)^{2} Z_{3}(m_{\sigma}, \Lambda_{\sigma} ,r) \right\} \nonumber \\
\label{eq:pot2}
\end{eqnarray}
where $i$ and $j$ are the labels of the channels and $C_{\pi}(i,j)$ and $C_{\sigma}(i,j)$ are the relevant coupling constants including the isospin factor.
The spin dependent operators, $\bm{\mathcal{O}}_{spin}$, $\bm{\mathcal{O}}_{ten}$ , and $\bm{\mathcal{O}}_{LS}$, and their expectation values are given in Appendix A.
The $r$-dependent functions $Y_{1}$, $H_{3}$, and $Z_{3}$ contain the cutoff parameters 
$\Lambda_{\pi}$ and $\Lambda_{\sigma}$. The explicit forms of the functions are given in Appendix B.
We will determine the coupling strength and the cutoff parameters later.

\subsection{Short range repulsion from Quark Cluster Model}
\label{sec:potqcm}

In a previous approach \cite{ref-1}, we considered exchanges of the vector mesons for the short range part of the $Y_{c}$N interaction.
They, however, do not provide enough repulsion at short distances and result in very deep bound states with compact wave functions.
As the wave function of the baryons overlap significantly at short distances, 
the quark exchange effect becomes important.
We here employ the quark cluster model (QCM) for the short-range potential \cite{ref-2}.

In QCM each baryon is a cluster determined made of three quarks. 
When the two clusters overlap at small $r$, the antisymmetrization among the (light) quarks induces a non-local interaction.
When the two baryons overlap completely  i.e., $r=0$, all the six quarks occupy the lowest energy orbit with a single center.
Such a state is approximately given by a product of the Gaussian wave functions with an appropriate symmetry according to the quantum numbers.
The strength of the repulsive potential at $r=0$ ($V_0$) is thus determined by the difference between the energy 
of the single-centered six-quark states and that of the individual baryons.
\begin{equation}
V_{0} \approx \langle 6q | H | 6q \rangle - 2 \langle 3q | H | 3q \rangle
\label{eq:qcm}
\end{equation}
It turns out \cite{ref-2} that the strengths are sensitive to and determined dominantly by the color-magnetic interaction (CMI) given by
\begin{equation}
V_{CM}=-\beta \sum_{i<j} (\bm{\sigma}_{i} \cdot \bm{\sigma}_{j})(\bm{\lambda}_{i} \cdot \bm{\lambda}_{j})
\label{eq:qcm0}
\end{equation}
where $\bm{\sigma}_{i}$ and $\bm{\lambda}_{i}$ are the spin and color operators of the $i$-th quark.
The expectation values of the color-magnetic operator for a five light quark systems can be computed by using the formula, 
\begin{equation}
\left< V_{CM} \right> = \beta \left[ 8N + \frac{4}{3}S(S+1) + 2C_{2} \left[ SU(3)_{c} \right] - 4C_{2} \left[ SU(6)_{cs} \right]  \right],
\label{eq:qcm1}
\end{equation}
where $N$ is total number of light quarks, $S$ is the total spin, 
and $C_{2} \left[ SU(g) \right]$ is the quadratic Casimir operator, whose value is specified by the Young diagram $[f_{1}, \ldots ,f_{g}]$, 
\begin{equation}
C_{2} \left[ SU(g) \right] = \frac{1}{2} \left[ \sum_{i} f_{i}(f_{i} -2i + g + 1)-\frac{N^{2}}{g} \right].
\end{equation}
The overall strength $\beta$ can be determined by the $\Delta (1232)$ - nucleon $(N)$ mass splitting, which comes also from $V_{CM}$.
From $\left< V_{CM} \right>_{\Delta} - \left< V_{CM} \right>_{N} = -16 \beta = 293$ MeV (exp.), we obtain $\beta \cong 18.2$ MeV. 
Table \ref{tb:pot2} shows the values of $V_{CM}$ for various two-baryon (six-quark) states.
These values are evaluated in the heavy quark limit, so that the heavy quark spin does not contribute.

\begin{table}[h!]
\centering
\begin{tabular}{c|c}\hline
System & $V_{0}$ [MeV] \\ \hline
$(NN)^{S=0}_{I=1}$ & 450 \\
$(NN)^{S=1}_{I=0}$ & 350 \\ \hline
$(\Lambda_cN)^{S=0}_{I=1/2}$ & 300 \\
$(\Lambda_cN)^{S=1}_{I=1/2}$ & 300 \\ \hline
\multicolumn{2}{c}{ }\\
\end{tabular}
\begin{tabular}{c}
$\left( \begin{array}{c}
\Sigma_cN \\ \Sigma_c^{*}N
\end{array} \right)^{S=0}_{I=1/2} = \left(
\begin{array}{cc}
100 & 0 \\
0 & 0 
\end{array} \right), \ \ \ 
\left( \begin{array}{c}
\Sigma_cN \\ \Sigma_c^{*}N
\end{array} \right)^{S=1}_{I=1/2} = \left(
\begin{array}{cc}
166.7 & -24.0 \\
-24.0 & 108.3 
\end{array} \right)$ \\
\end{tabular}
\caption{Expectation values in MeV of the color magnetic interaction (CMI) for the relevant channels 
in the heavy quark limit. For the correlated channels of $\Sigma_cN$ and $\Sigma_c^{*}N$, 
the values are given in the matrix form.}
\label{tb:pot2}
\end{table}

We assume, for simplicity, that the radial dependence of the QCM potential is given by a Gaussian,
\begin{equation}
V_{QCM}=V_{0}e^{-(r^{2}/b^{2})}.
\label{eq:qcm2}
\end{equation}
The range parameter $b$ is supposed to coincide with the extension of the quark wave functions in the baryon.
According to \cite{ref-3}, the typical values for the NN interaction are about $0.54 \sim 0.58$ fm.
For the $Y_{c}N$ systems, we use two typical values, $b=0.5$ and $0.6$ fm, and compute the results.

\subsection{Coulomb potential}
\label{sec:potcoul}

We also include the Coulomb potential between the charged baryons.
In this calculation, we consider only the two-baryon channels coupling to $\Lambda_{c}N (I=\frac{1}{2})$.
For the $\Lambda_{c}n$ system (total charge $Q=+1$), there is no Coulomb effect.
For the $Q=2$ channels, $\Lambda_{c}p$, $\Sigma_{c}^{+}p$ and $\Sigma_{c}^{++}n$, 
we assume that the mixing of $\Sigma_{c}^{+}p$ and $\Sigma_{c}^{++}n$ is not modified by the Coulomb interaction 
and make an approximation for the Coulomb potential with the combination, 
\begin{equation}
\Bigl| (\Sigma_{c}N)_{I=\frac{1}{2}, I_{3}=+\frac{1}{2}} \Bigr> = -\sqrt{\frac{1}{3}} \Bigl| \Sigma_{c}^{+}p \Bigr> +\sqrt{\frac{2}{3}} \Bigl| \Sigma_{c}^{++}n \Bigr>
\label{eq:coulomb_ch}
\end{equation}
Then the "effective" Coulomb potentials for the $Y_{c}N$ channels are given by
\begin{eqnarray}
V_{coulomb}^{\Lambda_{c}N}(r)=\frac{\alpha \hbar c}{r}, \nonumber \\
V_{coulomb}^{\Sigma_{c}N}(r)=\frac{1}{3}\frac{\alpha \hbar c}{r}, \nonumber \\
V_{coulomb}^{\Sigma_{c}^{*}N}(r)=\frac{1}{3}\frac{\alpha \hbar c}{r},
\label{eq:coulomb}
\end{eqnarray}
where $\alpha \sim 1/137$ is the fine structure constant.

\subsection{Determining the potential parameters}
\label{sec:potcom}

In the previous study using the OBE potential \cite{ref-1}, we found that the results are very sensitive to the cutoff parameters.
To remedy this problem, in the present approach, we adjust the parameters of the potential 
to reproduce the $NN$ interaction data using the same model. 
In doing so, we fix the $\pi$-baryon coupling constants and the short-range potential derived from the quark model. 
Then the cutoffs, $\Lambda_{\pi}$ and $\Lambda_{\sigma}$, and the sigma coupling constant, $C_{\sigma}$,  
are the parameters to be determined in the $NN$ system.
After that the results are generalized to the $Y_c N$ system.
In this method, we assume that the light mesons couple only to the light quarks 
and the cutoff parameters are common to $NN$ and $Y_c N$ systems. The details are given in Appendix B.

In searching the three undetermined parameters, $\Lambda_{\pi}$, $\Lambda_{\sigma}$ and $C_{\sigma}$, 
we use the following experimental values of the $NN$ interaction,
\begin{eqnarray}
NN (^{1}S_{0}) \  {\rm scattering \ length} = -23.7 {\rm fm}, \nonumber \\
NN (^{3}S_{1}) \  {\rm binding \ energy(deuteron)} = 2.22{\rm MeV}.
\label{eq:parafit1}
\end{eqnarray}
To get appropriate values, we restrict the possible range of parameters to be 
\begin{eqnarray}
\Lambda_{\pi} = 500 \sim 900 {\rm MeV}, \nonumber \\
\Lambda_{\sigma} = 900 \sim 1200 {\rm MeV}, \nonumber \\
C_{\sigma}(NN) = -64.0 \sim -225.0.
\label{eq:parafit2}
\end{eqnarray}

Unfortunately, we could not find a solution.
Therefore we relax the condition that $C_{\sigma}$ is independent of $J^\pi$ and take $C_{\sigma}(J^{\pi}=0^{+})$ and $C_{\sigma}(J^{\pi}=1^{+})$ independently.
Then we find solutions, which are given in Tables \ref{tb:paracal1} and \ref{tb:paracal2}.
In Table \ref{tb:paracal1}, we show parameters with a fixed $\Lambda_{\pi}=750$ MeV and varied $\Lambda_{\sigma}$  while Table \ref{tb:paracal2} lists the results with varied $\Lambda_{\pi}$ for a fixed $\Lambda_{\sigma}=1000$ MeV.
These two Tables also present the differences between $C_{\sigma}(J^{\pi}=0^{+})$ and $C_{\sigma}(J^{\pi}=1^{+})$, 
\begin{equation}
\Delta_{C} = | C_{\sigma}(J^{\pi}=0^{+}) - C_{\sigma}(J^{\pi}=1^{+}) |.
\end{equation}
From the results in the Tables, one finds that $\Delta_{C}$ is insensitive to the change of $\Lambda_{\sigma}$, 
but $\Delta_{C}$ increases rapidly for increasing $\Lambda_{\pi}$.
The features for $b=0.5$ fm and $0.6$ fm are qualitatively similar. 
Since the QCM repulsion for $b=0.5$ fm is weaker than that for $b=0.6$ fm, the resulting $C_{\sigma}$ is also smaller 
and $\Delta_{C}$ tends to be smaller for $b=0.5$ fm.
We suppose that the spin dependence of $C_{\sigma}$ is small, so that a small $\Delta_{C}$ is favoured.

\begin{table}[h]
\centering
\begin{tabular}{c||c|c||c||c|c||c} \hline
 & \multicolumn{3}{c||}{$b=0.6$[fm]} & \multicolumn{3}{c}{$b=0.5$[fm]} \\ \hline
$\Lambda_{\sigma}$[MeV] & $C_{\sigma}(0^{+})$ & $C_{\sigma}(1^{+})$ & $\Delta_{C}$ & $C_{\sigma}(0^{+})$ & $C_{\sigma}(1^{+})$ & $\Delta_{C}$ \\ \hline
900 & -213.16 & -161.29 & 51.87 & -176.89 & -132.25 & 44.64 \\
950 & -179.56 & -134.56 & 45.0 & -148.84 & -110.25 & 38.59 \\
1000 & -156.25 & -118.81 & 37.44 & -129.96 & -96.04 & 33.92 \\
1050 & -141.61 & -106.09 & 35.52 & -116.64 & -86.49 & 30.15 \\
1100 & -127.69 & -96.04 & 31.65 & -106.09 & -79.21 & 26.88 \\
1150 & -116.64 & -88.36 & 28.28 & -98.01 & -72.25 & 25.76 \\
1200 & -108.16 & -82.81 & 25.35 & -90.25 & -67.24 & 23.01 \\ \hline
\end{tabular}
\caption{The $NN$ 2-body parameters for $\Lambda_{\pi}=750$ MeV}
\label{tb:paracal1}
\end{table}

\begin{table}[h]
\centering
\begin{tabular}{c||c|c||c||c|c||c} \hline
 & \multicolumn{3}{c||}{$b=0.6$[fm]} & \multicolumn{3}{c}{$b=0.5$[fm]} \\ \hline
$\Lambda_{\pi}$[MeV] & $C_{\sigma}(0^{+})$ & $C_{\sigma}(1^{+})$ & $\Delta_{C}$ & $C_{\sigma}(0^{+})$ & $C_{\sigma}(1^{+})$ & $\Delta_{C}$ \\ \hline
500 & -148.84 & -139.24 & 9.6 & -118.81 & -121.0 & 2.19 \\ 
550 & -151.29 & -136.89 & 14.4 & -123.21 & -116.64 & 6.57 \\
600 & -153.76 & -134.56 & 19.2 & -125.44 & -114.49 & 10.95 \\
650 & -153.76 & -129.96 & 23.8 & -127.69 & -110.25 & 17.44 \\
700 & -156.25 & -125.44 & 30.81 & -127.69 & -104.04 & 23.65 \\
750 & -156.25 & -118.81 & 37.44 & -129.96 & -96.04 & 33.92 \\
800 & -156.25 & -110.25 & 48.51 & -129.96 & -88.36 & 41.6 \\
850 & -156.25 & -102.01 & 54.24 & -132.25 & -79.21 & 53.04 \\
900 & -156.25 & -92.16 & 66.6 & -132.25 & -70.56 & 61.69 \\ \hline
\end{tabular}
\caption{The $NN$ 2-body parameters for $\Lambda_{\sigma}=1000$ MeV}
\label{tb:paracal2}
\end{table}

\begin{figure}[!t]
\begin{tabular}{cc}
\begin{minipage}{0.5\hsize}
\begin{center}
\includegraphics[width=75mm]{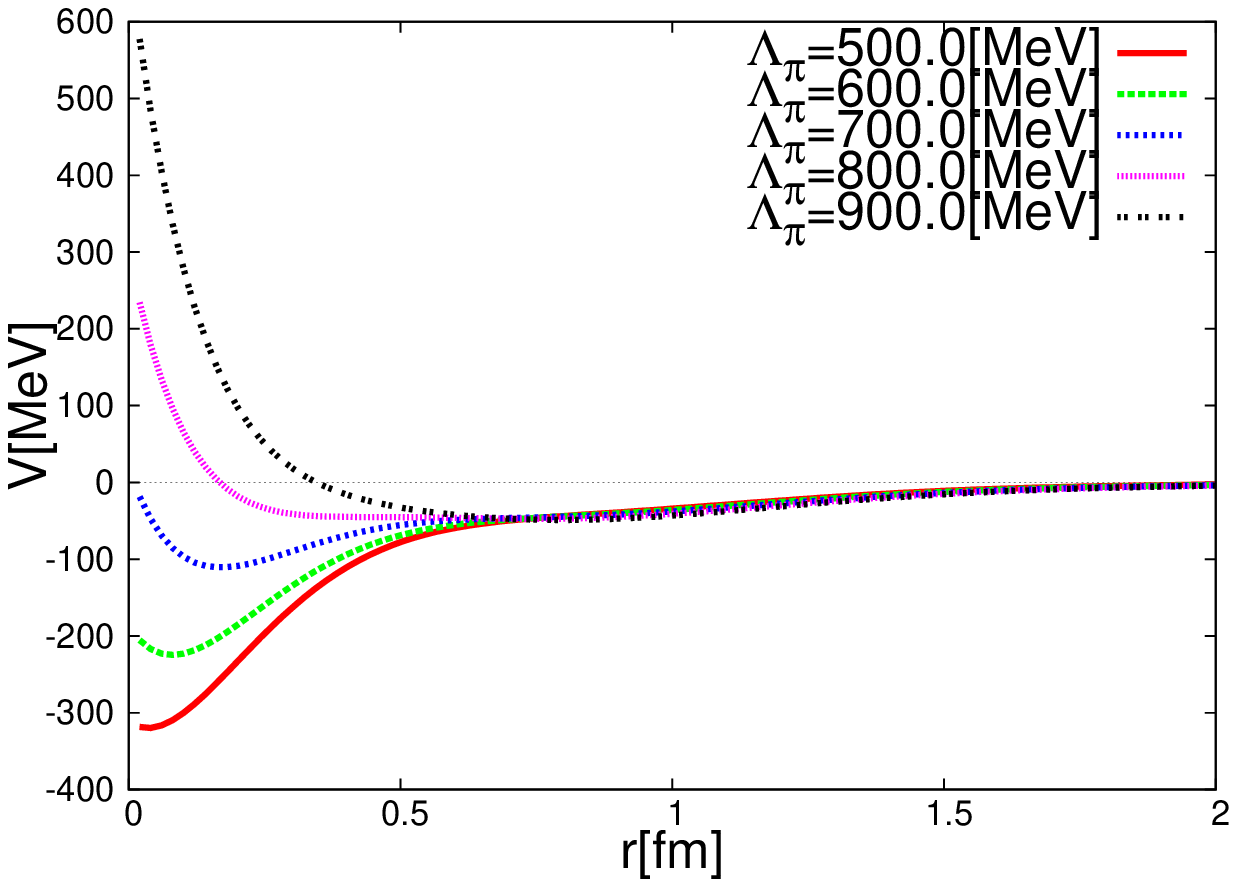}
\caption{NN potential $(J^{\pi}=0^{+})$ for $b$=0.6{fm} and $\Lambda_{\sigma}=1000$ MeV.}
\label{gr:reycn0p}
\end{center}
\end{minipage}
\begin{minipage}{0.5\hsize}
\begin{center}
\includegraphics[width=75mm]{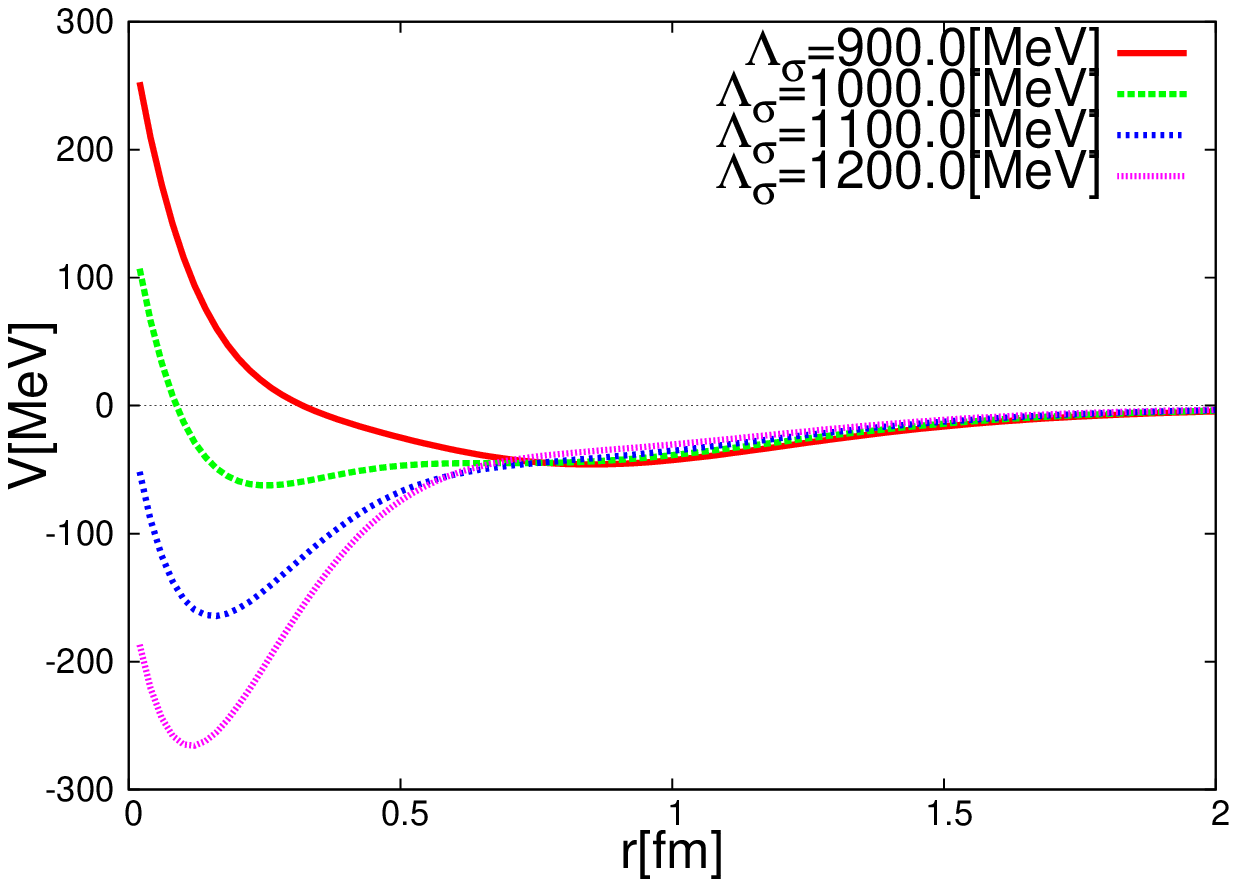}
\caption{NN potential $(J^{\pi}=0^{+})$ for $b$=0.6{fm} and $\Lambda_{\pi}=750$ MeV.}
\label{gr:reycn0s}
\end{center}
\end{minipage}
\end{tabular}
\end{figure}

The resulting $NN$ potentials for $J^{\pi}=0^{+}$ with the obtained parameters are plotted in Figs. \ref{gr:reycn0p} and \ref{gr:reycn0s}.
We find that some of the potentials (for small $\Lambda_{\pi}$) are strongly attractive at short distances and may not be appropriate for the current purpose, 
although they all reproduce the $^{1}S_{0}$ scattering length and the deuteron binding energy.

Therefore we choose the parameters with $\Lambda_{\pi}=750$ MeV and $\Lambda_{\sigma}=1000$ MeV 
and employ four sets of the most realistic potential parameters, given in Table \ref{tb:paracal3}.
We call these models ``CTNN'' potentials, 
as these parameter sets correspond to the NN experimental data.
The $Y_{c}N$ potentials derived for $J^{\pi}=0^{+}$ are shown in Figs. \ref{gr:potctnn-011} - \ref{gr:potctnn-023}.
In Appendix C, we show the $Y_{c}N$ potentials for $J^{\pi}=1^{+}$ and also
the individual components of the CTNN-a potential.

\begin{table}[!h]
\centering
\begin{tabular}{c||c|c} \hline
 & $C_{\sigma}$ & b[fm] \\ \hline
parameter a & -67.58 & 0.6 \\
parameter b & -77.5 & 0.6 \\
parameter c & -60.76 & 0.5 \\
parameter d & -70.68 & 0.5 \\ \hline
\end{tabular}
\caption{The CTNN potential parameters}
\label{tb:paracal3}
\end{table}

\begin{figure}[!t]
\begin{tabular}{cc}
\begin{minipage}{0.5\hsize}
\begin{center}
\includegraphics[width=75mm]{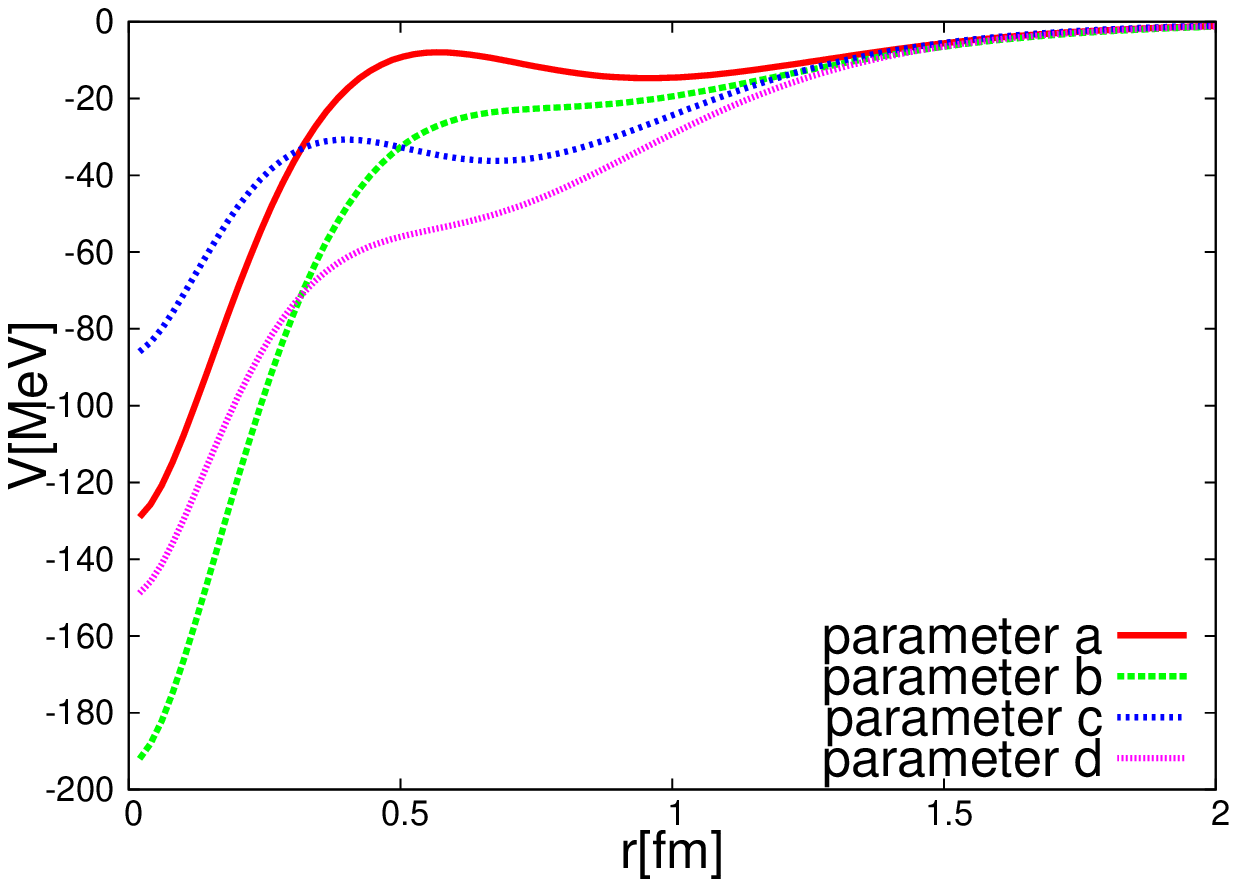}
\caption{$Y_{c}N$-CTNN potential for the $\Lambda_{c}N$ single channel.}
\label{gr:potctnn-011}
\end{center}
\end{minipage}
\begin{minipage}{0.5\hsize}
\begin{center}
\includegraphics[width=75mm]{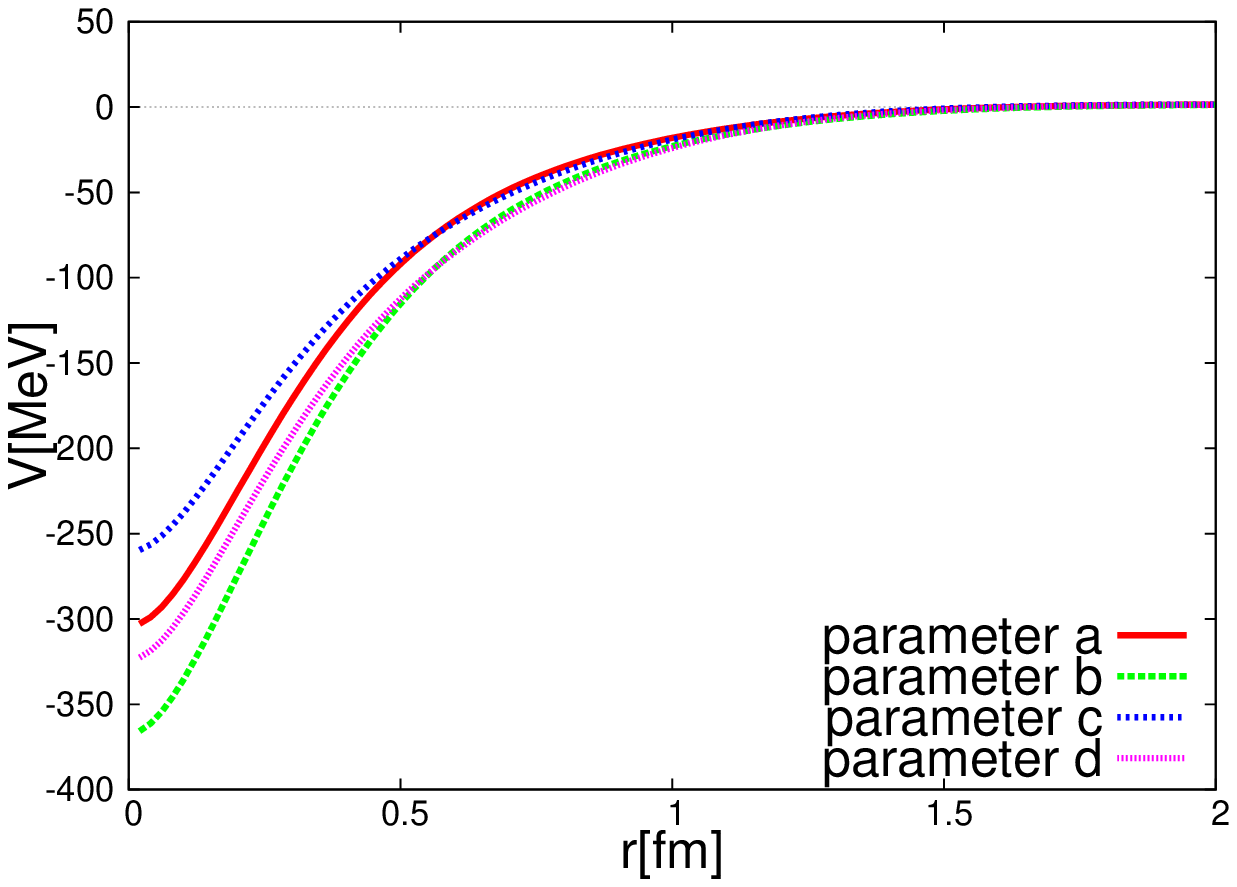}
\caption{$Y_{c}N$-CTNN potential for the $\Sigma_{c}N$ single channel.}
\label{gr:potctnn-022}
\end{center}
\end{minipage}
\\
\begin{minipage}{0.5\hsize}
\begin{center}
\includegraphics[width=75mm]{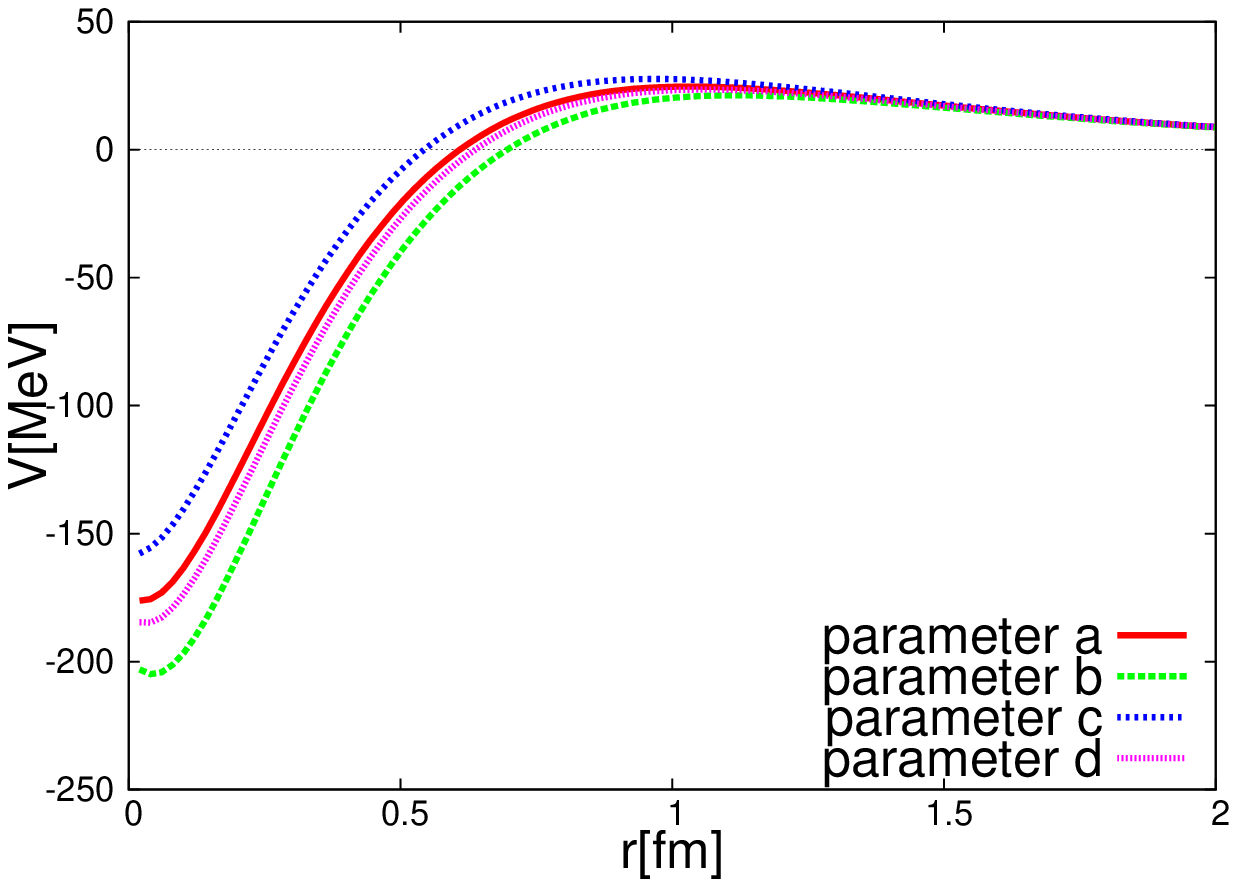}
\caption{$Y_{c}N$-CTNN potential for the $\Sigma_{c}^{*}N$ single channel.}
\label{gr:potctnn-033}
\end{center}
\end{minipage}
\begin{minipage}{0.5\hsize}
\begin{center}
\includegraphics[width=75mm]{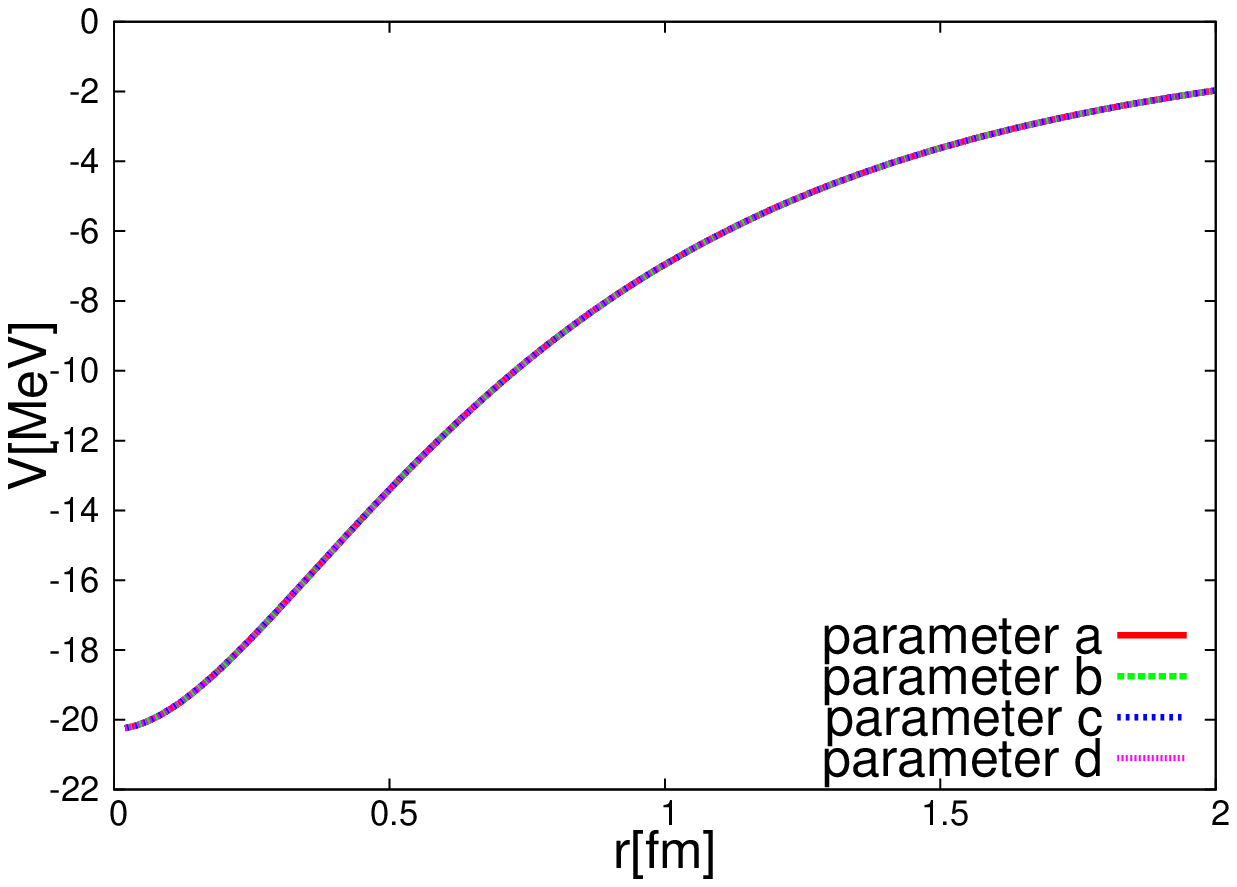}
\caption{$Y_{c}N$-CTNN potential for the $\Lambda_{c}N$-$\Sigma_{c}N$ channels.}
\label{gr:potctnn-012}
\end{center}
\end{minipage}
\\
\begin{minipage}{0.5\hsize}
\begin{center}
\includegraphics[width=75mm]{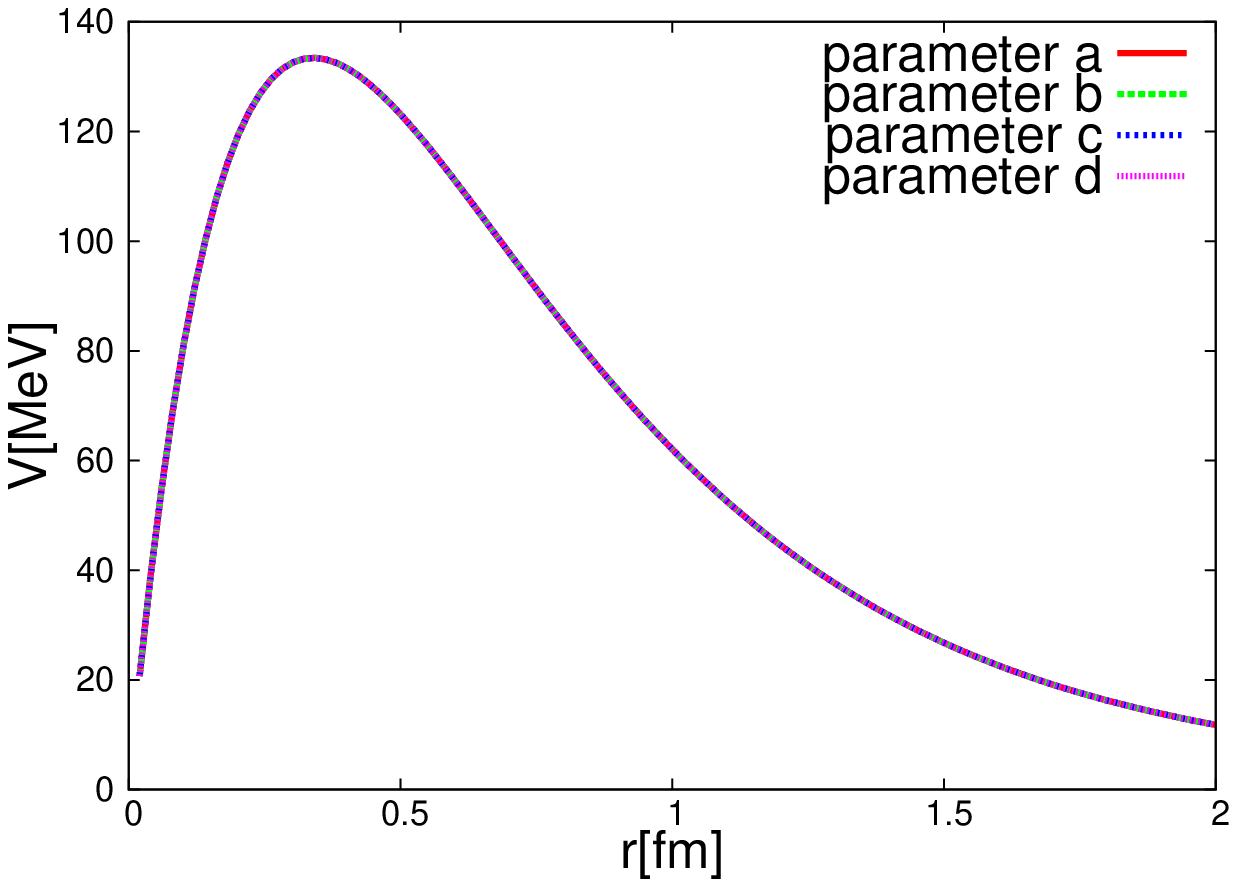}
\caption{$Y_{c}N$-CTNN potential for the $\Lambda_{c}N$-$\Sigma_{c}^{*}N$ channels.}
\label{gr:potctnn-013}
\end{center}
\end{minipage}
\begin{minipage}{0.5\hsize}
\begin{center}
\includegraphics[width=75mm]{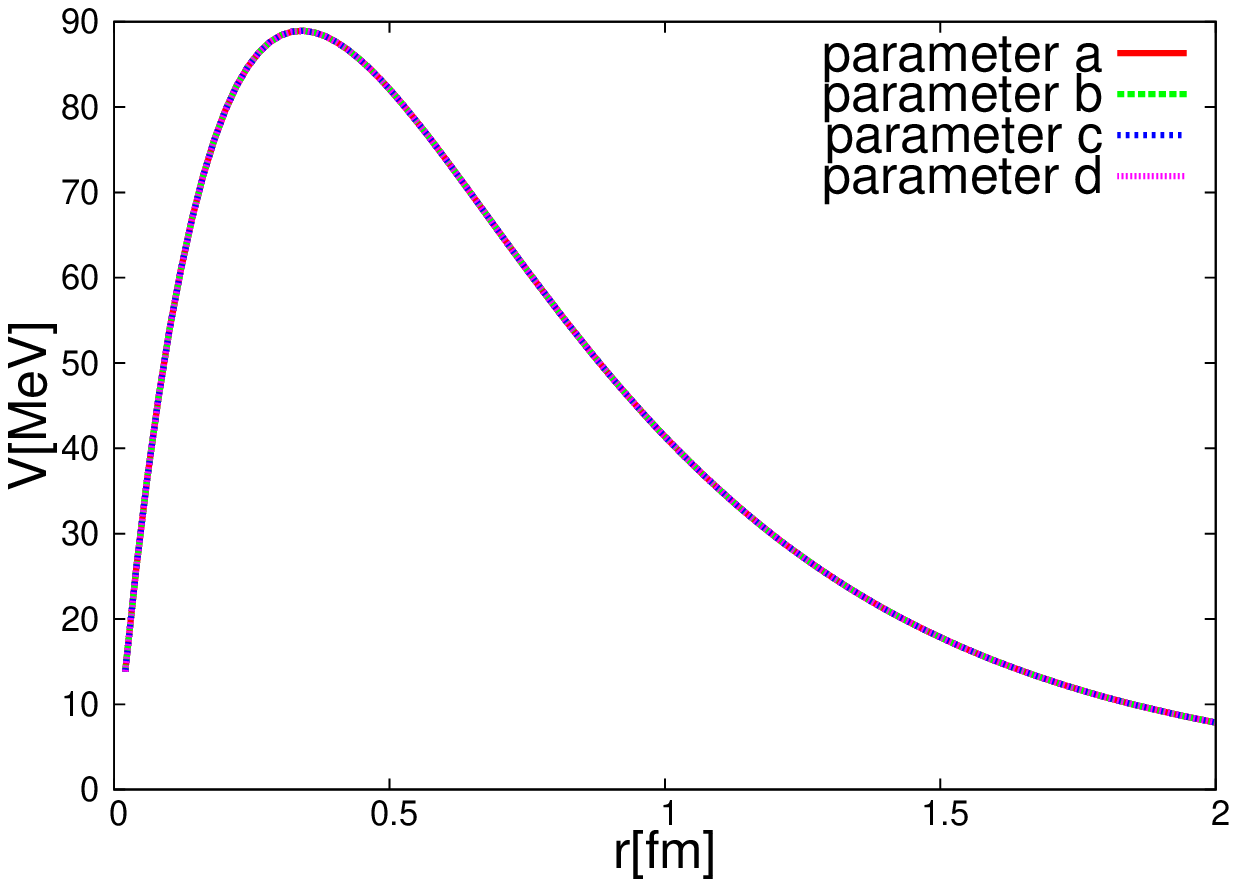}
\caption{$Y_{c}N$-CTNN potential for the $\Sigma_{c}N$-$\Sigma_{c}^{*}N$ channels.}
\label{gr:potctnn-023}
\end{center}
\end{minipage}
\end{tabular}
\end{figure}

\section{$Y_{c}N$ bound states}
\label{sec:potpara}

In solving the Schr\"odinger equation for the two-baryon system, we employ a variational method with Gaussian trial functions, 
the Gaussian Expansion Method \cite{ref-6}, where the radial wave functions are expanded by the basis states given by Gaussian functions 
with varied range parameters.
\begin{eqnarray}
\psi_{lm}(\bm{r}) = \sum_{n=1}^{n_{max}} c_{nl} \phi^{G}_{nlm}(\bm{r}) \nonumber \\
\phi^{G}_{nlm}(\bm{r}) = \phi^{G}_{nl}(r)Y_{lm}(\hat{\bm{r}}) \nonumber \\
\phi^{G}_{nl}(r) = N_{nl}r^{l}e^{-\nu_{n}r^{2}} \nonumber \\
\nu_{n} = \frac{1}{r_{n}} = \frac{1}{r_{1}a^{n-1}} 
\end{eqnarray}
This method has been applied to various bound state problems and was proved to give an accurate 
approximation to the eigenstate-energies and wave functions.

\begin{table}[t]
\centering
\begin{tabular}{r||c|c|c|c} \hline
$J^{\pi}=0^{+}$ & CTNN-a & CTNN-b & CTNN-c & CTNN-d \\ \hline
B.E. [MeV] & - & - & 1.72 $\times 10^{-3}$ & 1.37 \\
(+ Coulomb) &  &  &  & (0.56) \\
scattering length [fm] & -3.64 & -65.15 & 130.93 & 5.31 \\ \hline
probability($\Lambda_{c}N$)[\%] & - & - & 99.97 & 99.29 \\
probability($\Sigma_{c}N$)[\%] & - & - & 7.0 $\times 10^{-3}$ & 0.20 \\
probability($\Sigma_{c}^{*}N$)[\%] & - & - & 2.1 $\times 10^{-2}$ & 0.51 \\ \hline
\end{tabular}
\smallskip
\caption{
Binding energies of $\Lambda_{c}N$ ($J^{\pi}$ = $0^{+}$) for the CTNN potentials.
The probabilities of the $\Sigma_{c}N$ and $\Sigma_{c}^{*}N$ channels are also shown.}
\label{tb:becal0}
\end{table}

\begin{table}[t]
\centering
\begin{tabular}{r||c|c|c|c} \hline
$J^{\pi}=1^{+}$ & CTNN-a & CTNN-b & CTNN-c & CTNN-d \\ \hline
B.E. [MeV] & - & 1.67 $\times 10^{-4}$ & 1.91 $\times 10^{-2}$ & 1.56 \\
(+ Coulomb) &  &  &  & (0.72) \\
scattering length [fm] & -4.11 & 337.53 & 39.27 & 5.01 \\ \hline
probability($\Lambda_{c}N$)[\%] & - & 99.99 & 99.90 & 99.23 \\
probability($\Sigma_{c}N$)[\%] & - & 4.9 $\times 10^{-3}$ & 4.9 $\times 10^{-2}$ & 0.39 \\
(D-wave ($^{3}D_{1}$)) & - & 4.5 $\times 10^{-3}$ & 4.6 $\times 10^{-2}$ & 0.35 \\
probability($\Sigma_{c}^{*}N$)[\%] & - & 4.6 $\times 10^{-3}$ & 4.6 $\times 10^{-2}$ & 0.38 \\
(D-wave ($^{5}D_{1}$)) & - & 3.1 $\times 10^{-3}$ & 3.2 $\times 10^{-2}$ & 0.25 \\ \hline
\end{tabular}
\smallskip
\caption{
Binding energies of $\Lambda_{c}N$ ($J^{\pi}$ = $1^{+}$) for the CTNN potentials.
The probabilities of the coupled D-wave $\Lambda_{c}N$ and $\Sigma_{c}N$ and $\Sigma_{c}^{*}N$ channels are also shown.}
\label{tb:becal1}
\end{table}

The $Y_{c}N$ binding energies are given in Tables \ref{tb:becal0} and \ref{tb:becal1}.
The $\Lambda_{c}n$ system ($J^{\pi}=0^{+}$ or $1^{+}$) without the Coulomb potential has 
a bound state for the $Y_{c}N$-CTNN parameter b, c, and d.
The CTNN-b and c potentials allow very shallow bound states and the CTNN-d potential gives a larger binding energy.
The result indicates that there is a strong QCM repulsion at short distances for the CTNN-a and b parameters.
On the other hand, the $\Lambda_{c}p$ system with the Coulomb potential has a bound state only for the parameter CTNN-d.

As a cross check, we also calculate the $\Lambda_{c}N$ scattering lengths for the CTNN potentials.
These results are consistent with the binding energies (see Tables \ref{tb:becal0} and \ref{tb:becal1}.)

The above results show that the binding energies of the $J^{\pi}=1^{+}$ states are larger than those of $J^{\pi}=0^{+}$.
This is consistent with the previous study \cite{ref-1}.

Table \ref{tb:becal1} also shows the probabilities of the coupled channels.
For all the parameter sets and the total angular momenta, the probability of $\Lambda_{c}N$ is more than 99 \% .
In the case of $J^{\pi}=0^{+}$, the probabilities of $\Sigma_{c}^{*}N$ ($^{5}D_{0}$) are larger than that of $\Sigma_{c}N$ ($^{1}S_{0}$).
On the other hand, in the case of $J^{\pi}=1^{+}$, the respective total probabilities of $\Sigma_{c}N$ and $\Sigma_{c}^{*}N$ are almost equal.
Looking closely, $\Sigma_{c}N (^{3}S_{1})$ and $\Sigma_{c}^{*}N (^{5}D_{1})$ contribute largely to the results.
These observations can be explained with the contributions of the strong tensor force between the $\Lambda_{c}N$ ($L=0$) and $\Sigma_{c}N$ / $\Sigma_{c}^{*}N$ ($L=2$) channels.

\section{$\Lambda_{c} NN$ systems}
\label{sec:3b}

\subsection{Effects of channel couplings}
\label{sec:comp1ch}

\begin{table}[t]
\centering
\begin{tabular}{c||c|c||c|c|c|c|c|c} \hline
 & \multicolumn{2}{c|}{$\Lambda_{c}N-\Sigma_{c}N-\Sigma_{c}^{*}N$} & \multicolumn{2}{c|}{$\Lambda_{c}N$} & \multicolumn{2}{c|}{$\Lambda_{c}N-\Sigma_{c}N$} & \multicolumn{2}{c}{$\Lambda_{c}N-\Sigma_{c}^{*}N$} \\ \hline
$J^{\pi}$ & $0^{+}$ & $1^{+}$ & $0^{+}$ & $1^{+}$ & $0^{+}$ & $1^{+}$ & $0^{+}$ & $1^{+}$ \\ \hline
CTNN-a & -3.63 & -4.10 & -1.11 & -1.11 & -1.16 & -2.07 & -3.13 & -2.09  \\
CTNN-b & -63.25 & 398.67 & -2.62 & -2.62 & -2.78 & --6.74 & -20.84 & -7.00  \\
CTNN-c & 139.07 & 39.96 & -3.01 & -3.01 & -3.19 & -8.61 & -48.56 & -9.00  \\
CTNN-d & 5.32 & 5.02 & -28.59 & -28.59 & -44.65 & 9.79 & 6.01 & 9.36  \\ \hline
B.E. & 1.37 & 1.56 & - & - & - & 0.36 & 1.04 & 0.39  \\ \hline
\end{tabular}
\smallskip
\caption{Scattering lengths in fm, and two-body binding energies in MeV 
for the cases that the coupled channels are reduced to $\Lambda_{c}N$ only, $\Lambda_{c}N$-$\Sigma_{c}N$ or $\Lambda_{c}N$-$\Sigma_{c}^{*}N$}
\label{tb:sl_p}
\end{table}

In the previous section, we find that the probabilities of the $\Lambda_{c}N$ component are almost 100\% and the mixings of $\Sigma_{c}N$ or $\Sigma_{c}^{*}N$ are small.
However, effects of the $\Sigma_{c}N - \Sigma_{c}^{*}N$ channel coupling are important in binding the $\Lambda_{c}N$ system,
because the single channel calculations of $\Lambda_{c}N$ do not show any binding solutions for all the CTNN-a $\sim$ d potentials. 

In Table \ref{tb:sl_p}, we show the scattering lengths obtained for partially coupled systems, i.e., 
$\Lambda_{c}N$, $\Lambda_{c}N-\Sigma_{c}N$, and $\Lambda_{c}N-\Sigma_{c}^{*}N$. 
It is found that only the CTNN-d potential allows bound states in the $\Lambda_{c}N-\Sigma_{c}^{*}N$ ($0^{+}$ and $1^{+}$) and the $\Lambda_{c}N-\Sigma_{c}N$ ($1^{+}$) systems.
In particular, effect of the $\Sigma_{c}^{*}N$ ($0^{+}$) channel is large because no bound state is found without this channel.
This indicates that the tensor force from the one-pion exchange, which induces the coupling between $\Sigma_{c}^{*}N$ ($^{5}D_{0}$) and $\Lambda_{c}N$ ($^{1}S_{0}$), is significant.
Similarly, the contributions of the tensor force and $\Sigma_{c}N-\Sigma_{c}^{*}N$ couplings are found to be important for the $J^{\pi}=1^{+}$ state.

\subsection{Effective $\Lambda_{c}N$ potential}
\label{sec:eff}

In order to apply the obtained potentials to many-body systems with $\Lambda_{c}$, we construct effective single-channel $\Lambda_{c}N$ potentials, 
in which the effects of the $\Sigma_{c}N-\Sigma_{c}^{*}N$ couplings are absorbed in phenomenological parameters.
We assume two-range Gaussian forms, 
\begin{eqnarray}
V_{\rm eff}^{0+} = V_{1}^{0+}e^{-\frac{\bm{r}^{2}}{b_{1}^{2}}} + V_{2}^{0+}e^{-\frac{\bm{r}^{2}}{b_{2}^{2}}}. \nonumber \\
V_{\rm eff}^{1+} = V_{1}^{1+}e^{-\frac{\bm{r}^{2}}{b_{1}^{2}}} + V_{2}^{1+}e^{-\frac{\bm{r}^{2}}{b_{2}^{2}}}.
\label{eq:ycneff1}
\end{eqnarray}
Here, $b_{1}$ and $b_{2}$ are range parameters, and $V_{1}^{JP}$ and $V_{2}^{JP}$ are the strength parameters.
They are determined from the CTNN potentials.
For simplicity, we choose the same $b_{1}$ and $b_{2}$ for $J^{\pi}=0^{+}$ and $1^{+}$, 
and rewrite Eq. (\ref{eq:ycneff1}) to be
\begin{equation}
V_{{\rm eff}_{YcN}} = \left[ V_{r}^{1} + \bm{\sigma}_{\Lambda_{c}} \cdot \bm{\sigma} V_{s}^{1} \right] e^{-\frac{\bm{r}^{2}}{b_{1}^{2}}} + \left[ V_{r}^{2} + \bm{\sigma}_{\Lambda_{c}} \cdot \bm{\sigma} V_{s}^{2} \right] e^{-\frac{\bm{r}^{2}}{b_{2}^{2}}}, 
\label{eq:ycneff2}
\end{equation}
\begin{equation}
\begin{array}{rl}
V_{r}^{i} & = \frac{1}{4}(V_{i}^{0+} + 3 V_{i}^{1+}), \\
V_{s}^{i} & = \frac{1}{4}(V_{i}^{1+} - V_{i}^{0+}). \\
\end{array}
\label{eq:ycneff3}
\end{equation}

Now, we use only the CTNN-d $Y_{c}N$ potential to construct the effective one. 
By reproducing the binding energies and the scattering lengths in Tables \ref{tb:becal0} and \ref{tb:becal1}, 
we search values of the parameters, in Eq. (\ref{eq:ycneff1}). In doing so, we assume that the first term ($V_{1}$ of Eq. (\ref{eq:ycneff1})) is an attractive potential like OBEP 
and the second term ($V_{2}$) is repulsive like the QCM repulsion.
Accordingly, $b_{2}$ is taken to be the radius of the quark wave function, $b_{2}=0.5$fm.
On the other hand, $b_{1}=0.9$ fm is chosen to represent a typical range of the meson exchange potential which couples $\Sigma_{c}N$ and $\Sigma_{c}^{*}N$ to $\Lambda_{c}N$.

\begin{figure}
\begin{center}
\includegraphics[width=100mm]{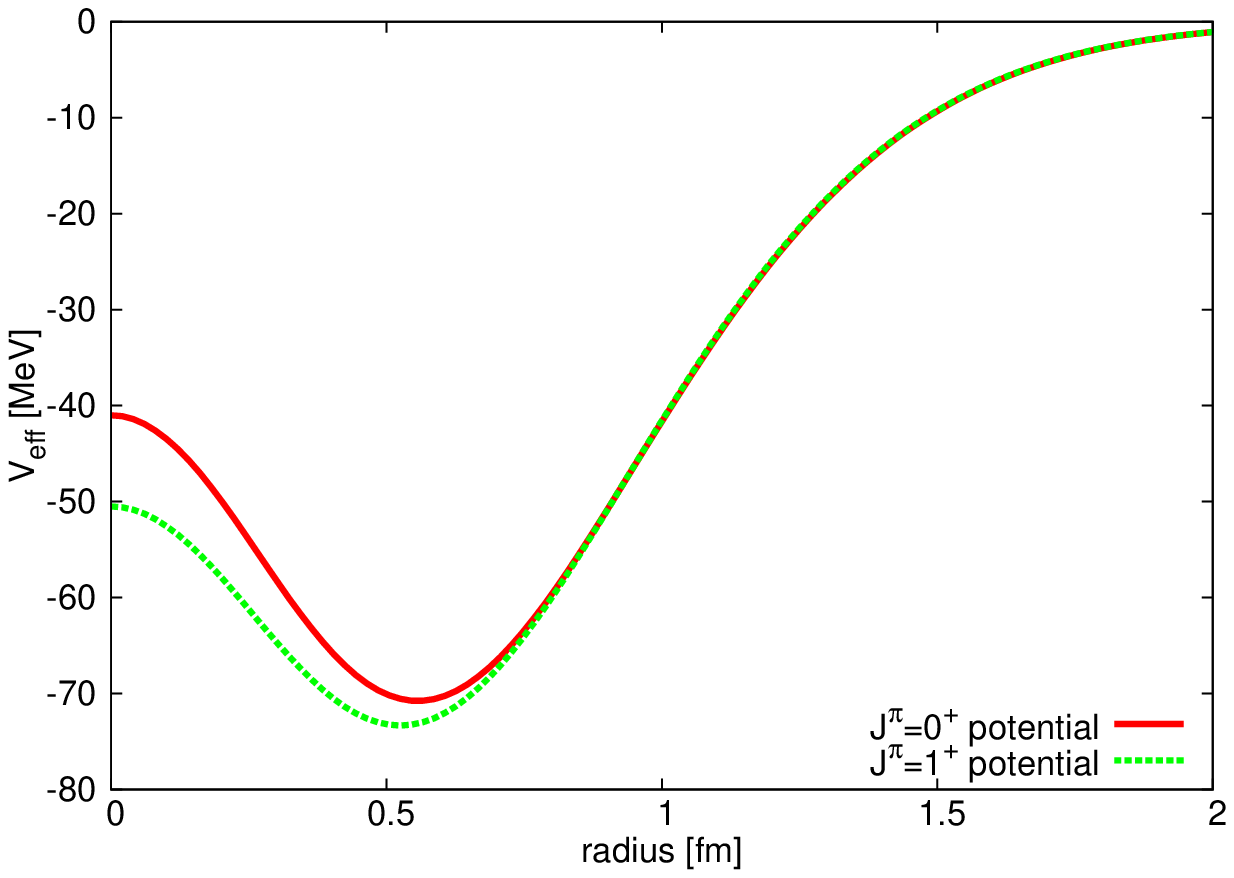}
\caption{$\Lambda_{c}N$ effective potentials for $J^{\pi}=0^{+}$ and $1^{+}$.}
\label{gr:reycn50}
\end{center}
\end{figure}

The remaining parameters of the effective potential are chosen as
\begin{equation}
V_{1}^{0+}=-150.0\rm{[MeV]}, \ V_{2}^{0+}=109.0\rm{[MeV]}, \ V_{1}^{1+}=-149.0\rm{[MeV]}, \ V_{2}^{1+}=98.5\rm{[MeV]}. \\
\end{equation}
Fig. \ref{gr:reycn50} illustrates the shape of the effective potentials.

Alternatively, one can express the effective potential with the strengths of the spin-independent terms and the spin-spin terms, 
\begin{equation}
V_{r}^{1}=-149.25\rm{[MeV]}, \ V_{s}^{1}=0.25\rm{[MeV]}, \ V_{r}^{2}=101.125\rm{[MeV]}, \ V_{s}^{2}=-2.625\rm{[MeV]}. \\
\end{equation}
It is obvious that the spin dependent potential is weak. This is a consequence of the heavy quark spin symmetry \cite{ref-0,ref-l,ref-m,ref-n}.

\subsection{$\Lambda_{c}NN$ bound states}

\begin{figure}
\begin{center}
\includegraphics[width=150mm,bb=0 0 600 200]{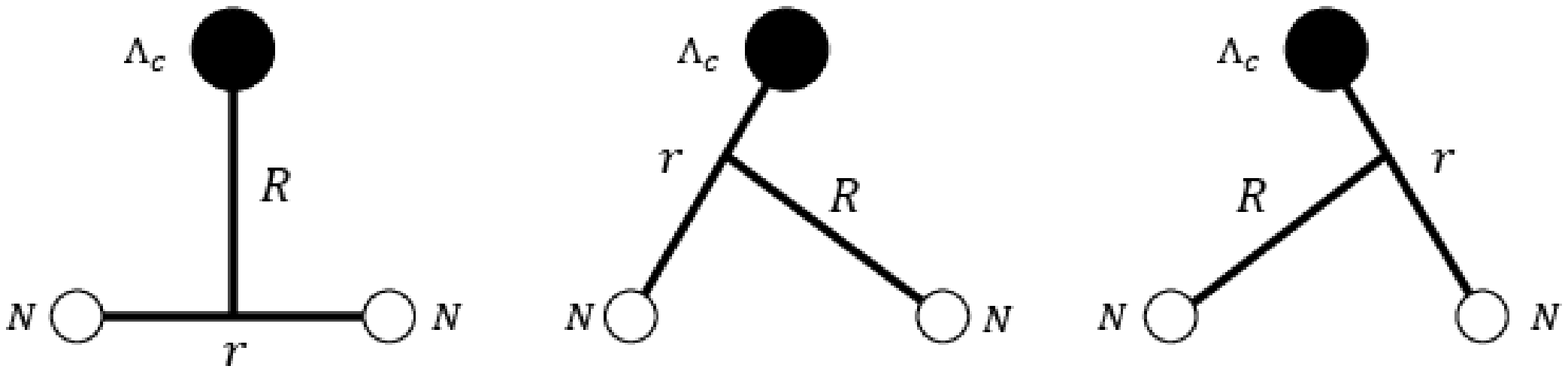}
\caption{Jacobi coordinates for $\Lambda_{c}NN$ system.}
\label{gr:ja}
\end{center}
\end{figure}

Next, we study the $\Lambda_{c}NN$ three-body system by using the above $\Lambda_{c}N$ effective potential.
In the calculation, we adopt the Jacobi coordinates shown in Fig. \ref{gr:ja}.
The angular momenta, $l$ and $L$ are defined corresponding to $r$ and $R$, respectively.
On the other hand, we define the spin for two nucleons as $S_{NN}$, and the total spin of the systems as $S_{tot}$.
For the isoscalar case, the binding energy is calculated from the threshold of $\Lambda_{c}$ plus the deuteron,
while for the case $I=1$, the calculation is from the threshold of the bound $\Lambda_{c}N$ (binding energy=1.56 MeV) plus a nucleon.

In this study, we assume that the orbital angular momenta, $l$ and $L$, are both zero.
Then $S_{NN}$ and $I$ are related and possible $J^{\pi}$ are 
\begin{equation}
\begin{array}{rcl}
I=0 & \cdots & S_{NN}=1, \ \ J^{\pi}=\frac{1}{2} \ \rm{or} \ \frac{3}{2}, \\
I=1 & \cdots & S_{NN}=0, \ \ J^{\pi}=\frac{1}{2}. \\
\end{array}
\end{equation}

There are various potential models for the $NN$ system \cite{ref-5a,ref-5b,ref-5c,ref-5d,ref-5}. 
Here we employ the Minnesota potential \cite{ref-5}.
It describes the deuteron only with S-wave component, as the D-wave contribution is effectively included in the central potential,
\begin{equation}
V_{ij}(r) = \left(  V_{R} + \frac{1}{2}(1+P^{\sigma}_{ij})V_{t} + \frac{1}{2}(1-P^{\sigma}_{ij})V_{s} \right) \left( \frac{1}{2}u + \frac{1}{2}(2-u)P^{r}_{ij} \right)
\label{eq:miNN1}
\end{equation}
\begin{equation}
V_{R}(r) = V_{0R}e^{-\kappa_{R}r^{2}}, \ V_{t}(r) = V_{0t}e^{-\kappa_{t}r^{2}}, \ V_{s}(r) = V_{0s}e^{-\kappa_{s}r^{2}}
\label{eq:miNN2}
\end{equation}
\begin{eqnarray}
V_{0R} = 200.0[{\rm MeV}], \ \kappa_{R} = 1.487[{\rm fm}^{-2}], \nonumber \\
V_{0t} = -178.0[{\rm MeV}], \ \kappa_{t} = 0.639[{\rm fm}^{-2}], \nonumber \\
V_{0s} = -91.85[{\rm MeV}], \ \kappa_{s} = 0.465[{\rm fm}^{-2}].
\label{eq:miNN3}
\end{eqnarray}

\begin{figure}
\begin{center}
\includegraphics[width=150mm,bb=0 0 700 600]{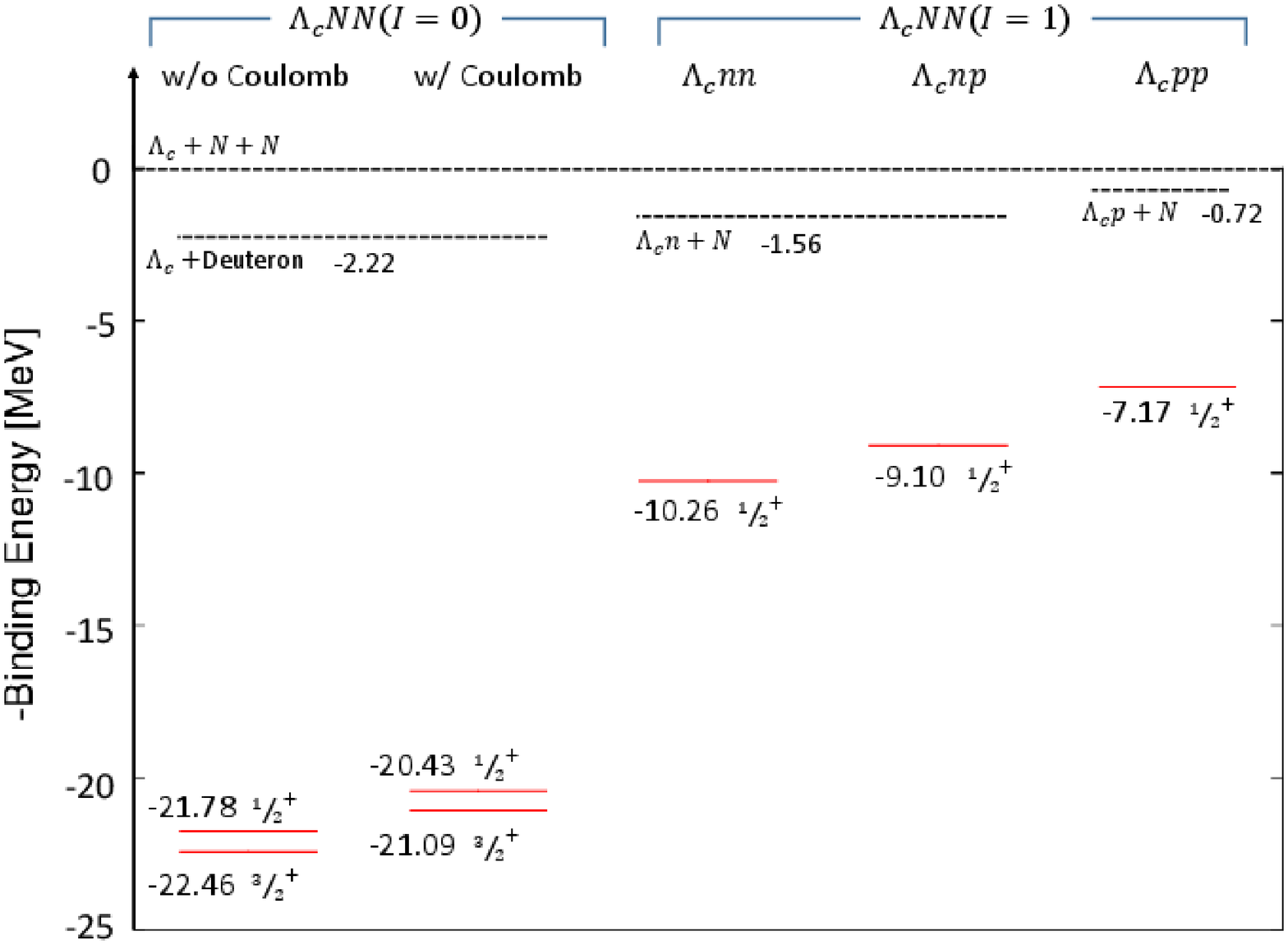}
\caption{$\Lambda_{c}NN$ binding energies. The dotted lines are the $\Lambda_{c}$-deuteron threshold that is $2.22$ MeV below the $\Lambda_{c}NN$ threshold, 
the threshold of the ($\Lambda_{c}n$) $1^{+}$ bound state plus a nucleon, i.e., $1.56$ MeV below $\Lambda_{c}NN$, 
and the threshold of the ($\Lambda_{c}p$) $1^{+}$ bound state plus a nucleon, i.e., $0.72$ MeV below $\Lambda_{c}NN$}
\label{gr:3b-be11}
\end{center}
\end{figure} 

We again use GEM \cite{ref-6} to solve the three-body problem.
The binding energies of the $\Lambda_{c}NN$ three-body system with the $\Lambda_{c}N$ effective potential are shown in Fig. \ref{gr:3b-be11}.
The left two plots show the results for $I=0$ without and with the Coulomb potential, respectively.
The others represent the $I=1$ results for $\Lambda_{c}nn$, $\Lambda_{c}np$, and $\Lambda_{c}pp$. 
For each system, we find a bound state.
Because the spin-spin interaction between $\Lambda_{c} $ and $N$ is weak ( Table \ref{tb:becal1} ), 
the binding energies of $J^{\pi}=\frac{1}{2}^{+}$ and $\frac{3}{2}^{+}$ are close 
to each other for the case of $I=0$.
These results agree well with the heavy quark spin symmetry \cite{ref-0}.
In the present calculation, we find that the $\frac{3}{2}^{+}$ state is lower in energy than the $\frac{1}{2}^{+}$ state by about $0.7$ MeV.
These results are consistent with a recent three-body calculation \cite{ref-01}, which also favors the $\frac{3}{2}^{+}$ $\Lambda_{c}NN$ as the lowest state.
Effect of the Coulomb potential between $\Lambda_{c}$ and proton is smaller than the difference between the $I=0$ and $1$ results.

\begin{table}[!h]
\begin{center}
\smallskip
\begin{tabular}{c|cccc} \hline
 & Binding energy [MeV] & (From threshold) & $r$ [fm] & $R$ [fm] \\ \hline
$\Lambda_{c}np$ w/o Coulomb $J^{\pi}=\frac{1}{2}^{+}$ & 21.78 & 19.56 & 1.91 & 1.34 \\
$\Lambda_{c}np$ w/o Coulomb $J^{\pi}=\frac{3}{2}^{+}$ & 22.46 & 20.24 & 1.90 & 1.32 \\ \hline
$\Lambda_{c}np$ w/ Coulomb $J^{\pi}=\frac{1}{2}^{+}$ & 20.43 & 18.21 & 1.93 & 1.36 \\
$\Lambda_{c}np$ w/ Coulomb $J^{\pi}=\frac{3}{2}^{+}$ & 21.09 & 18.87 & 1.91 & 1.34 \\ \hline
\end{tabular}
\caption{$\Lambda_{c}NN$ binding energies and particle distances for the $I=0$ bound states}
\label{tb:3brms11}
\end{center}
\end{table}

\begin{table}[!h]
\begin{center}
\smallskip
\begin{tabular}{c|cccc} \hline
 & Binding energy [MeV] & (From threshold) & $r$ [fm] & $R$ [fm] \\ \hline
$\Lambda_{c}nn$ $J^{\pi}=\frac{1}{2}^{+}$ & 10.26 & 8.70 & 2.62 & 1.64 \\
$\Lambda_{c}np$ $J^{\pi}=\frac{1}{2}^{+}$ & 9.10 & 7.54 & 2.67 & 1.68 \\
$\Lambda_{c}pp$ $J^{\pi}=\frac{1}{2}^{+}$ & 7.17 & 6.35 & 2.78 & 1.75 \\ \hline
\end{tabular}
\caption{$\Lambda_{c}NN$ binding energies and particle distances for the $I=1$ bound states}
\label{tb:3brms12}
\end{center}
\end{table}

Tables \ref{tb:3brms11} and \ref{tb:3brms12} show the mean distances in the bound state, 
where $r$ is the root mean square (rms) distance between the nucleons, and $R$ is rms distance between $\Lambda_{c}$ and the center of $NN$.
These Tables illustrate that the mean NN separations, $r$'s, are significantly smaller than the mean distance of $p-n$ in the deuteron, about 3.8 fm.
Namely, the $\Lambda_{c}$ attraction to the nucleon makes the $\Lambda_{c}NN$ system shrink.
It is also observed that $R$ is smaller than $r$.
Thus we can draw an intuitive picture that the nucleons go around $\Lambda_{c}$ sitting at the center in the $\Lambda_{c}NN$ system.
This property is also seen in the hypernuclear system, but the attractive force in the charmed nuclear system is stronger than that in the hypernuclear system.

We may check whether these results depend on the choice of the $NN$ potential.
Here we have used the Minnesota potential, which is known to have a weak repulsive core.
For comparison, we replace the Minnesota potential by the AV8$^{\prime}$ potential \cite{ref-8} 
and apply it to the $I=1$ $\Lambda_{c}nn$ system.
For the $I=0$ $\Lambda_{c}NN$ systems, the comparison is not appropriate because the AV8$^{\prime}$ contains the tensor force which make the D-wave components couple.

Then we obtain the binding energy BE$=11.93$ [MeV] and the rms distances as $r=2.41$ [fm] and $R=1.54$ [fm].
These results can be compared to the first line of Table \ref{tb:3brms12}.
One sees that they are consistent with each other.
Thus we conclude that the qualitative behavior of the $\Lambda_{c}NN$ bound state is independent of the choice of the $NN$ potential.

\section{Conclusion}
\label{sec:conc}

We have examined the interactions between the ground-state charmed baryons $\Lambda_c$, $\Sigma_c$ and $\Sigma_c^*$ and the nucleon $N$.
Potential models are proposed, which are composed of the long-range one-boson exchange (OBE) force 
and the short-range quark exchange force
based on the quark cluster model (QCM).
We also include the Coulomb potential between the charged baryons.
The parameters in the model, the cutoffs and the coupling constant of $\sigma$ meson, are determined so that the $NN$ data are reproduced in the same model. 
By fitting the deuteron binding energy and the S-wave scattering lengths, four parameter sets are obtained.

We have applied these potentials to the two-body $\Lambda_cN-\Sigma_cN-\Sigma_c^{*}N$ coupled system.
The coupled-channel Schr\"odinger equation is solved to a good precision by the Gaussian expansion method.
We find shallow bound $\Lambda_cN$ states both in $J^{\pi}=0^+$ and $1^+$ in two of the four parameter sets.
Those parameters are the cases that the ranges of the QCM repulsion are small and thus the quark exchange effect is weaker than the other two.
The difference between the $J^{\pi}=0^+$ ($\Lambda_c N(^1S_0)$) and $1^+$ ($\Lambda_c N(^3S_1)$) systems is small, 
which is a consequence of the heavy quark spin symmetry.

It is not surprising to have a bound state in these channels, because the $Y_c N$ interactions are similar to those of $YN$ except for the $K$ exchange
part, while the kinetic energy is suppressed by the heavier mass. Furthermore, the effect of the channel coupling to 
$\Sigma_c^* N$ is significant due to the strong tensor force coming from the one-pion exchange.
We then conclude that shallow $\Lambda_c N$ bound states may exist.

Encouraged by the possible existence of the two-body bound states, we further consider three-body $\Lambda_c NN$ bound states.
This is the lightest ``charmed nucleus'', which corresponds to the hyper-triton ($\Lambda pn$) bound state in the strangeness sector.
In order to simplify the calculation, in the present approach, we first construct an effective one-channel ($\Lambda_c N$) potential 
from the fully coupled ($\Lambda_c N-\Sigma_c N-\Sigma^*_c N$) calculation.
Using the effective one-channel potential, we have solved the three-body Schr\"odinger equation with the GEM and have obtained bound states
with the binding energy about 20-22 MeV for $I=0$, and 7-10 MeV for $I=1$ from the $\Lambda_{c} NN$ threshold.
The corresponding wave functions indicate that the $\Lambda_c$ baryon makes the size of the $NN$ system significantly smaller owing to attractive force.
In order to confirm these results, further studies with the full and explicit couplings with the $\Sigma_cN$ and $\Sigma^*_cN$ channels
will be necessary.

\section*{Acknowledgment}

S. M. is supported by the RIKEN Junior Research Associate Program.
This work is supported in part by JSPS KAKENHI Grant Nos. 25247036, and 24250294.
Y. R. L. was supported by the \textquotedblleft Program for Promoting the Enhancement of Research Universities \textquotedblright at Tokyo Institute of Technology in 2014 
and partly by NNSFC(No. 11275115).
This work is also supported by the Research Abroad and Invitational Program for the 
Promotion of International Joint Research, Category (C) and the International Physics 
Leadership Program at Tokyo Tech.

\vfill\pagebreak

\newpage

\appendix
\def\thesection{Appendix \Alph{section}}

\section{Spin Matrix Elements of $Y_{c}N$ OBE potential}
\label{sec:appa}

In this Appendix, we present the definitions of the spin dependent operators in the potentials.
$\bm{\mathcal{O}}_{spin}$ is the spin-spin operator between $Y_{c}$ and $N$:
\begin{equation}
\bm{\mathcal{O}}_{spin} = \bm{\mathcal{O}}_1\cdot \bm{\sigma}_2,
\end{equation}
where $\bm{\sigma}_2$ is the Pauli matrix of the nucleon spin, and $\bm{\mathcal{O}}_1$ is the spin operator of the charmed baryon,
\begin{equation}
\bm{\mathcal{O}}_1 = \left\{ 
\begin{array}{cl}
\bm{\sigma}_1 & \rm{for} \ \Lambda_{c} \ \rm{and} \ \Sigma_{c} \\
\bm{\bar{\Sigma}}_1 & \rm{for \ the \ transition \ from} \ \Lambda_{c} \ \rm{and} \ \Sigma_{c} \ \rm{to} \ \Sigma_{c}^{*} \\
\bm{\Sigma}_1 & \rm{for} \ \Sigma_{c}^{*} \\
\end{array} \right.
\end{equation}

The transition spin $\bm{\bar{\Sigma}}$ is defined as $u^{\mu} \equiv \bm{\bar{\Sigma}} \Phi$, where $u^{\mu}$ is the Rarita Schwinger field, 
and $\Phi$ is the spin wave functions of $\Sigma_{c}^{*}$, 
\begin{equation}
\Phi(3/2) = \left( \begin{array}{c}1 \\ 0 \\ 0 \\ 0 \\ \end{array} \right), \ 
\Phi(1/2) = \left( \begin{array}{c}0 \\ 1 \\ 0 \\ 0 \\ \end{array} \right), \ 
\Phi(-1/2) = \left( \begin{array}{c}0 \\ 0 \\ 1 \\ 0 \\ \end{array} \right), \ 
\Phi(-3/2) = \left( \begin{array}{c}0 \\ 0 \\ 0 \\ 1 \\ \end{array} \right). \ 
\end{equation}
Then we calculate the transition spin explicitly, 
\begin{equation}
\begin{array}{rcl}
\bm{\bar{\Sigma}}^\dag & = & -\frac{1}{\sqrt{2}}\left( \bm{\bar{\Sigma}}_{x}^\dag + i\bm{\bar{\Sigma}}_{y}^\dag \right) + \frac{1}{\sqrt{2}}\left( \bm{\bar{\Sigma}}_{x}^\dag - i\bm{\bar{\Sigma}}_{y}^\dag \right) + \bm{\bar{\Sigma}}_{z}^\dag \\
 & = & \left(\begin{array}{cccc}1&0&0&0\\0&\sqrt{\frac13}&0&0\end{array}\right)+\left(\begin{array}{cccc}0&\sqrt\frac23&0&0\\0&0&\sqrt\frac23&0\end{array}\right)+\left(\begin{array}{cccc}0&0&\sqrt{\frac13}&0\\0&0&0&1\end{array}\right) \\
\end{array}
\end{equation}
\begin{equation}
\bm{\bar{\Sigma}} \bm{\bar{\Sigma}}^\dag=-I_{4\times4}, \quad 
\bm{\bar{\Sigma}}^\dag \bm{\bar{\Sigma}}=-2I_{2\times2}.
\end{equation}
With $\hat{\bm{e}}(\lambda=+1)=-\frac{1}{\sqrt2}(1,i,0)$, $\hat{\bm{e}}(\lambda=-1)=\frac{1}{\sqrt2}(1,-i,0)$, $\hat{\bm{e}}(\lambda=0)=(0,0,1)$, and $S_{t\mu}^\dag=(0,\vec{S}_t^\dag)$, one has
\begin{eqnarray}
\bm{\bar{\Sigma}}(1,+1)=\left(\begin{array}{cc}1&0\\0&\sqrt{\frac13}\\0&0\\0&0\end{array}\right),\qquad
\bm{\bar{\Sigma}}(1,-1)=\left(\begin{array}{cc}0&0\\0&0\\\sqrt{\frac13}&0\\0&1\end{array}\right),\qquad
\bm{\bar{\Sigma}}(1,0)=\left(\begin{array}{cc}0&0\\\sqrt{\frac23}&0\\0&\sqrt{\frac23}\\0&0\end{array}\right).
\label{eq:trs}
\end{eqnarray}

Next, we define the spin operator of $\Sigma_{c}^{*}$, 
\begin{equation}
\begin{array}{c}
\bm{\Sigma} = -S_{t\mu}^{\dag}\bm{\sigma}S_{t}^{\mu} = (S_{t}^{\dag})^{j}\bm{\sigma}(S_{t})^{j}, \\
\bm{S}(\Sigma_{c}^{*}) = \frac{3}{2}\bm{\Sigma}.\\
\end{array}
\end{equation}
With Eq. (\ref{eq:trs}), the explicit matrices are
\begin{equation}
\begin{array}{c}
\bm{\sigma}_{rs}(1,+1) = -\frac{1}{\sqrt{2}}\left( \bm{\Sigma}_{x} + i\bm{\Sigma}_{y} \right) =-\left( 
\begin{array}{cccc}
0 & \sqrt{\frac{2}{3}} & 0 & 0 \\
0 & 0 & \frac{2\sqrt{2}}{3} & 0 \\
0 & 0 & 0 & \sqrt{\frac{2}{3}} \\
0 & 0 & 0 & 0 \\
\end{array} \right), \\
\bm{\sigma}_{rs}(1,-1) = \frac{1}{\sqrt{2}}\left( \bm{\Sigma}_{x} - i\bm{\Sigma}_{y} \right) = -\left( 
\begin{array}{cccc}
0 & 0 & 0 & 0 \\
\sqrt{\frac{2}{3}} & 0 & 0 & 0 \\
0 & \frac{2\sqrt{2}}{3} & 0 & 0 \\
0 & 0 & \sqrt{\frac{2}{3}} & 0 \\
\end{array} \right), \\
\bm{\sigma}_{rs}(1,0) = \bm{\Sigma}_{z} = -\left( 
\begin{array}{cccc}
1 & 0 & 0 & 0 \\
0 & \frac{1}{3} & 0 & 0 \\
0 & 0 & -\frac{1}{3} & 0 \\
0 & 0 & 0 & -1 \\
\end{array} \right). \\
\end{array}
\end{equation}

The tensor operators can be defined similarly as follows
\begin{eqnarray}
\rm{In} \  \Lambda_{c}  \ \rm{and} \  \Sigma_{c}  \ \rm{channels  \ : \ } \bm{\mathcal{O}}_{ten} = \frac{3(\bm{\sigma}_1 \cdot \bm{r})(\bm{\sigma}_2 \cdot \bm{r})}{r^2} - \bm{\sigma}_1 \cdot \bm{\sigma}_2, \nonumber \\
\rm{In} \  \Lambda_{c} \to \Sigma_{c}^{*}  \ \rm{and}  \ \Sigma_{c} \to \Sigma_{c}^{*}  \ \rm{channels  \ : \ } \bm{\mathcal{O}}_{ten} = \frac{3(\bm{\bar{\Sigma}} \cdot \bm{r})(\bm{\sigma}_2 \cdot \bm{r})}{r^2} - \bm{\bar{\Sigma}} \cdot \bm{\sigma}_2, \nonumber \\
\rm{In} \  \Sigma_{c}^{*}   \ \rm{diagonal  \ channels  \ : \ } \bm{\mathcal{O}}_{ten} = \frac{3(\bm{\Sigma} \cdot \bm{r})(\bm{\sigma}_2 \cdot \bm{r})}{r^2}-\bm{\Sigma} \cdot \bm{\sigma}_2,
\end{eqnarray}

The spin-orbit operator $\bm{\mathcal{O}}_{LS}$ is defined as
\begin{equation}
\bm{\mathcal{O}}_{LS} = \bm{L}\cdot \bm{\sigma}_2.
\end{equation}
$\bm{L}\cdot \bm{\mathcal{O}}_1$ is not included in the potential for this calculation.

\newpage

\subsection{$I=\frac12$, $J^{\pi}=0^+$ coupled system}

Tables \ref{tb:op1}-\ref{tb:op3} give the matrix elements of $\bm{\mathcal{O}}_{spin}$, $\bm{\mathcal{O}}_{ten}$, and $\bm{\mathcal{O}}_{LS}$ for the channels in $I=1/2$, $J^{\pi}=0^{+}$.
We label the relevant channels by $i$ and $j$, and tabulate the $ij$ component of the matrix elements $\left< \bm{\mathcal{O}} \right>_{ij}$.

\begin{table}[htb]
\begin{center}
\begin{tabular}{c|ccc}\hline
\backslashbox{i}{j} &  $\Lambda_cN(^1S_0)$   &   $\Sigma_cN(^1S_0)$   &    $\Sigma_c^*N(^5D_0)$ \\\hline
$\Lambda_cN(^1S_0)$ & $-3$ & $-3$  & 0\\
$\Sigma_cN(^1S_0)$  & $-3$ & $-3$  & 0\\
$\Sigma_c^*N(^5D_0)$& 0    & 0     & 1\\\hline
\end{tabular}
\caption{The matrix elements of the spin-spin operators $\left< \bm{\mathcal{O}}_{spin} \right>_{ij}$ for the $I=\frac12$, $J^{\pi}=0^+$ coupled system.}
\label{tb:op1}
\end{center}
\end{table}

\begin{table}[htb]
\begin{center}
\begin{tabular}{c|ccc}\hline
\backslashbox{i}{j} &  $\Lambda_cN(^1S_0)$   &   $\Sigma_cN(^1S_0)$   &    $\Sigma_c^*N(^5D_0)$ \\\hline
$\Lambda_cN(^1S_0)$ & 0 & 0  & $-\sqrt6$\\
$\Sigma_cN(^1S_0)$  & 0 & 0  & $-\sqrt6$\\
$\Sigma_c^*N(^5D_0)$& $-\sqrt6$    & $-\sqrt6$     & $-2$\\\hline
\end{tabular}
\caption{The matrix elements of the tensor operators $\left< \bm{\mathcal{O}}_{ten} \right>_{ij}$ for the $I=\frac12$, $J^{\pi}=0^+$ coupled system.}
\label{tb:op2}
\end{center}
\end{table}

\begin{table}[htb]
\begin{center}
\begin{tabular}{c|ccc}\hline
\backslashbox{i}{j} &  $\Lambda_cN(^1S_0)$   &   $\Sigma_cN(^1S_0)$   &    $\Sigma_c^*N(^5D_0)$ \\\hline
$\Lambda_cN(^1S_0)$ & 0 & 0  & 0\\
$\Sigma_cN(^1S_0)$  & 0 & 0  & 0\\
$\Sigma_c^*N(^5D_0)$& 0 & 0     & $-3$\\\hline
\end{tabular}
\caption{The matrix elements of the orbital-spin operators $\left< \bm{\mathcal{O}}_{LS} \right>_{ij}$ for the $I=\frac12$, $J^{\pi}=0^+$ coupled system.}
\label{tb:op3}
\end{center}
\end{table}

\newpage

\subsection{$I=\frac12$, $J^{\pi}=1^+$ coupled system}

For $I=1/2$, $J^{\pi}=1^{+}$, the matrix elements are given in Tables \ref{tb:op4}-\ref{tb:op6}.

\begin{table}[htb]
\begin{center}
\small
\begin{tabular}{c|ccccccc}\hline
\backslashbox{i}{j} & $\Lambda_cN(^3S_1)$&$\Sigma_cN(^3S_1)$&$\Sigma_c^*N(^3S_1)$&$\Lambda_cN(^3D_1)$&$\Sigma_cN(^3D_1)$&$\Sigma_c^*N(^3D_1)$&$\Sigma_c^*N(^5D_1)$ \\\hline
$\Lambda_cN(^3S_1)$ & 1 & 1  & $-\sqrt\frac83$ & 0 & 0 & 0 & 0\\
$\Sigma_cN(^3S_1)$  & 1 & 1  & $-\sqrt\frac83$ & 0 & 0 & 0 & 0\\
$\Sigma_c^*N(^3S_1)$& $-\sqrt\frac83$&$-\sqrt\frac83$&$-\frac53$&0&0&0&0 \\
$\Lambda_c N(^3D_1)$& 0 & 0  & 0 & 1 & 1 &$-\sqrt\frac83$&0\\
$\Sigma_cN(^3D_1)$  & 0 & 0  & 0 & 1 & 1 &$-\sqrt\frac83$&0\\
$\Sigma_c^*N(^3D_1)$& 0 & 0  & 0 &$-\sqrt\frac83$&$-\sqrt\frac83$&$-\frac53$&0\\
$\Sigma_c^*N(^5D_1)$& 0 & 0  & 0 &0 & 0&0&1\\\hline
\end{tabular}
\normalsize
\caption{The matrix elements of the spin-spin operators $\left< \bm{\mathcal{O}}_{spin} \right>_{ij}$ for the $I=\frac12$, $J^{\pi}=1^+$ coupled system.}
\label{tb:op4}
\end{center}
\end{table}

\begin{table}[htb]
\begin{center}
\small
\begin{tabular}{c|ccccccc}\hline
\backslashbox{i}{j} & $\Lambda_cN(^3S_1)$&$\Sigma_cN(^3S_1)$&$\Sigma_c^*N(^3S_1)$&$\Lambda_cN(^3D_1)$&$\Sigma_cN(^3D_1)$&$\Sigma_c^*N(^3D_1)$&$\Sigma_c^*N(^5D_1)$ \\\hline
$\Lambda_cN(^3S_1)$ &0&0&0&$\sqrt8$&$\sqrt8$&$\frac{1}{\sqrt3}$&$\sqrt3$ \\
$\Sigma_cN(^3S_1)$  &0&0&0&$\sqrt8$&$\sqrt8$&$\frac{1}{\sqrt3}$&$\sqrt3$ \\
$\Sigma_c^*N(^3S_1)$&0&0&0&$\frac{1}{\sqrt3}$&$\frac{1}{\sqrt3}$&$-\frac{\sqrt2}{3}$&$-\sqrt2$  \\
$\Lambda_c N(^3D_1)$&$\sqrt8$&$\sqrt8$&$\frac{1}{\sqrt3}$&$-2$&$-2$&$-\frac{1}{\sqrt6}$&$\sqrt\frac32$   \\
$\Sigma_cN(^3D_1)$  &$\sqrt8$&$\sqrt8$&$\frac{1}{\sqrt3}$&$-2$&$-2$&$-\frac{1}{\sqrt6}$&$\sqrt\frac32$\\
$\Sigma_c^*N(^3D_1)$&$\frac{1}{\sqrt3}$&$\frac{1}{\sqrt3}$&$-\frac{\sqrt2}{3}$&$-\frac{1}{\sqrt6}$&$-\frac{1}{\sqrt6}$&$\frac13$&$-1$ \\
$\Sigma_c^*N(^5D_1)$&$\sqrt3$&$\sqrt3$&$-\sqrt2$&$\sqrt\frac32$&$\sqrt\frac32$&$-1$&$-1$\\\hline
\end{tabular}
\normalsize
\caption{The matrix elements of the tensor operators $\left< \bm{\mathcal{O}}_{ten} \right>_{ij}$ for the $I=\frac12$, $J^{\pi}=1^+$ coupled system.}
\label{tb:op5}
\end{center}
\end{table}

\begin{table}[htb]
\begin{center}
\small
\begin{tabular}{c|ccccccc}\hline
\backslashbox{i}{j} & $\Lambda_cN(^3S_1)$&$\Sigma_cN(^3S_1)$&$\Sigma_c^*N(^3S_1)$&$\Lambda_cN(^3D_1)$&$\Sigma_cN(^3D_1)$&$\Sigma_c^*N(^3D_1)$&$\Sigma_c^*N(^5D_1)$ \\\hline
$\Lambda_cN(^3S_1)$ & 0 & 0  & 0 & 0 & 0 & 0 & 0\\
$\Sigma_cN(^3S_1)$  & 0 & 0  & 0 & 0 & 0 & 0 & 0\\
$\Sigma_c^*N(^3S_1)$& 0 & 0  & 0 & 0 & 0 & 0 & 0 \\
$\Lambda_c N(^3D_1)$& 0 & 0  & 0 &$-3$&$-3$& 0 & 0 \\
$\Sigma_cN(^3D_1)$  & 0 & 0  & 0 &$-3$&$-3$& 0 & 0 \\
$\Sigma_c^*N(^3D_1)$&0&0&0& 0 & 0 &$\frac32$&$-\frac32$\\
$\Sigma_c^*N(^5D_1)$&0&0&0& 0 & 0 &$-\frac32$&$-\frac52$\\\hline
\end{tabular}
\normalsize
\caption{The matrix elements of the orbital-spin operators $\left< \bm{\mathcal{O}}_{LS} \right>_{ij}$ for the $I=\frac12$, $J^{\pi}=1^+$ coupled system.}
\label{tb:op6}
\end{center}
\end{table}

\newpage

\section{Definitions of the radial functions and coupling constants}
\label{sec:appb}

\subsection{The explicit forms of the radial functions}

The Yukawa potential functions in Eq. (\ref{eq:pot2}) are defined as
\begin{equation}
Y(x) = \frac{e^{-x}}{x}, \nonumber
\end{equation}
\begin{equation}
Z(x) = (\frac{1}{x} + \frac{1}{x^{2}})Y(x), \nonumber \\
\end{equation}
\begin{equation}
H(x) = (1 + \frac{3}{x} + \frac{3}{x^{2}})Y(x), \nonumber \\
\end{equation}
\begin{equation}
Y_{1}(m, \Lambda ,r) = Y(mr) - \left( \frac{\Lambda}{m} \right) Y(\Lambda r) - \frac{\Lambda^{2} - m^{2}}{2m\Lambda}e^{-\Lambda r}, \nonumber \\
\end{equation}
\begin{equation}
Y_{3}(m, \Lambda ,r) = Y(mr) - \left( \frac{\Lambda}{m} \right) Y(\Lambda r) - \frac{(\Lambda^{2} - m^{2})\Lambda}{2m^{3}}e^{-\Lambda r}, \nonumber \\
\end{equation}
\begin{equation}
Z_{3}(m, \Lambda ,r) = Z(mr) - \left( \frac{\Lambda}{m} \right)^{3}Z(\Lambda r) - \frac{(\Lambda^{2} - m^{2})\Lambda}{2m^{3}}Y(\Lambda r), \nonumber \\
\end{equation}
\begin{equation}
H_{3}(m, \Lambda ,r) = H(mr) - \left( \frac{\Lambda}{m} \right)^{3}H(\Lambda r) - \frac{(\Lambda^{2} - m^{2})\Lambda}{2m^{3}}Y(\Lambda r) - \frac{(\Lambda^{2} - m^{2})\Lambda}{2m^{3}}e^{-\Lambda r}. \\
\label{eq:pot1}
\end{equation}
In Eq. (\ref{eq:pot1}), $\Lambda$ is a cutoff parameter introduced in the monopole type form factor
\begin{equation}
F(q)=\frac{\Lambda^{2}-m^{2}}{\Lambda^{2}-q^{2}},
\end{equation}
where $m$ is the mass of the exchanged meson and $q$ is the 4-dimensional momentum of the meson.
We use the following values of the meson masses: $m_{\pi}=137.27$[MeV], \ $m_{\sigma}=600.0$[MeV].


\subsection{The coupling constants for the meson-exchange potentials}

The coupling constants are given in Tables \ref{tb:cp} and \ref{tb:cs},

\begin{table}[!h]
\begin{tabular}{cc}
\begin{minipage}{0.5\hsize}
\begin{center}
\begin{tabular}{c|ccc}
$C_{\pi}$  &  $\Lambda_cN$   &   $\Sigma_cN$   &    $\Sigma_c^*N$ \\ \hline
$\Lambda_cN$ & 0 & $(-\frac{\sqrt{6}}{2}g_{2}g_{A})$ & $-(\frac{\sqrt{6}}{2}g_{4}g_{A})$ \\
$\Sigma_cN$  &  & $(-g_{1}g_{A})$ & $(-g_{3}g_{A})$ \\
$\Sigma_c^*N$ &  &  & $(g_{5}g_{A})$ \\
\end{tabular}
\caption{$C_{\pi}$ for the $Y_{c}N$ systems}
\label{tb:cp}
\end{center}
\end{minipage}
\begin{minipage}{0.5\hsize}
\begin{center}
\begin{tabular}{c|ccc}
$C_{\sigma}$  &  $\Lambda_cN$   &   $\Sigma_cN$   &    $\Sigma_c^*N$ \\ \hline
$\Lambda_cN$ & $(2l_{B}h_{\sigma})$ &  &  \\
$\Sigma_cN$  &  & $(-l_{S}h_{\sigma})$  &  \\
$\Sigma_c^*N$ &  &  & $(-l_{S}h_{\sigma})$ \\
\end{tabular}
\caption{$C_{\sigma}$ for the $Y_{c}N$ systems}
\label{tb:cs}
\end{center}
\end{minipage}
\end{tabular}
\end{table}
with the following values \cite{ref-af,ref-ag,ref-ah,ref-ai,ref-aj,ref-ak}:
\begin{table}[!h]
\begin{center}
\begin{tabular}{c}
$g_{2}=-0.598$ ,\  $g_{4}=0.999$ ,\  $g_{1}=\frac{\sqrt{8}}{3}g_{4}$ ,\  $g_{3}=\sqrt{\frac{2}{3}}g_{4}$ ,\  $g_{5}=-\sqrt{2}g_{4}$ , \\
$g_{A}=1.25$ ,\  $l_{B}=-3.1$ ,\  $l_{S}=-2l_{B}$,
\end{tabular}
\end{center}
\end{table}
and $h_{\sigma}$ is a free parameter to be determined in sect. \ref{sec:potcom}
The explicit values of the coupling constants are given in Table \ref{tb:cpp} and \ref{tb:css}
For the NN potential, we note that the function $Y_{3}(m, \Lambda ,r)$ is used in place of $Y_{1}(m, \Lambda ,r)$ in the $\pi$ exchange potential
and we adopt the coupling constants given in Tables \ref{tb:ncp} and \ref{tb:ncs}.
\begin{table}[!t]
\begin{tabular}{cc}
\begin{minipage}{0.5\hsize}
\begin{center}
\begin{tabular}{c|ccc}
$C_{\pi}$  &  $\Lambda_cN$   &   $\Sigma_cN$   &    $\Sigma_c^*N$ \\ \hline
$\Lambda_cN$ & 0 & $0.92$ & $-1.53$ \\
$\Sigma_cN$  &  & $-1.18$ & $-1.02$ \\
$\Sigma_c^*N$ &  &  & $-1.77$ \\
\end{tabular}
\caption{$Y_{c}N$ potential $C_{\pi}$ values}
\label{tb:cpp}
\end{center}
\end{minipage}
\begin{minipage}{0.5\hsize}
\begin{center}
\begin{tabular}{c|ccc}
$C_{\sigma}$  &  $\Lambda_cN$   &   $\Sigma_cN$   &    $\Sigma_c^*N$ \\ \hline
$\Lambda_cN$ & $(-6.2h_{\sigma})$ &  &  \\
$\Sigma_cN$  &  & $(-6.2h_{\sigma})$  &  \\
$\Sigma_c^*N$ &  &  & $(-6.2h_{\sigma})$ \\
\end{tabular}
\caption{$Y_{c}N$ potential $C_{\sigma}$ values}
\label{tb:css}
\end{center}
\end{minipage}
\\
\vspace{3cm} \\
\begin{minipage}{0.5\hsize}
\begin{center}
\begin{tabular}{c|cc}
$C_{\pi}$  &  $NN(I=0)$   &   $NN(I=1)$  \\ \hline
$NN(I=0)$ & $\frac{1}{2}g_{A}^2$ &  \\
$NN(I=1)$ &  & $\frac{3}{2}g_{A}^2$ \\
\end{tabular}
\caption{$C_{\pi}$ for the $NN$ systems}
\label{tb:ncp}
\end{center}
\end{minipage}
\begin{minipage}{0.5\hsize}
\begin{center}
\begin{tabular}{c|cc}
$C_{\sigma}$  &  $NN(I=0)$   &   $NN(I=1)$  \\ \hline
$NN(I=0)$ & $-h_{\sigma}^{2}$ &  \\
$NN(I=1)$ &  & $-h_{\sigma}^{2}$ \\
\end{tabular}
\caption{$C_{\sigma}$ for the $NN$ systems}
\label{tb:ncs}
\end{center}
\end{minipage}
\end{tabular}
\end{table}

\clearpage

\section{$Y_{c}N$ potentials}
\label{sec:appc}

Here we show the shapes of the $Y_{c}N$ potentials in the present study.
The potentials for $J^{\pi}=0^{+}$ are given in Figs. \ref{gr:potctnn-011} - \ref{gr:potctnn-023} in sect. \ref{sec:potcom}.

\subsection{Components of the $Y_{c}n$ CTNN-a potential ($J^{\pi}=0^{+}$)}
\label{appc:1}

\begin{figure}[!h]
\begin{tabular}{cc}
\begin{minipage}{0.5\hsize}
\begin{center}
\includegraphics[width=75mm]{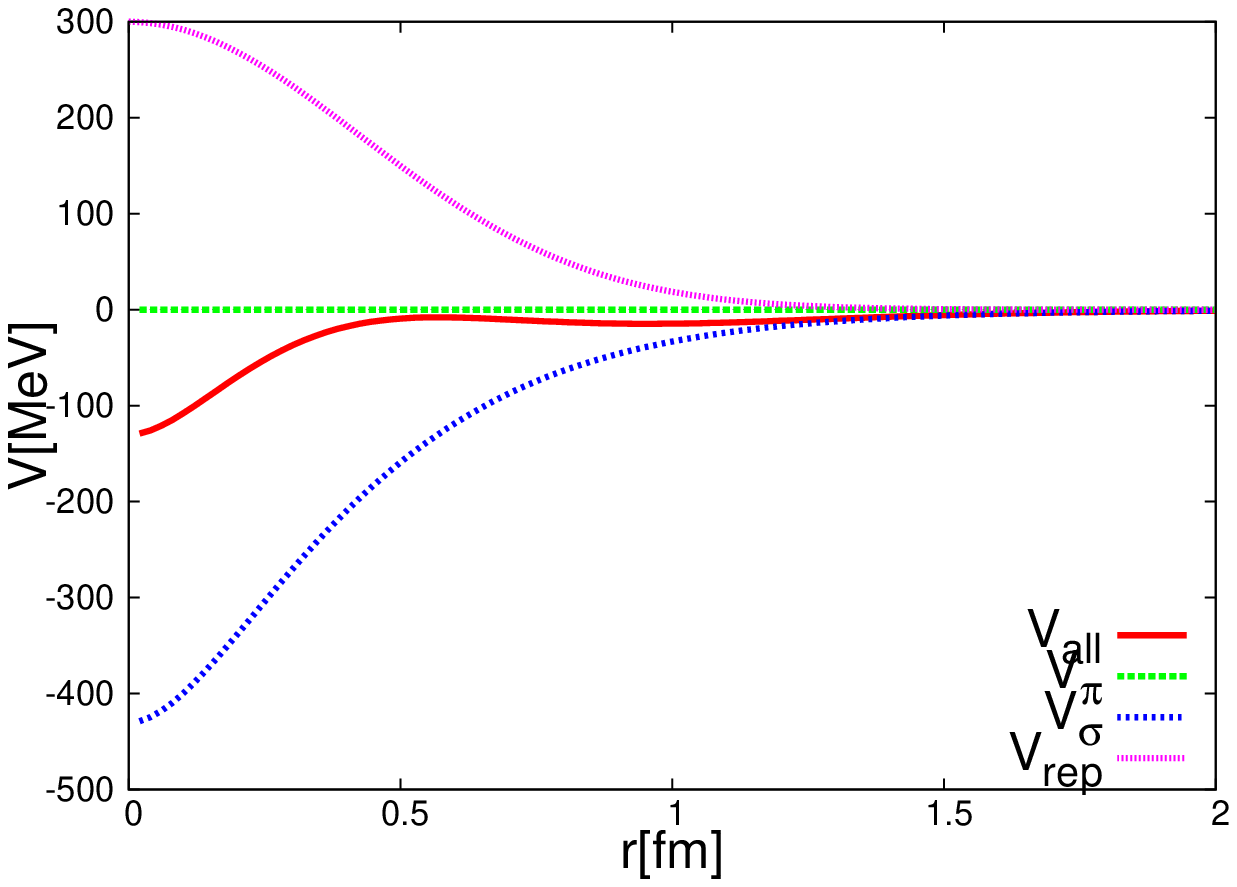}
\caption{$Y_{c}N$-CTNN potential for $\Lambda_{c}N$ single channel}
\label{grp:potctnn-011}
\end{center}
\end{minipage}
\begin{minipage}{0.5\hsize}
\begin{center}
\includegraphics[width=75mm]{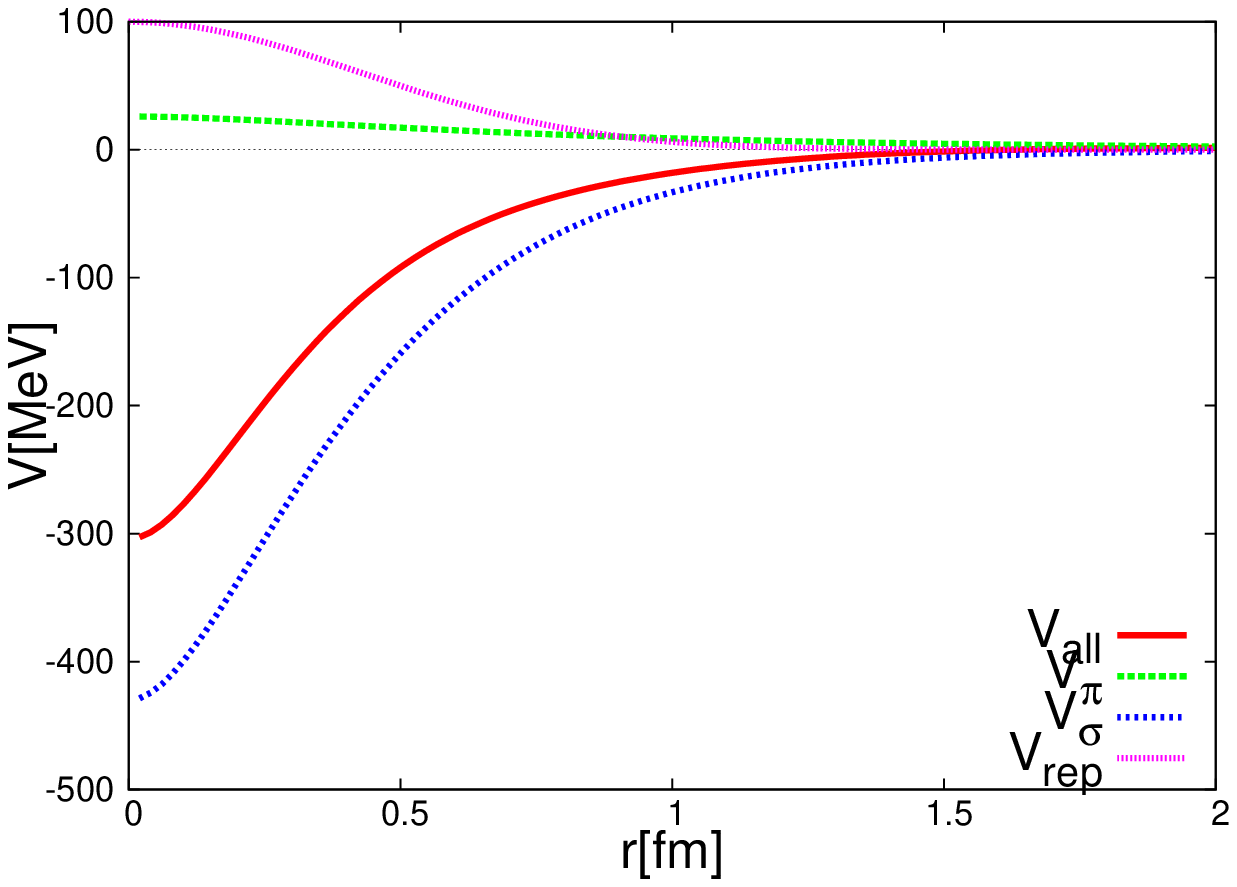}
\caption{$Y_{c}N$-CTNN potential for $\Sigma_{c}N$ single channel}
\label{grp:potctnn-022}
\end{center}
\end{minipage}
\end{tabular}
\end{figure}

\begin{figure}[!h]
\begin{center}
\includegraphics[width=75mm]{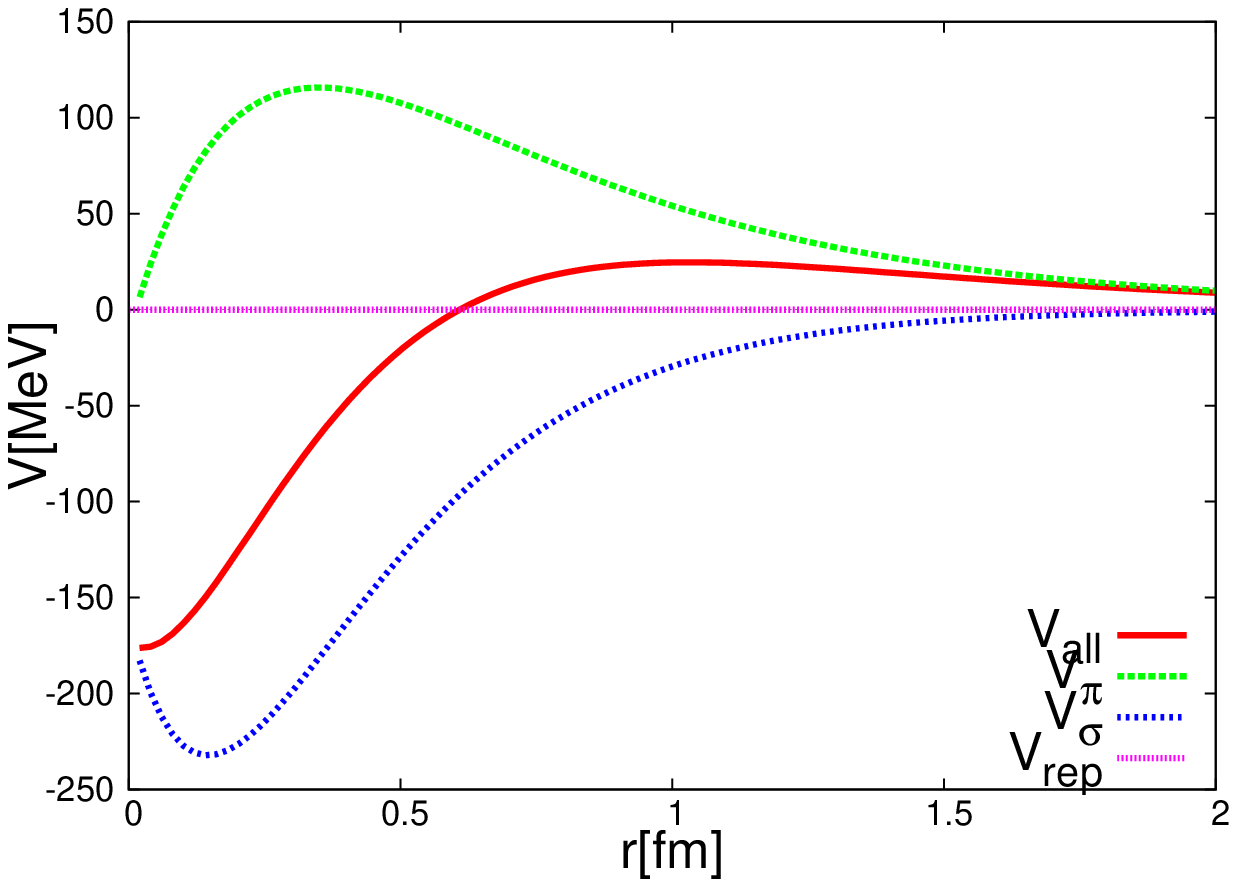}
\caption{$Y_{c}N$-CTNN potential for $\Sigma_{c}^{*}N$ single channel}
\label{grp:potctnn-033}
\end{center}
\end{figure}

The individual contributions of one-pion exchange, one $\sigma$ exchange, and QCM repulsion are given 
in Figs. \ref{grp:potctnn-011} - \ref{grp:potctnn-033} for $J^{\pi}=0^{+}$ CTNN-a potential.
One sees that one $\sigma$ exchange is strongly attractive, so that even with the QCM repulsion the total potential becomes attractive.
The similar behavior is also observed for CTNN-b,c,d potentials.


\clearpage

\subsection{$Y_{c}N$ potentials($J^{\pi}=1^{+}$)}
\label{appc:2}

\begin{figure}[!h]
\begin{tabular}{cc}
\begin{minipage}{0.5\hsize}
\begin{center}
\includegraphics[width=75mm]{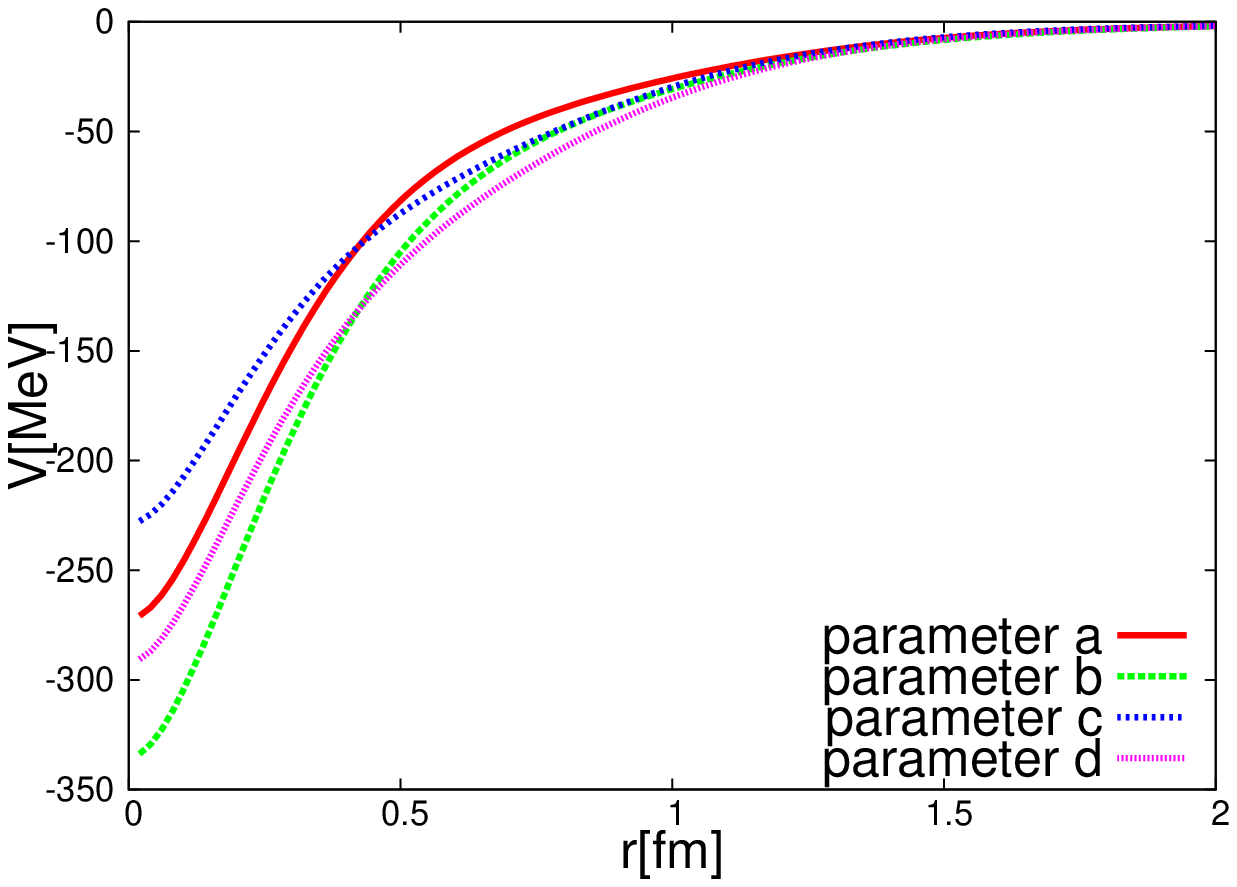}
\caption{$\Sigma_{c}N(^{3}S_{1})$ diagonal potential.}
\label{gr:potctnn-122}
\end{center}
\end{minipage}
\begin{minipage}{0.5\hsize}
\begin{center}
\includegraphics[width=75mm]{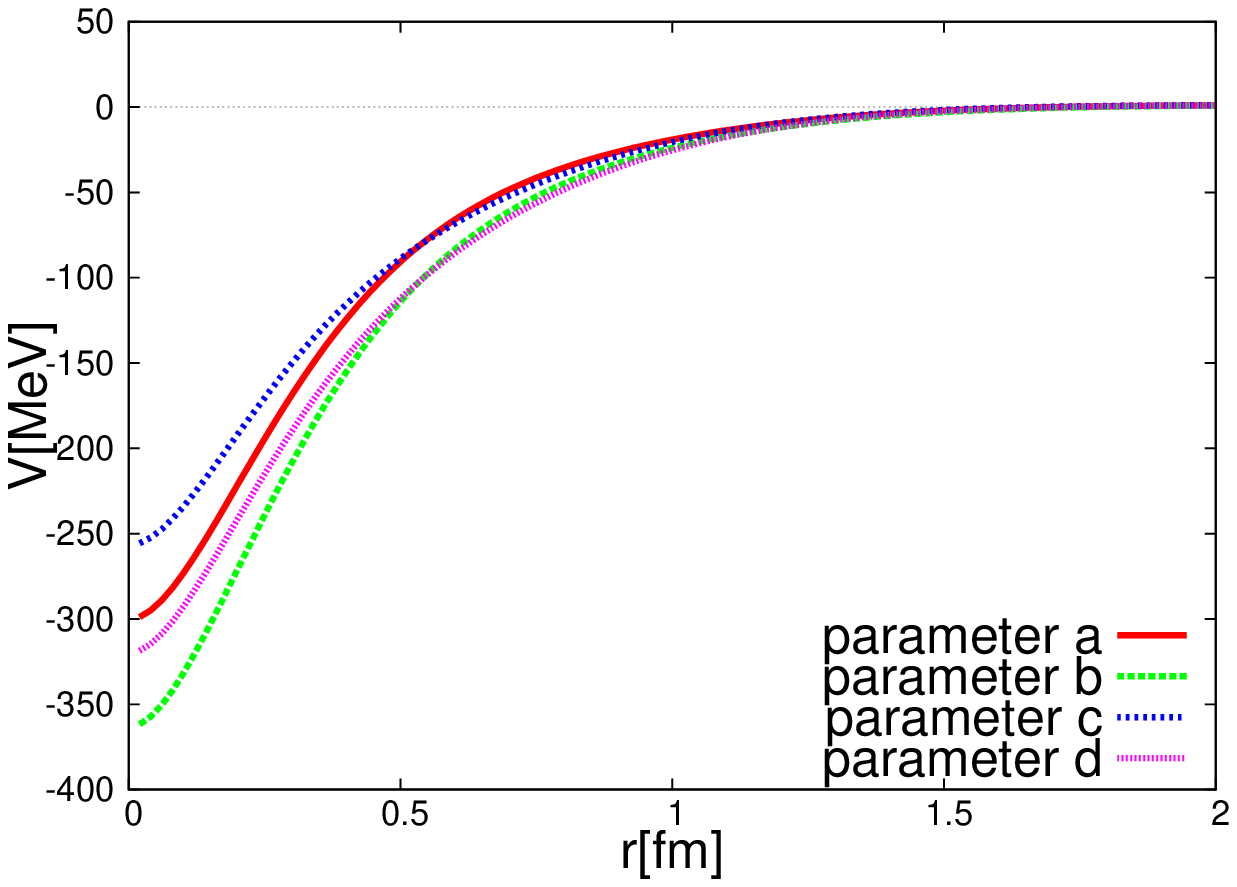}
\caption{$\Sigma_{c}^{*}N(^{3}S_{1})$ diagonal potential.}
\label{gr:potctnn-133}
\end{center}
\end{minipage}
\\
\begin{minipage}{0.5\hsize}
\begin{center}
\includegraphics[width=75mm]{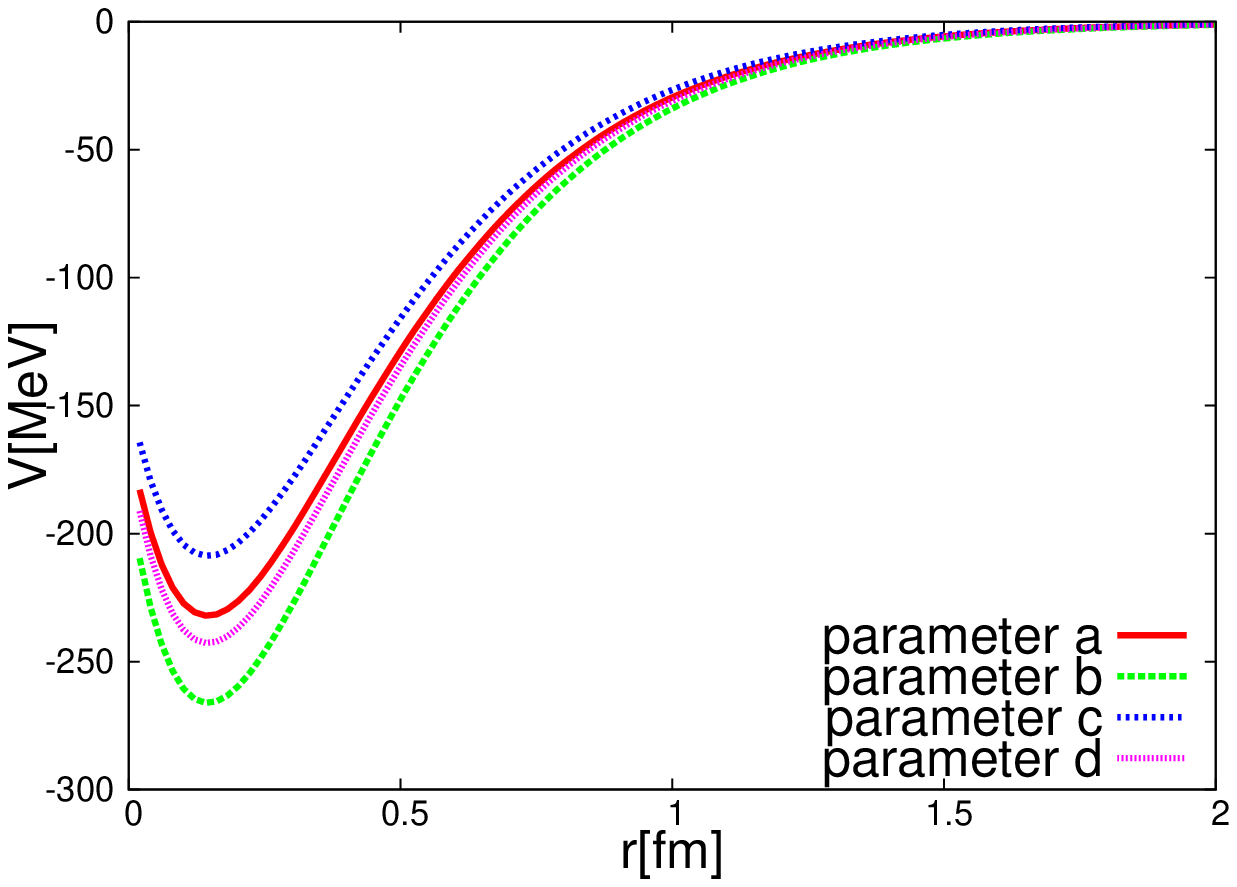}
\caption{$\Lambda_{c}N(^{3}D_{1})$ diagonal potential.}
\label{gr:potctnn-144}
\end{center}
\end{minipage}
\begin{minipage}{0.5\hsize}
\begin{center}
\includegraphics[width=75mm]{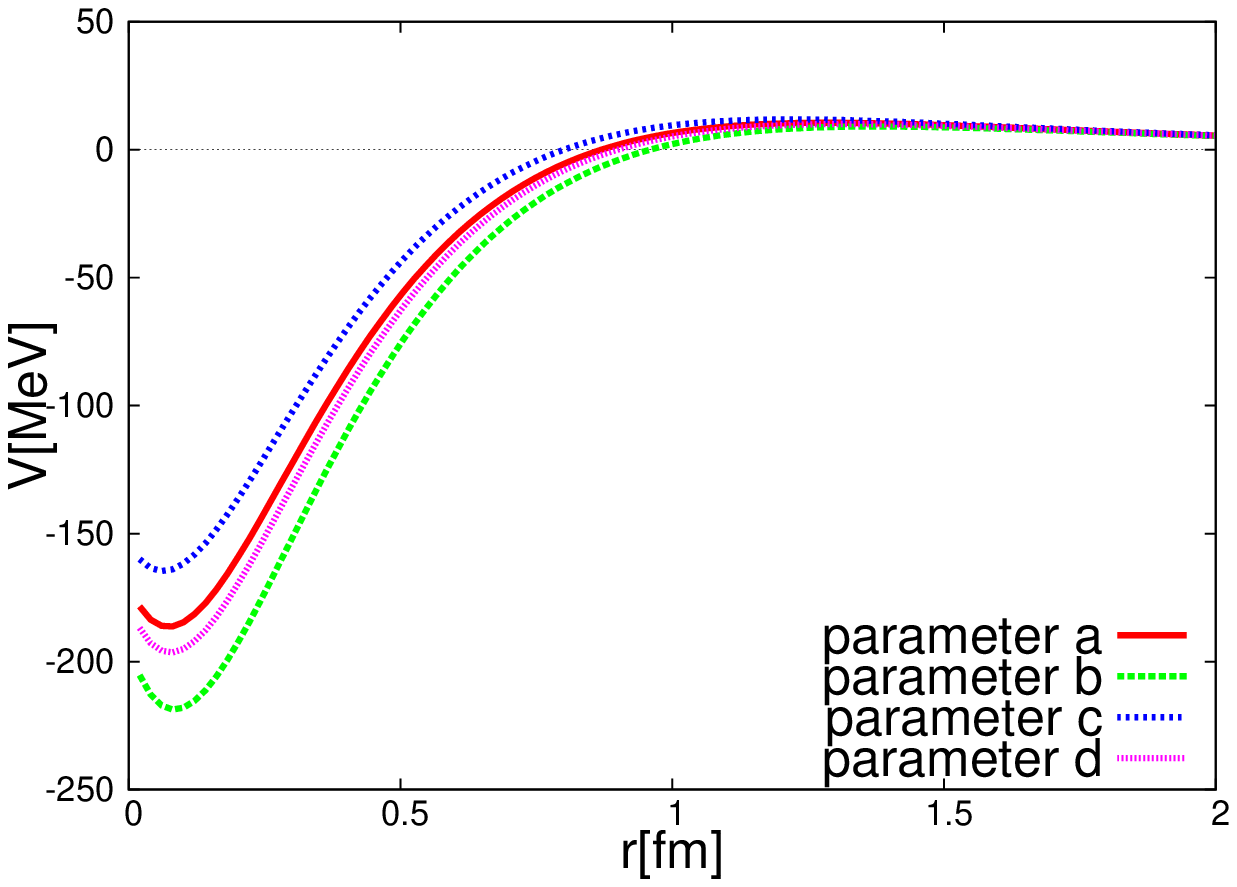}
\caption{$\Sigma_{c}N(^{3}D_{1})$ diagonal potential.}
\label{gr:potctnn-155}
\end{center}
\end{minipage}
\\
\begin{minipage}{0.5\hsize}
\begin{center}
\includegraphics[width=75mm]{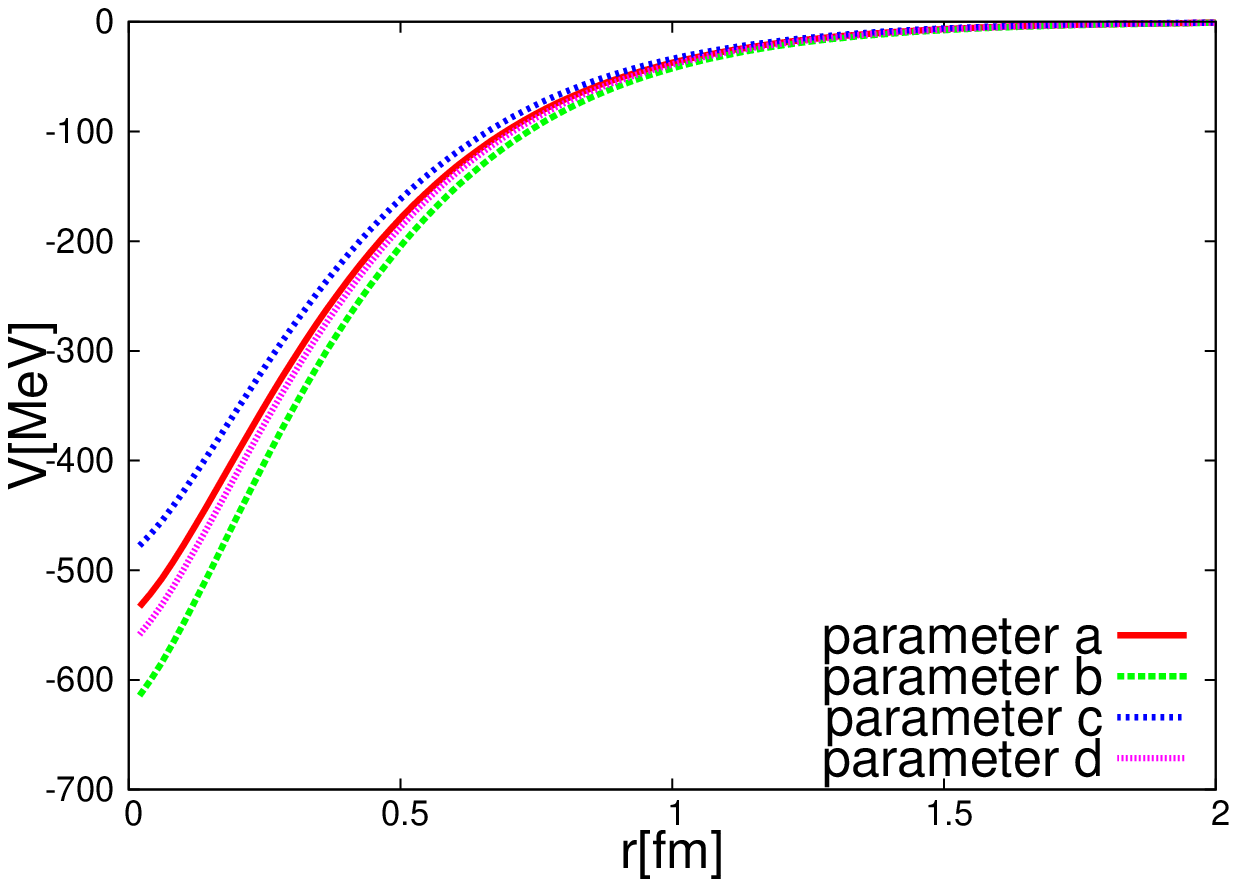}
\caption{$\Sigma_{c}^{*}N(^{3}D_{1})$ diagonal potential.}
\label{gr:potctnn-166}
\end{center}
\end{minipage}
\begin{minipage}{0.5\hsize}
\begin{center}
\includegraphics[width=75mm]{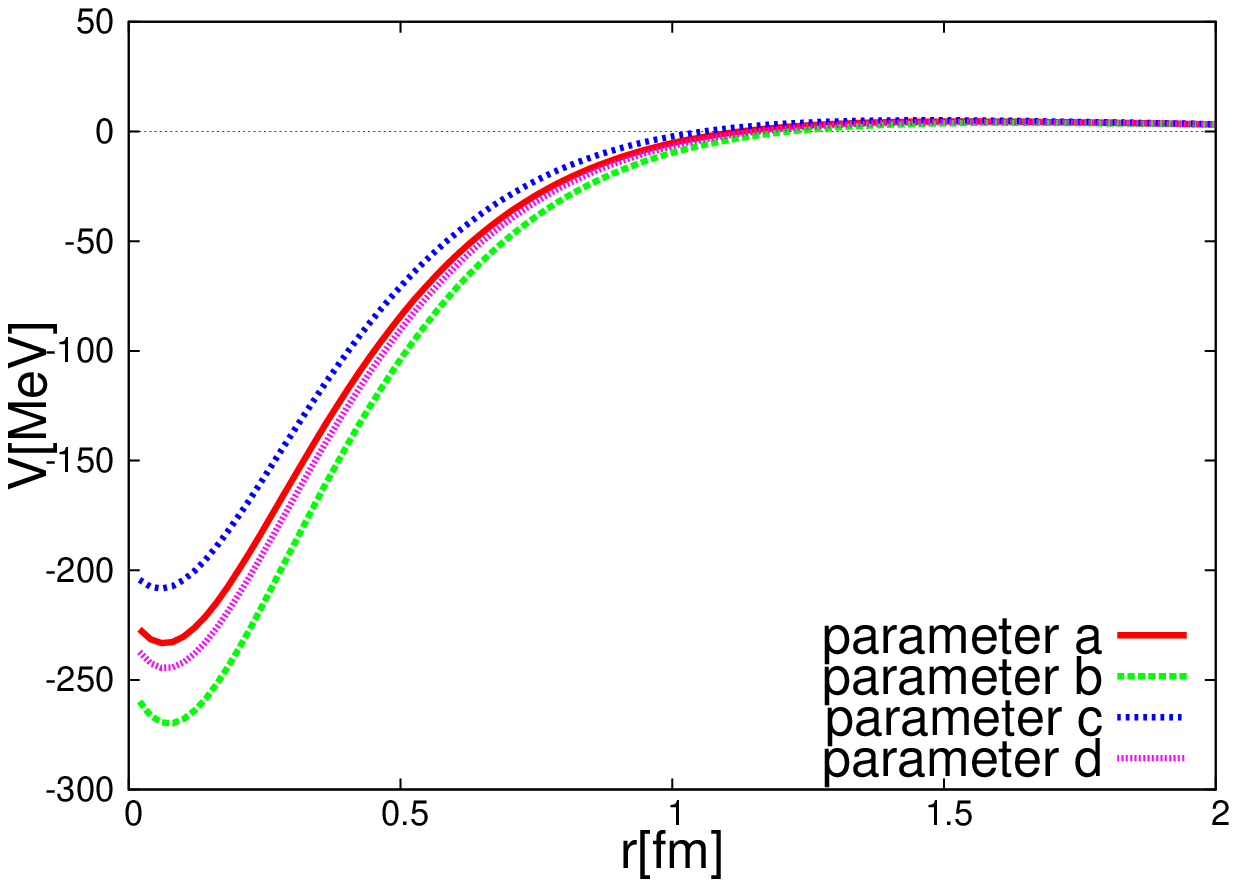}
\caption{$\Sigma_{c}^{*}N(^{5}D_{1})$ diagonal potential.}
\label{gr:potctnn-177}
\end{center}
\end{minipage}
\end{tabular}
\end{figure}

\begin{figure}[!h]
\begin{tabular}{cc}
\begin{minipage}{0.5\hsize}
\begin{center}
\includegraphics[width=75mm]{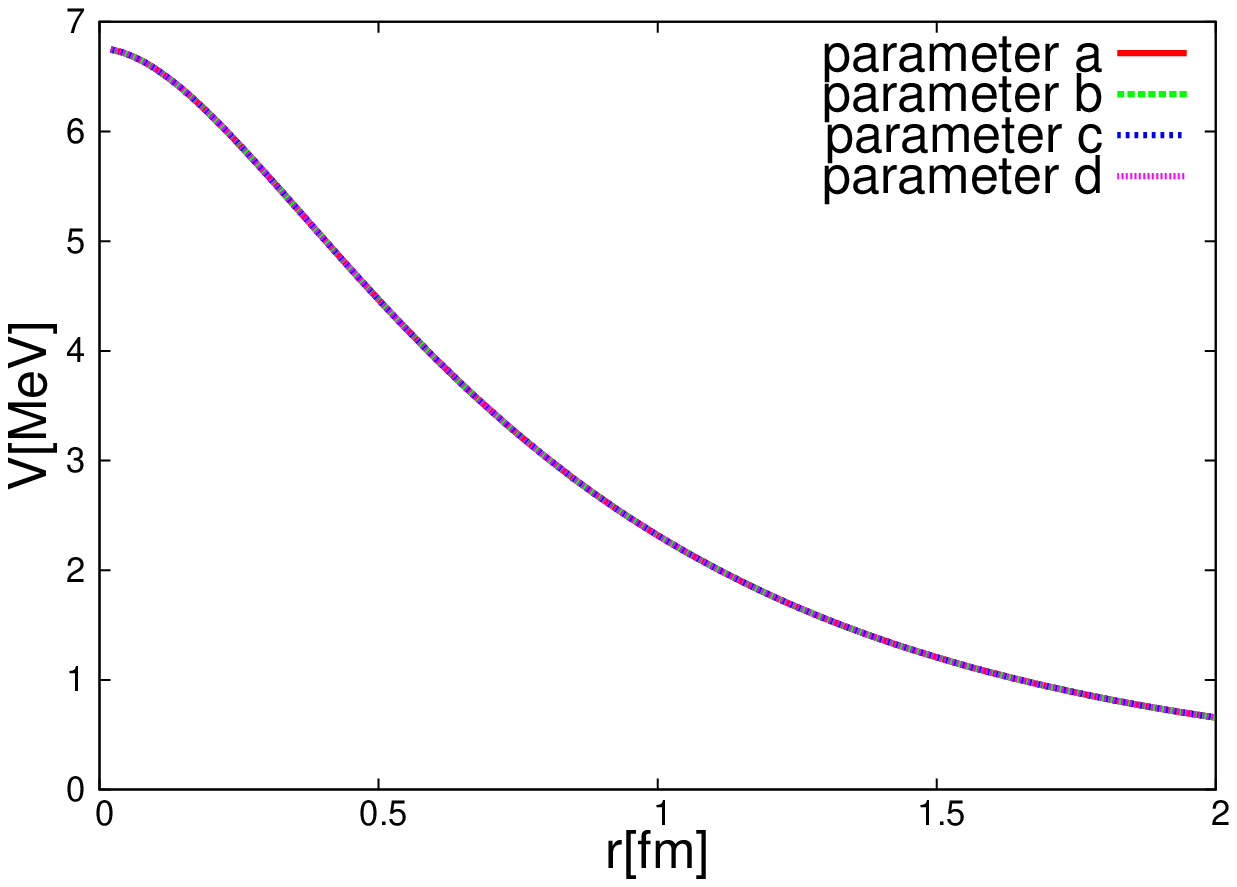}
\caption{$\Lambda_{c}N(^{3}S_{1})$-$\Sigma_{c}N(^{3}S_{1})$.}
\label{gr:potctnn-112}
\end{center}
\end{minipage}
\begin{minipage}{0.5\hsize}
\begin{center}
\includegraphics[width=75mm]{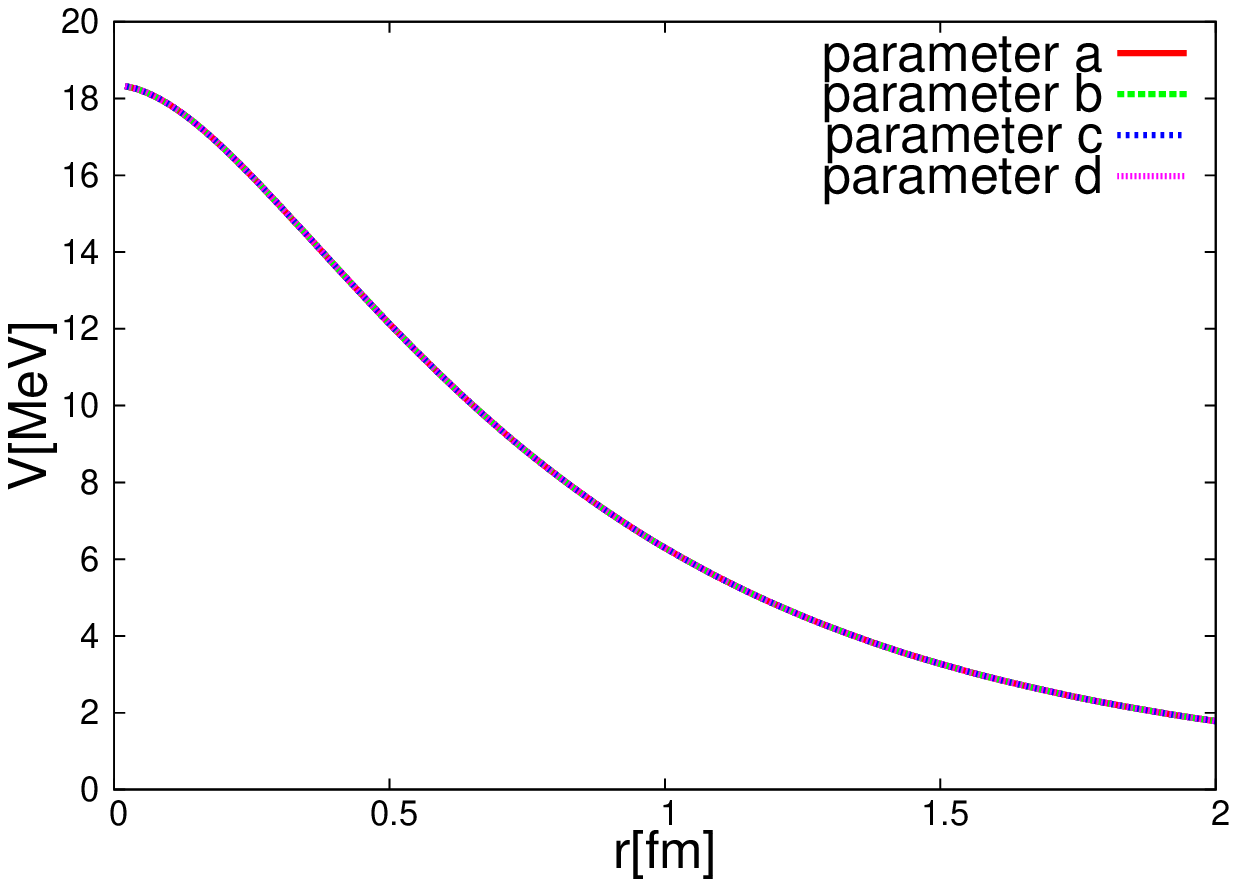}
\caption{$\Lambda_{c}N(^{3}S_{1})$-$\Sigma_{c}^{*}N(^{3}S_{1})$.}
\label{gr:potctnn-113}
\end{center}
\end{minipage}
\\
\begin{minipage}{0.5\hsize}
\begin{center}
\includegraphics[width=75mm]{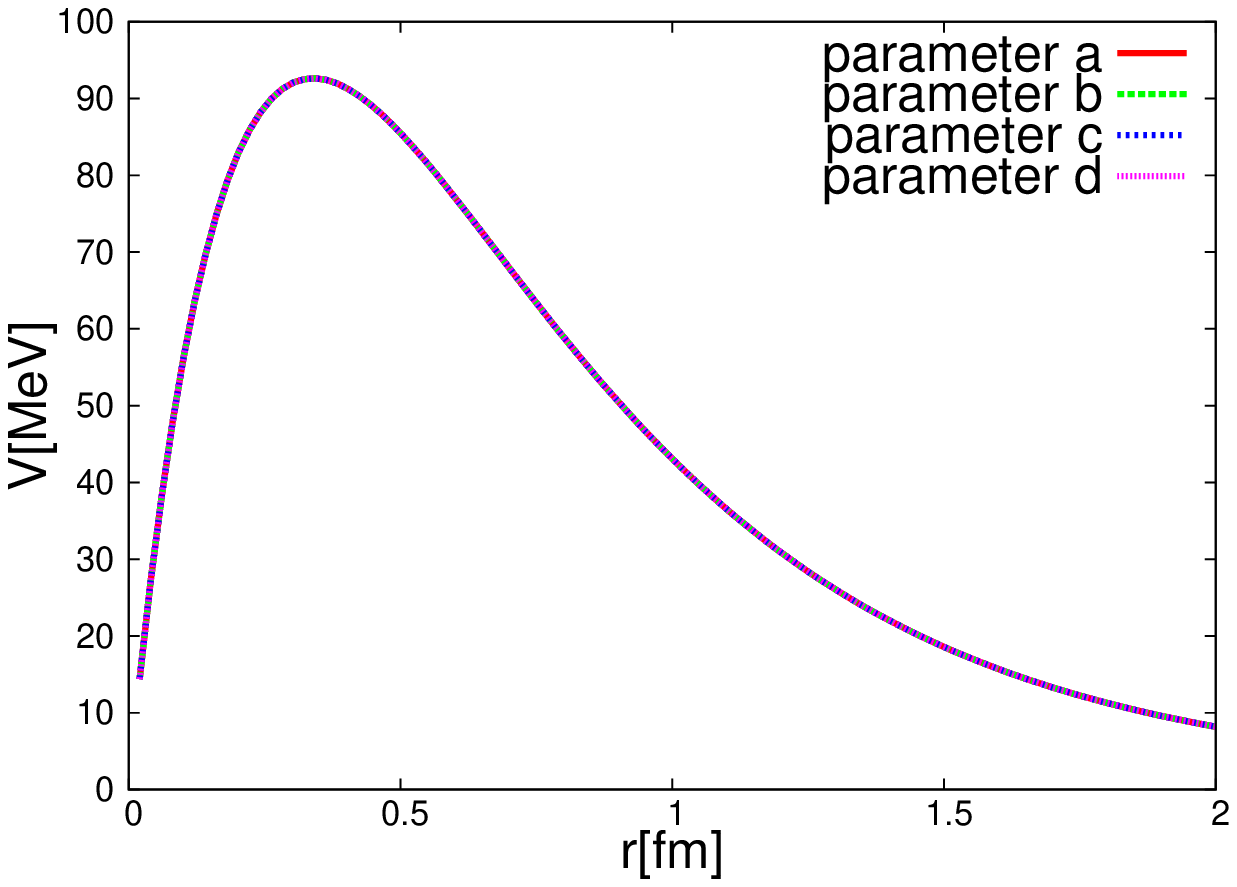}
\caption{$\Lambda_{c}N(^{3}S_{1})$-$\Sigma_{c}N(^{3}D_{1})$.}
\label{gr:potctnn-115}
\end{center}
\end{minipage}
\begin{minipage}{0.5\hsize}
\begin{center}
\includegraphics[width=75mm]{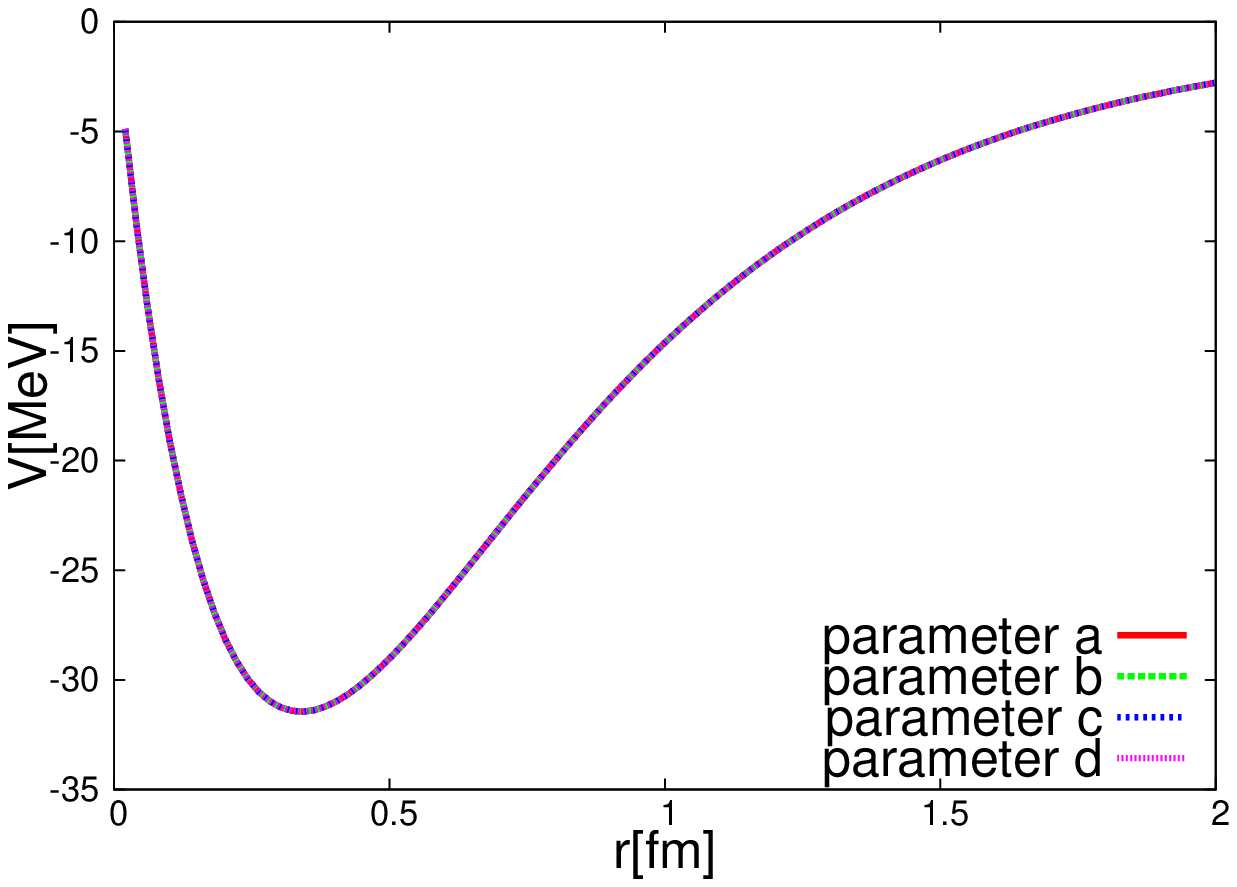}
\caption{$\Lambda_{c}N(^{3}S_{1})$-$\Sigma_{c}^{*}N(^{3}D_{1})$.}
\label{gr:potctnn-116}
\end{center}
\end{minipage}
\\
\begin{minipage}{0.5\hsize}
\begin{center}
\includegraphics[width=75mm]{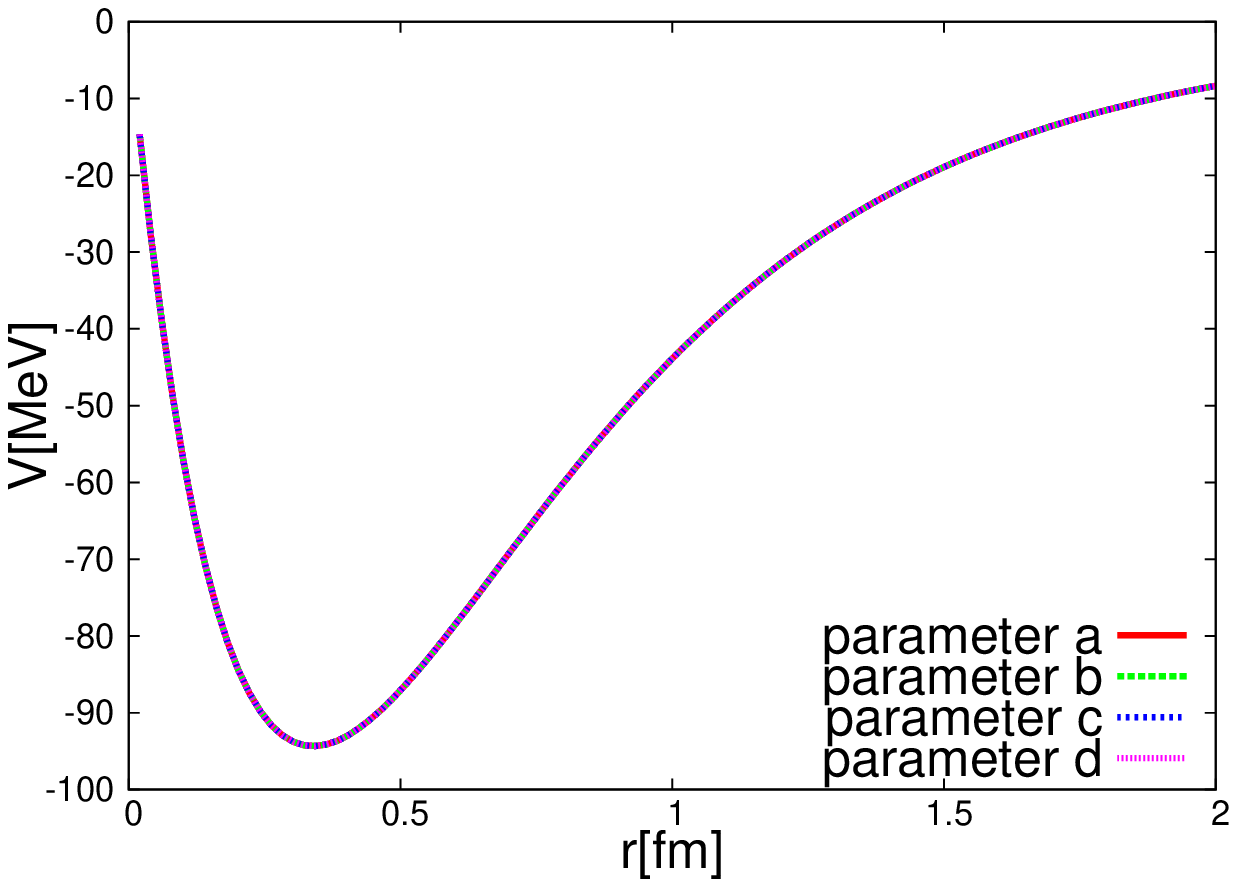}
\caption{$\Lambda_{c}N(^{3}S_{1})$-$\Sigma_{c}^{*}N(^{3}D_{1})$.}
\label{gr:potctnn-117}
\end{center}
\end{minipage}
\begin{minipage}{0.5\hsize}
\begin{center}
\includegraphics[width=75mm]{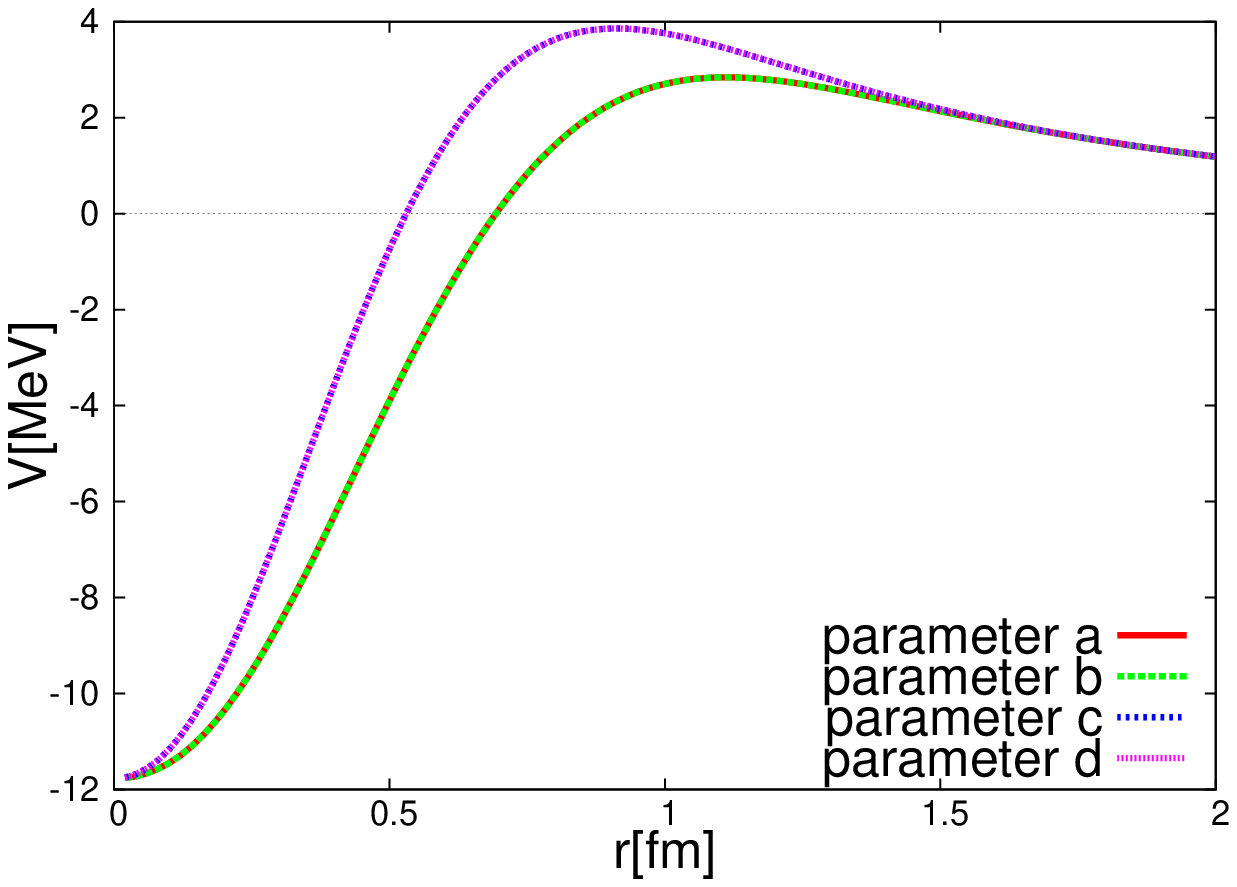}
\caption{$\Sigma_{c}N(^{3}S_{1})$-$\Sigma_{c}^{*}N(^{3}S_{1})$.}
\label{gr:potctnn-123}
\end{center}
\end{minipage}
\end{tabular}
\end{figure}

\begin{figure}[!h]
\begin{tabular}{cc}
\begin{minipage}{0.5\hsize}
\begin{center}
\includegraphics[width=75mm]{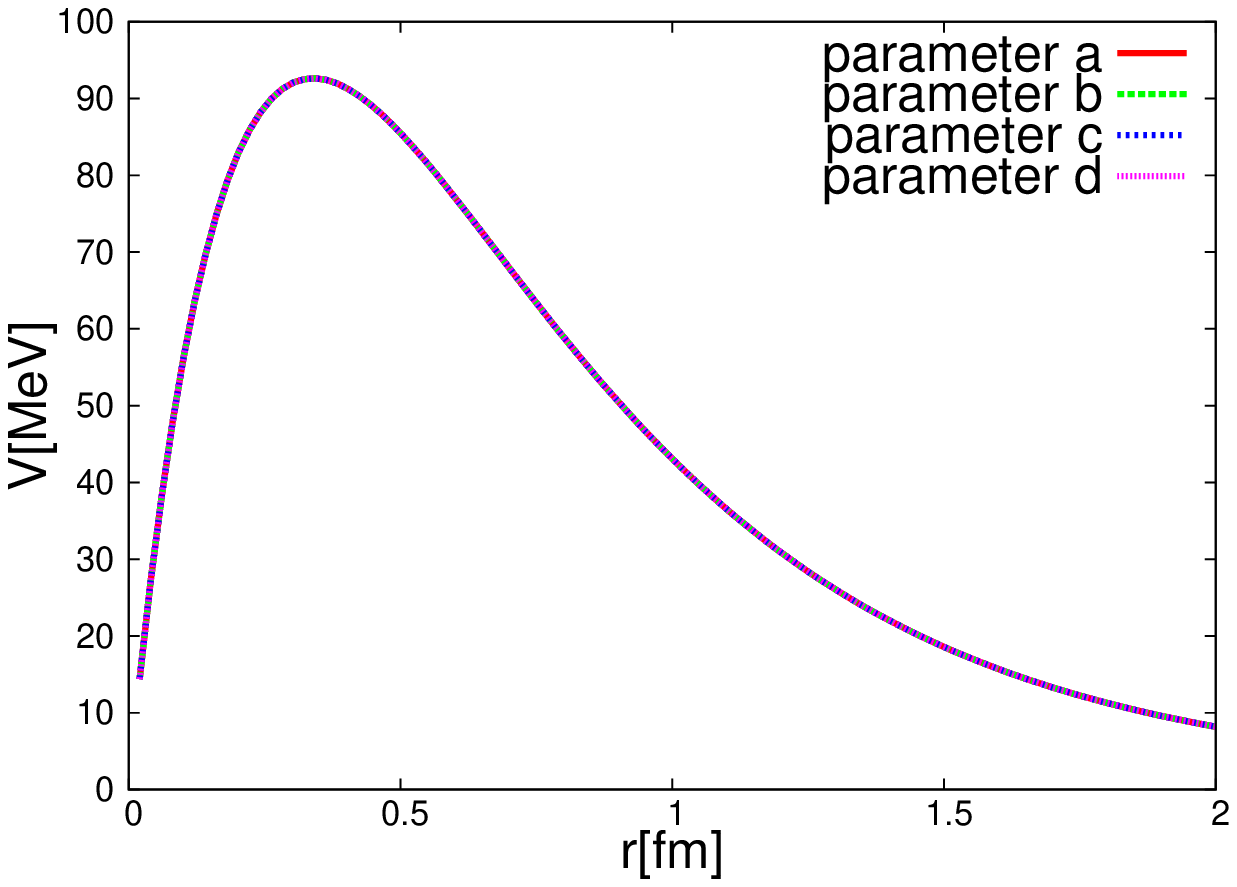}
\caption{$\Sigma_{c}N(^{3}S_{1})$-$\Lambda_{c}N(^{3}D_{1})$.}
\label{gr:potctnn-124}
\end{center}
\end{minipage}
\begin{minipage}{0.5\hsize}
\begin{center}
\includegraphics[width=75mm]{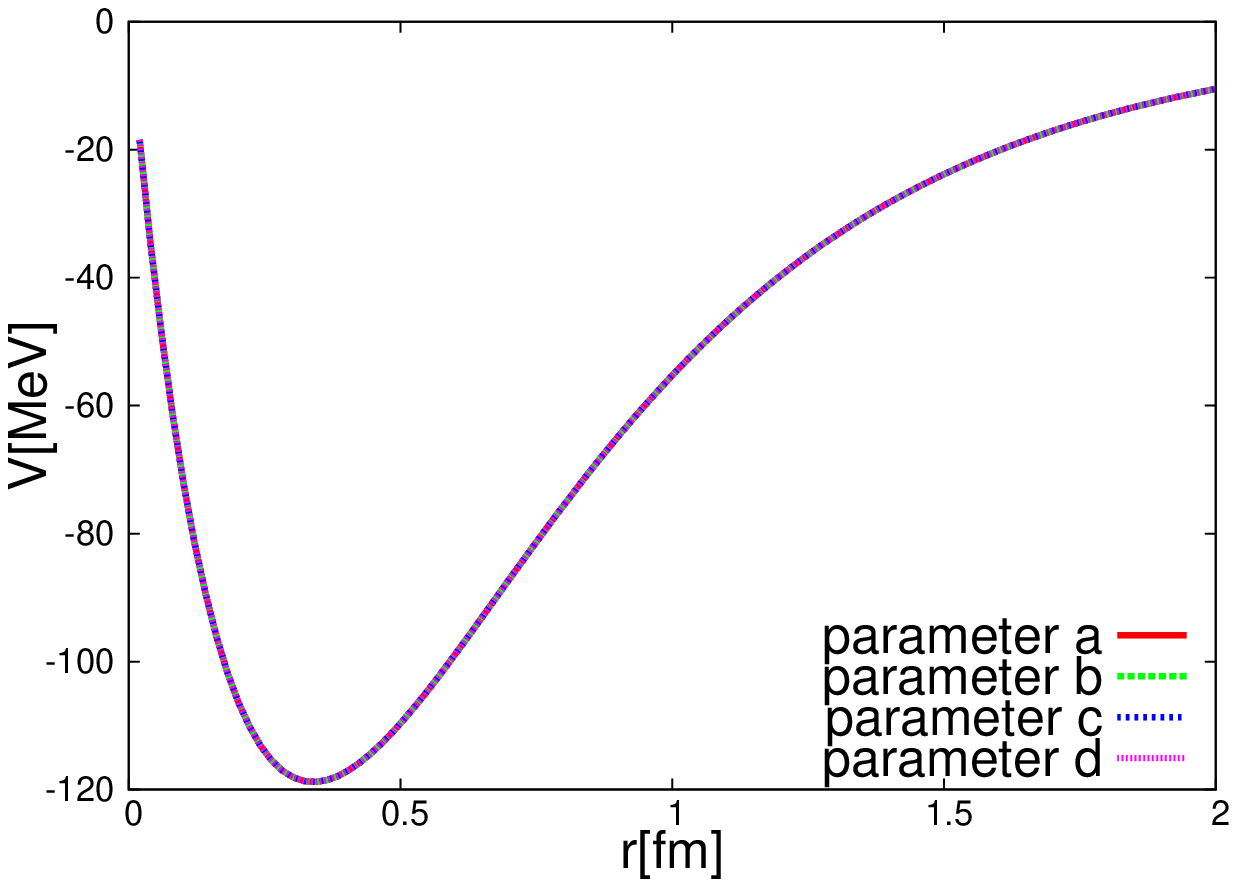}
\caption{$\Sigma_{c}N(^{3}S_{1})$-$\Sigma_{c}N(^{3}D_{1})$.}
\label{gr:potctnn-125}
\end{center}
\end{minipage}
\\
\begin{minipage}{0.5\hsize}
\begin{center}
\includegraphics[width=75mm]{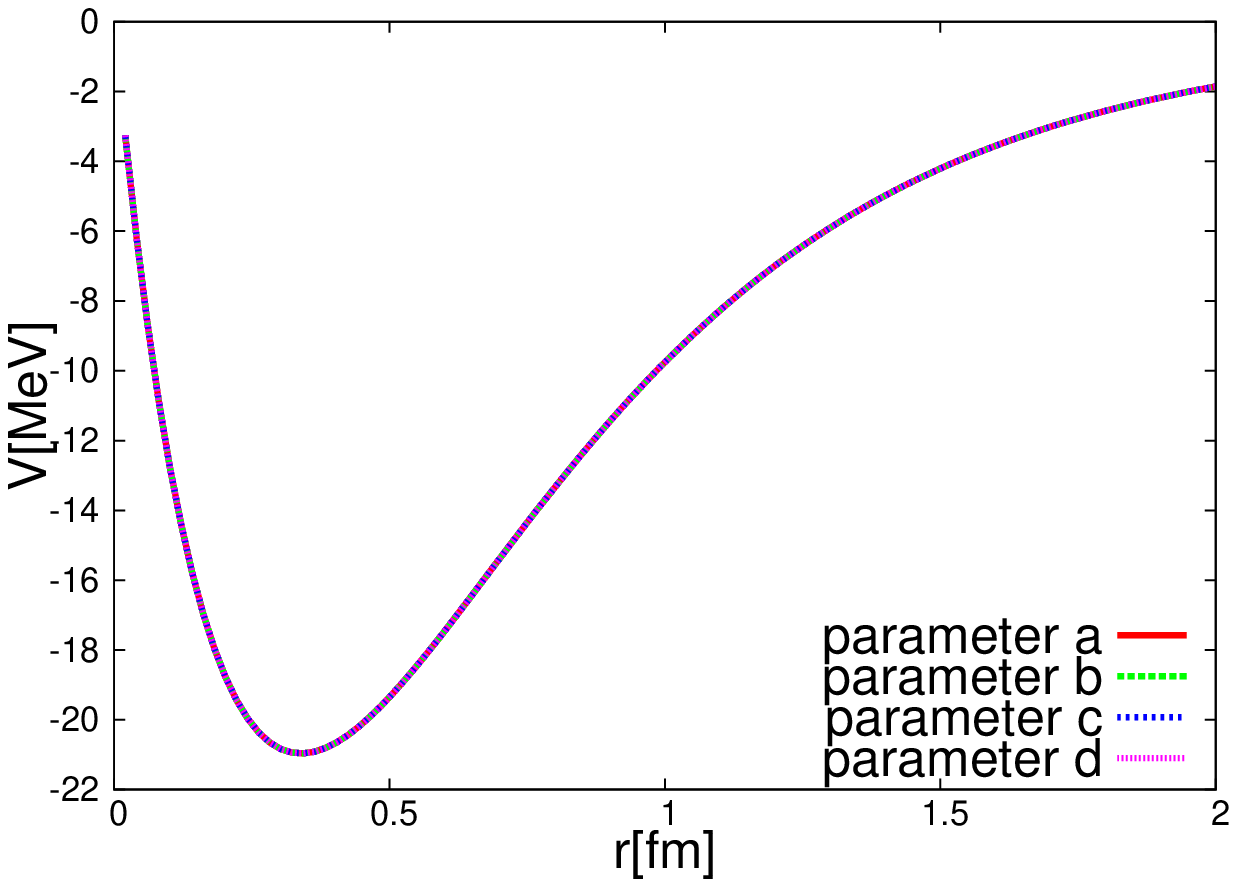}
\caption{$\Sigma_{c}N(^{3}S_{1})$-$\Sigma_{c}^{*}N(^{3}D_{1})$.}
\label{gr:potctnn-126}
\end{center}
\end{minipage}
\begin{minipage}{0.5\hsize}
\begin{center}
\includegraphics[width=75mm]{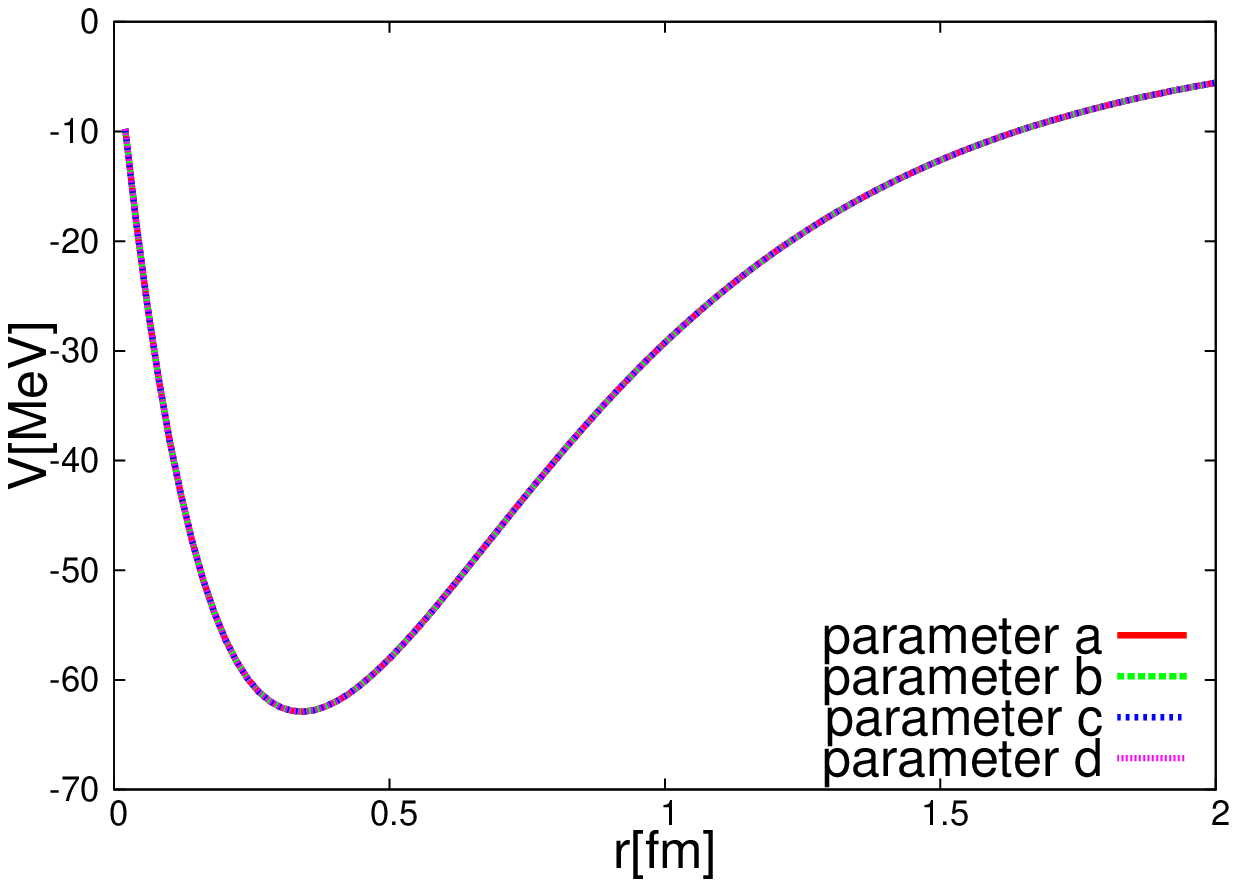}
\caption{$\Sigma_{c}N(^{3}S_{1})$-$\Sigma_{c}^{*}N(^{5}D_{1})$.}
\label{gr:potctnn-127}
\end{center}
\end{minipage}
\\
\begin{minipage}{0.5\hsize}
\begin{center}
\includegraphics[width=75mm]{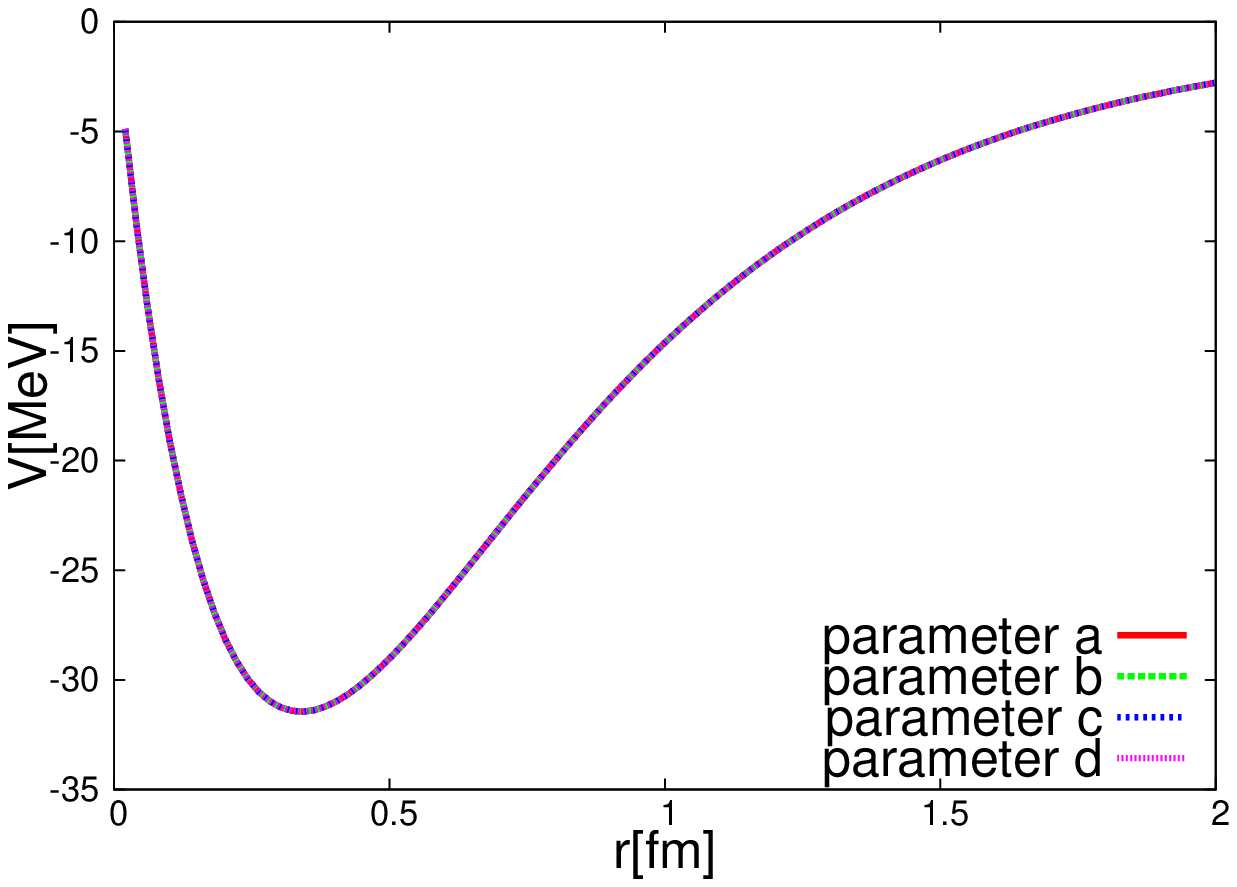}
\caption{$\Sigma{c}^{*}N(^{3}S_{1})$-$\Lambda_{c}N(^{3}D_{1})$.}
\label{gr:potctnn-134}
\end{center}
\end{minipage}
\begin{minipage}{0.5\hsize}
\begin{center}
\includegraphics[width=75mm]{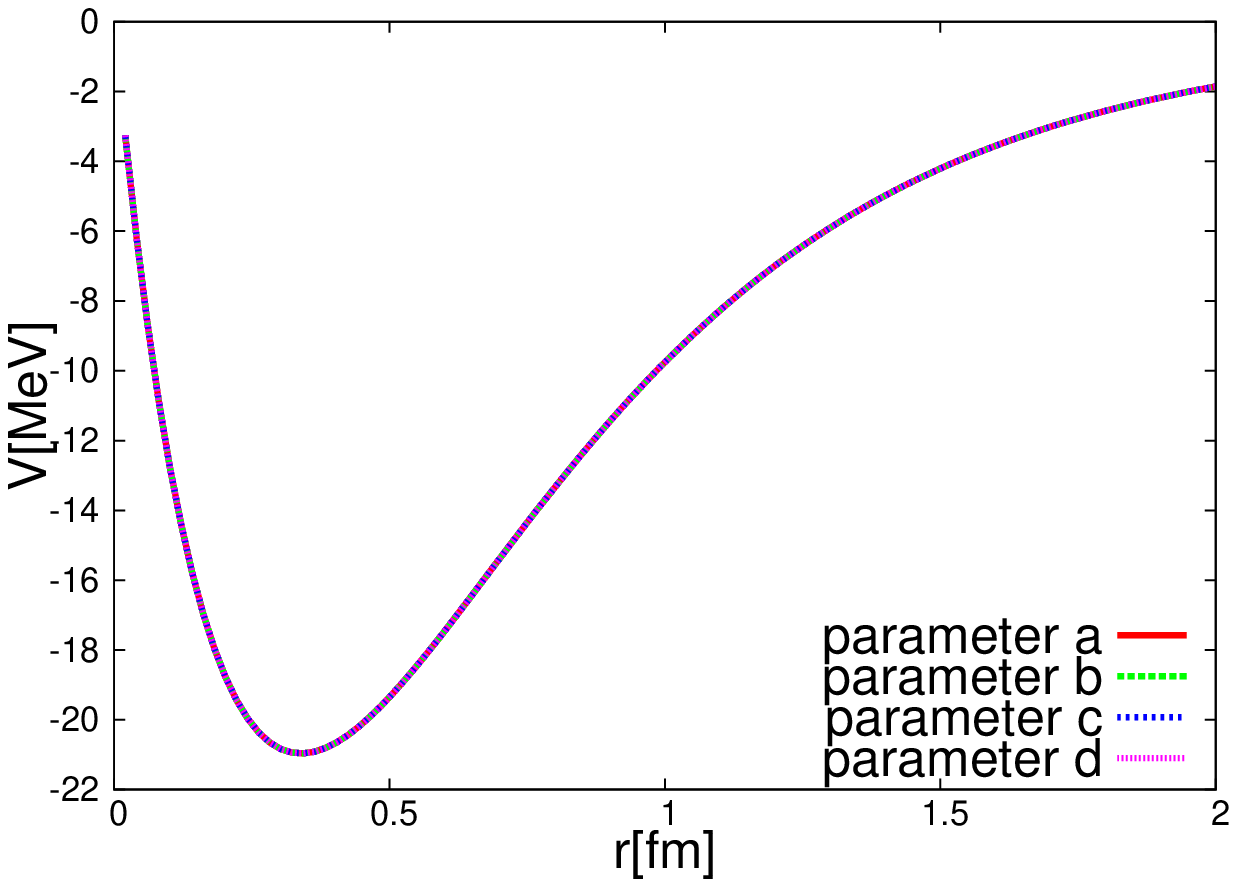}
\caption{$\Sigma{c}^{*}N(^{3}S_{1})$-$\Sigma_{c}N(^{3}D_{1})$.}
\label{gr:potctnn-135}
\end{center}
\end{minipage}
\end{tabular}
\end{figure}

\begin{figure}[!h]
\begin{tabular}{cc}
\begin{minipage}{0.5\hsize}
\begin{center}
\includegraphics[width=75mm]{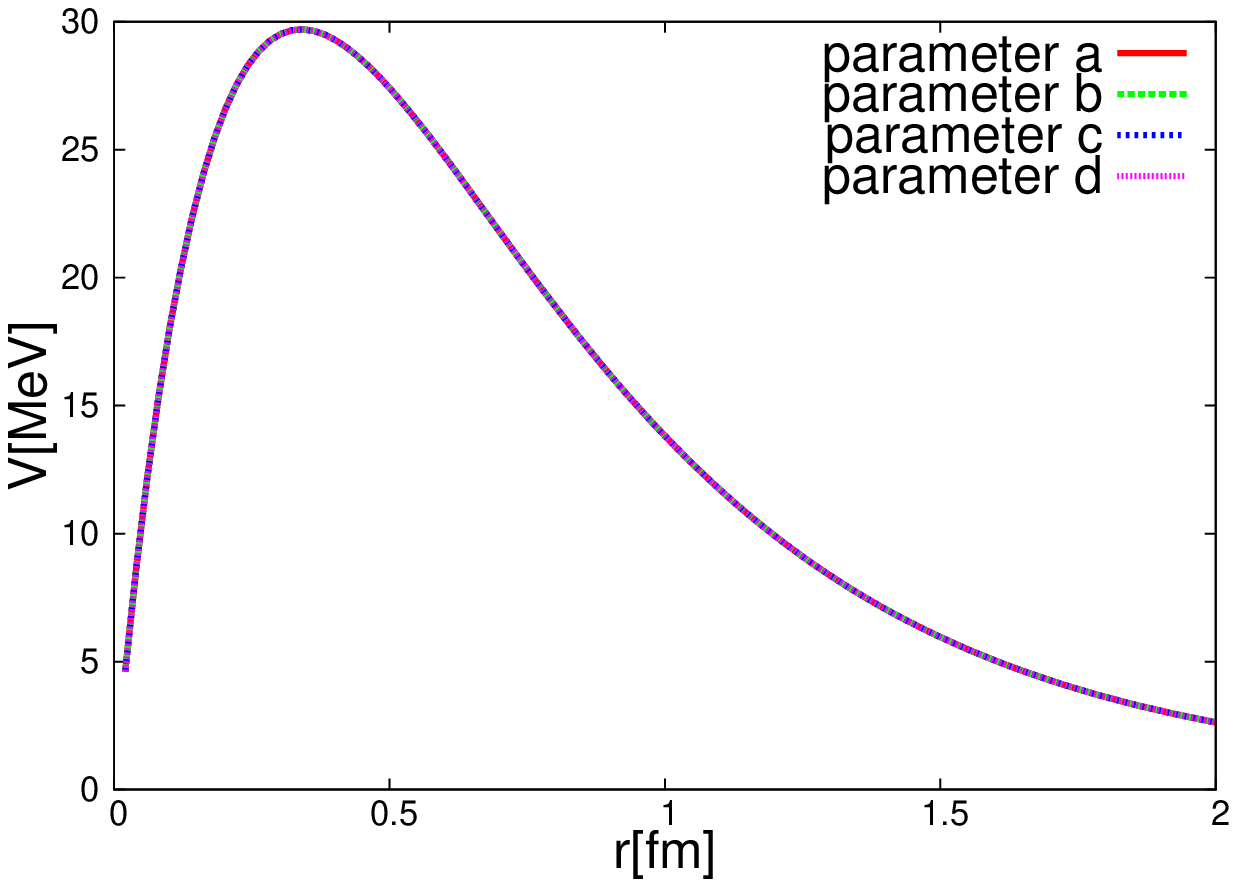}
\caption{$\Sigma{c}^{*}N(^{3}S_{1})$-$\Sigma_{c}^{*}N(^{3}D_{1})$.}
\label{gr:potctnn-136}
\end{center}
\end{minipage}
\begin{minipage}{0.5\hsize}
\begin{center}
\includegraphics[width=75mm]{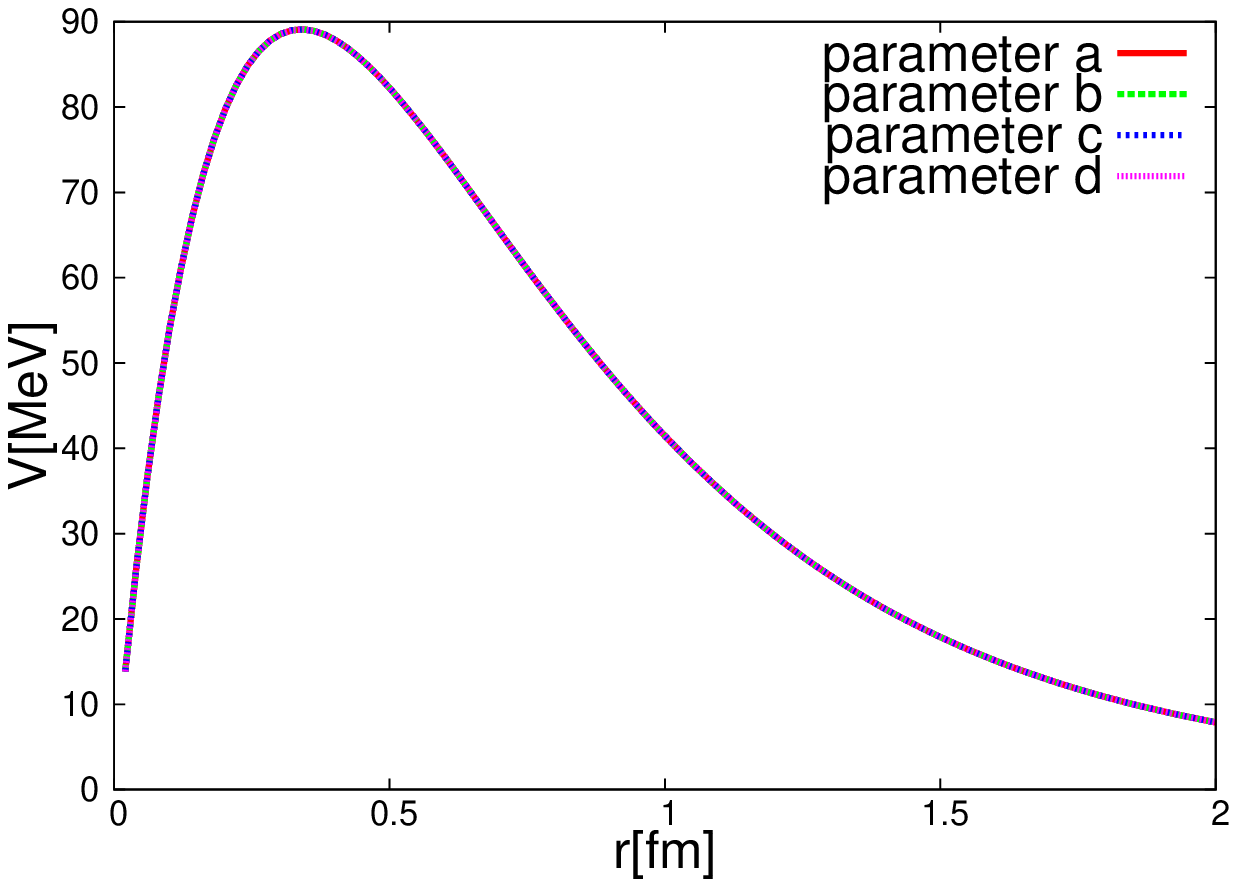}
\caption{$\Sigma{c}^{*}N(^{3}S_{1})$-$\Sigma_{c}^{*}N(^{5}D_{1})$.}
\label{gr:potctnn-137}
\end{center}
\end{minipage}
\\
\begin{minipage}{0.5\hsize}
\begin{center}
\includegraphics[width=75mm]{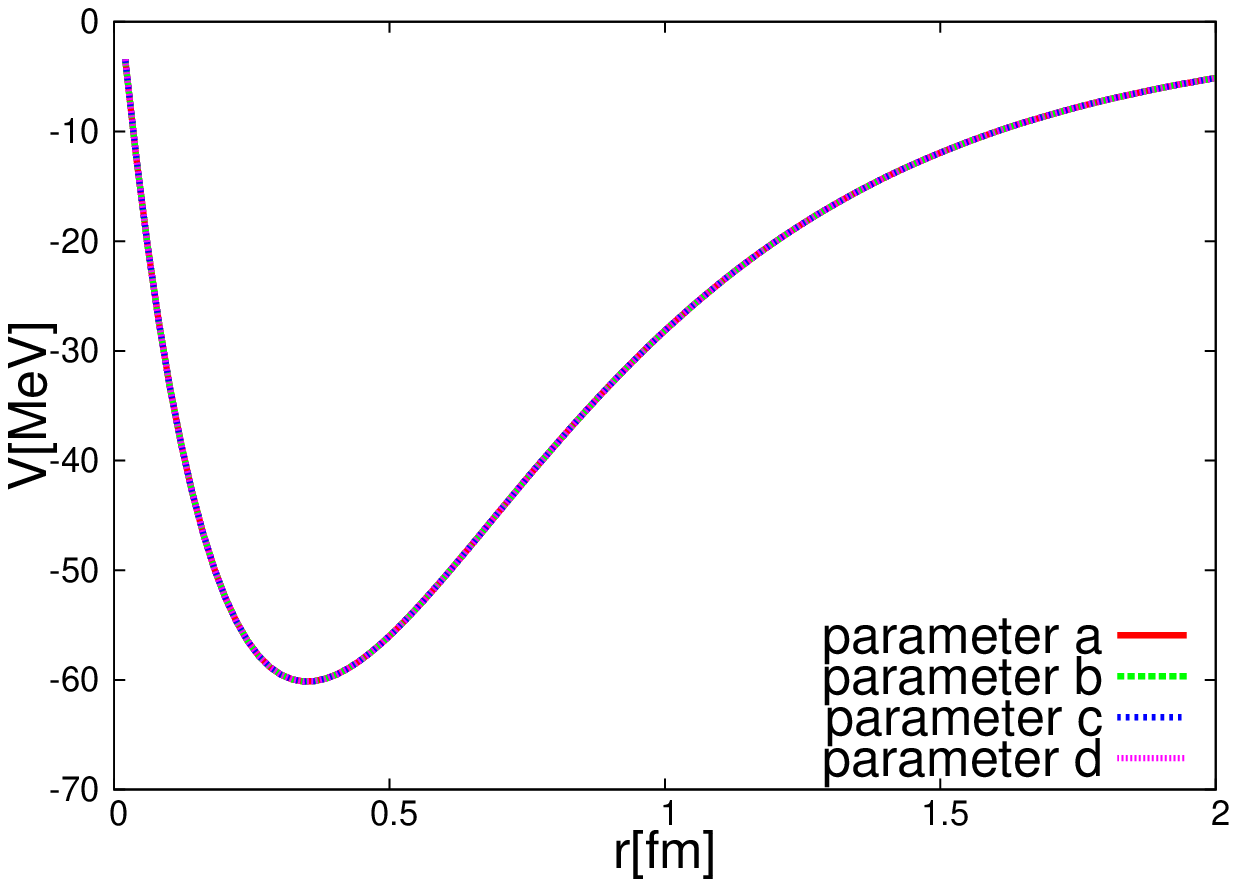}
\caption{$\Lambda{c}N(^{3}D_{1})$-$\Sigma_{c}N(^{3}D_{1})$.}
\label{gr:potctnn-145}
\end{center}
\end{minipage}
\begin{minipage}{0.5\hsize}
\begin{center}
\includegraphics[width=75mm]{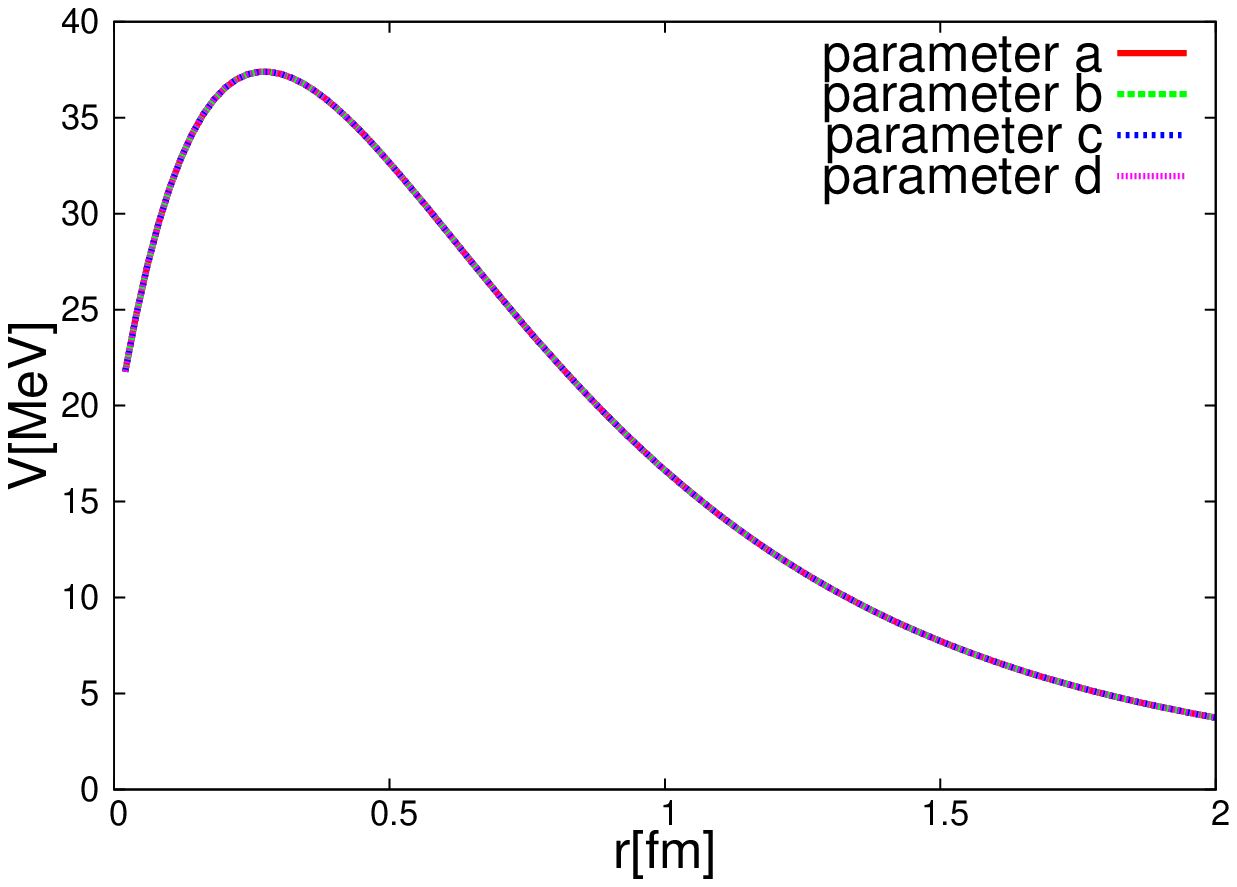}
\caption{$\Lambda{c}N(^{3}D_{1})$-$\Sigma_{c}^{*}N(^{3}D_{1})$.}
\label{gr:potctnn-146}
\end{center}
\end{minipage}
\\
\begin{minipage}{0.5\hsize}
\begin{center}
\includegraphics[width=75mm]{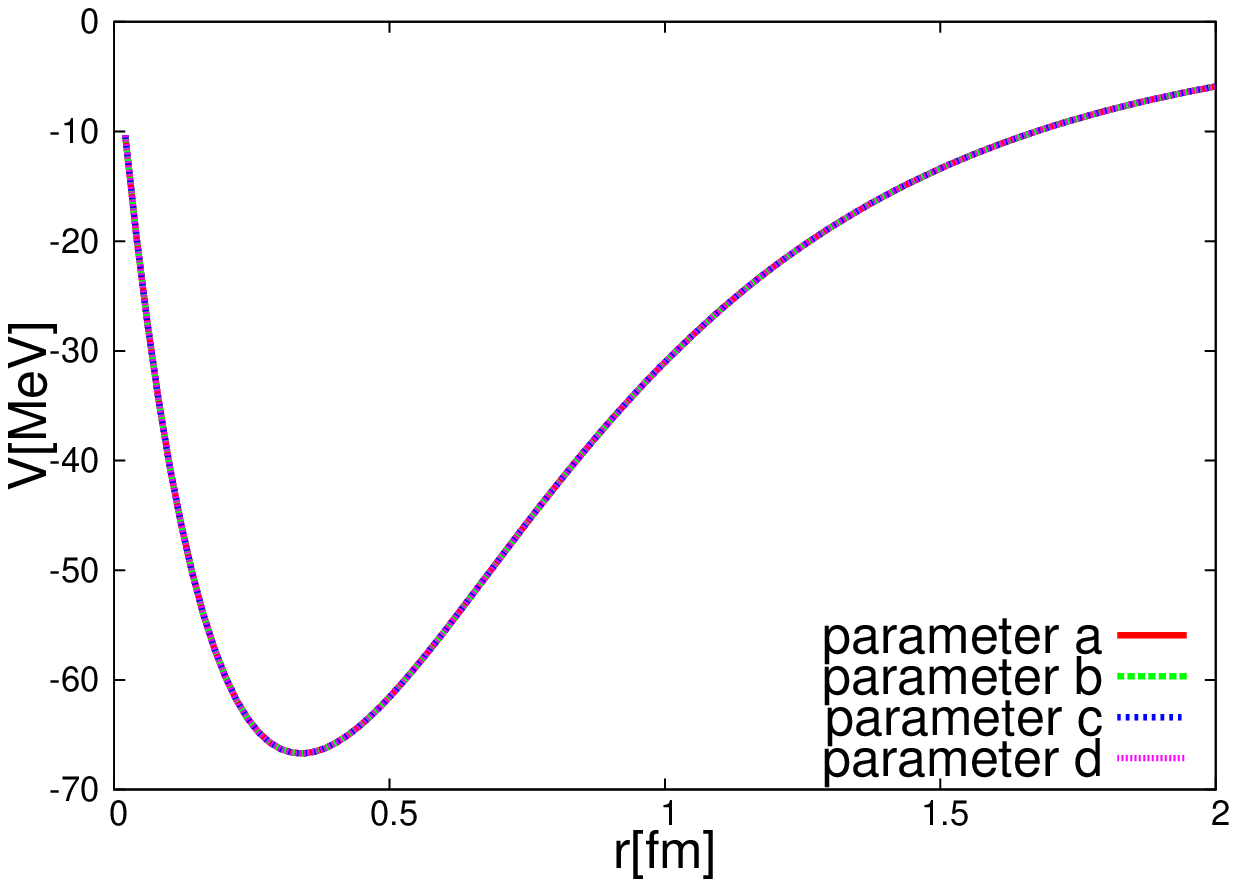}
\caption{$\Lambda{c}N(^{3}D_{1})$-$\Sigma_{c}^{*}N(^{5}D_{1})$.}
\label{gr:potctnn-147}
\end{center}
\end{minipage}
\begin{minipage}{0.5\hsize}
\begin{center}
\includegraphics[width=75mm]{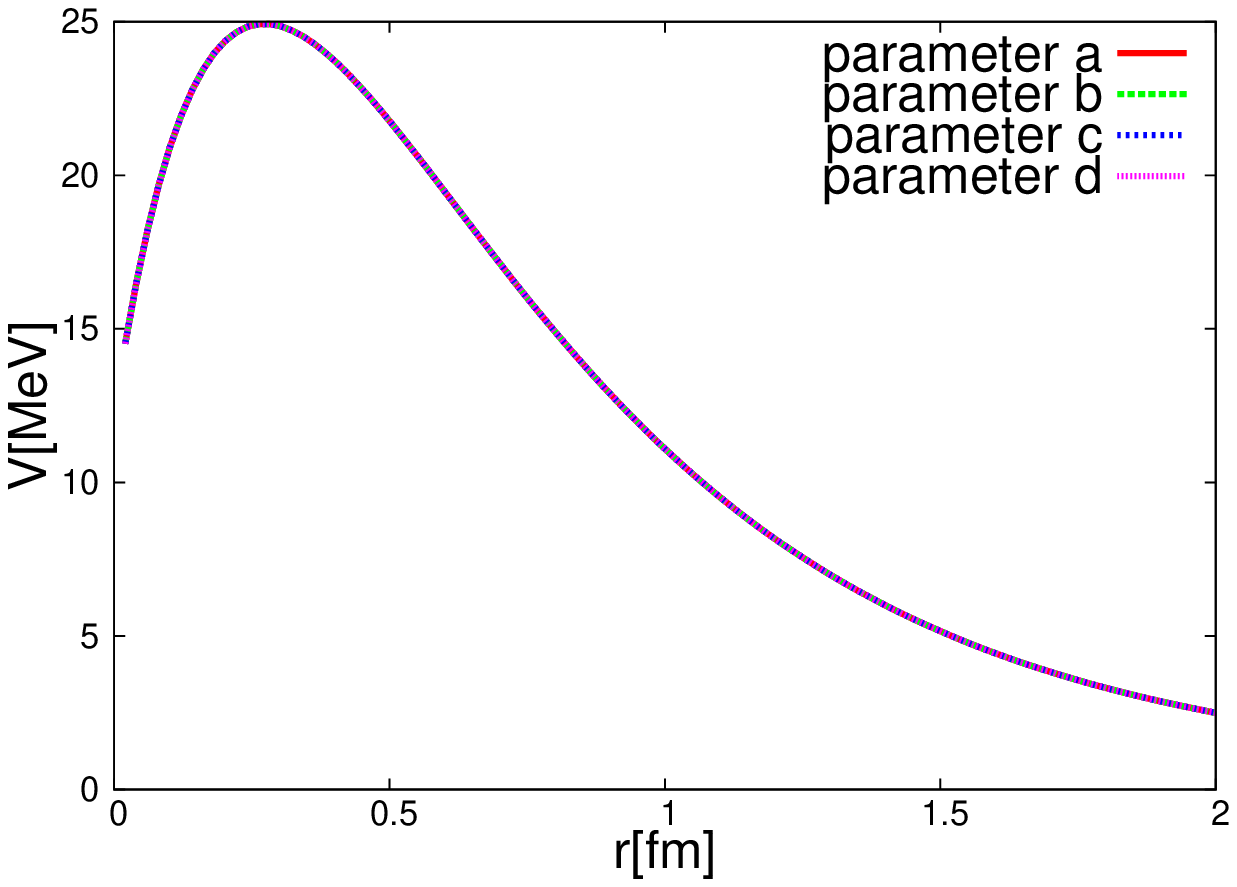}
\caption{$\Sigma{c}N(^{3}D_{1})$-$\Sigma_{c}^{*}N(^{3}D_{1})$.}
\label{gr:potctnn-156}
\end{center}
\end{minipage}
\\
\begin{minipage}{0.5\hsize}
\begin{center}
\includegraphics[width=75mm]{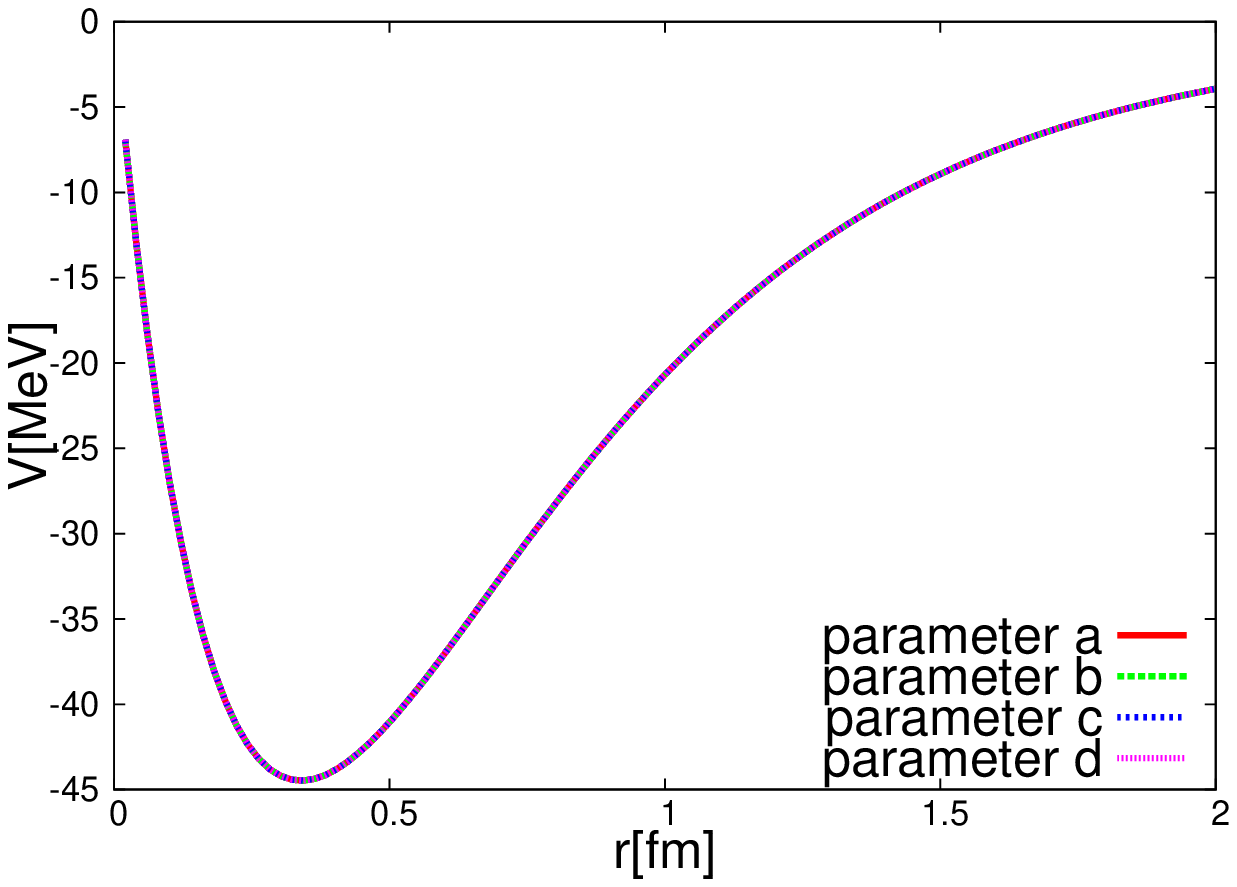}
\caption{$\Sigma{c}N(^{3}D_{1})$-$\Sigma_{c}^{*}N(^{5}D_{1})$.}
\label{gr:potctnn-157}
\end{center}
\end{minipage}
\begin{minipage}{0.5\hsize}
\begin{center}
\includegraphics[width=75mm]{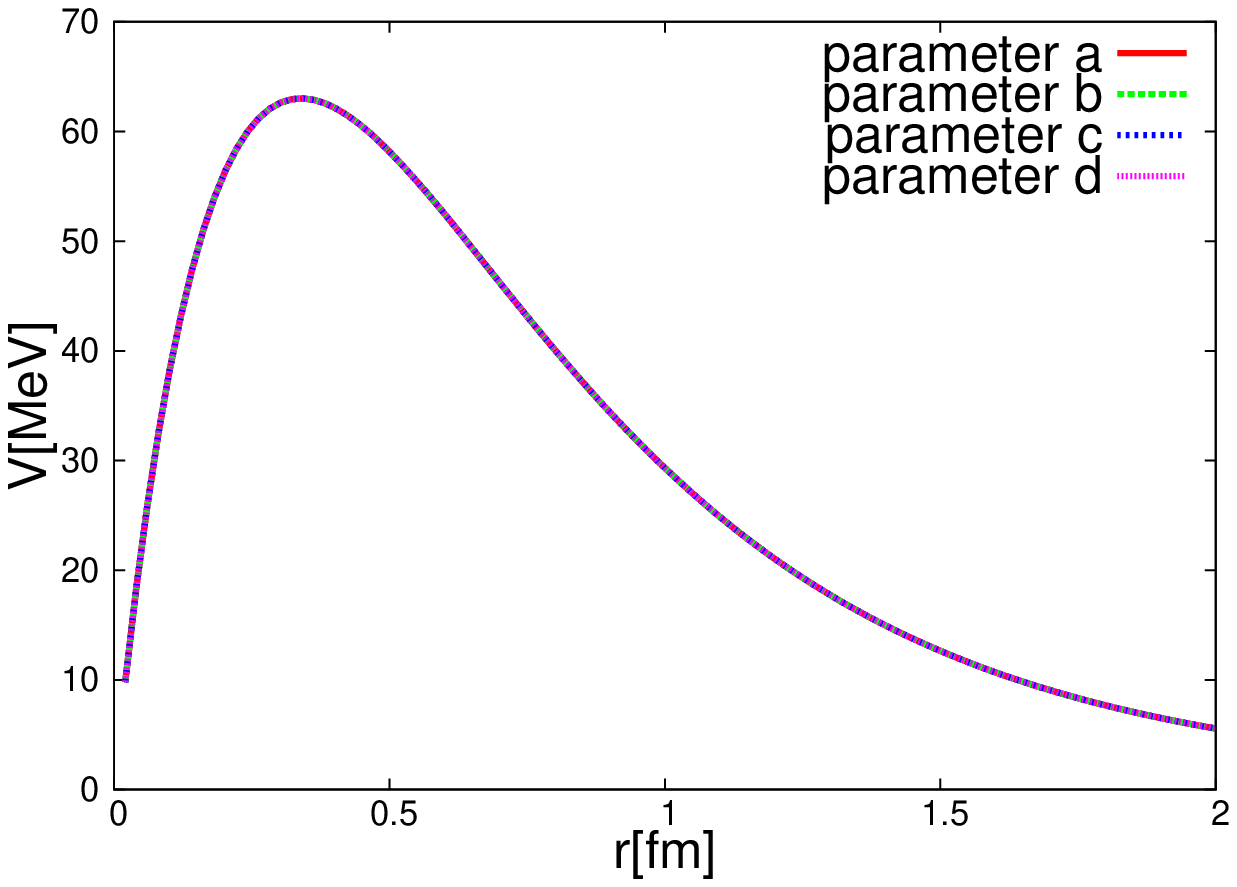}
\caption{$\Sigma{c}^{*}N(^{3}D_{1})$-$\Sigma_{c}^{*}N(^{5}D_{1})$.}
\label{gr:potctnn-167}
\end{center}
\end{minipage}
\end{tabular}
\end{figure}

Figs. \ref{gr:potctnn-122} - \ref{gr:potctnn-167} show the $J^{\pi}=1^{+} Y_{c}N$ potentials.
Four lines correspond to the four choices of the parameter sets, a $\sim$ d.
The potential for the diagonal $\Lambda_{c}N (^{3}S_{1})$ channel is not shown here as it coincides with that for $\Lambda_{c}N (^{1}S_{0})$.
The off-diagonal potential $\Lambda_{c}N (^{3}S_{1} - ^{3}D_{1})$ is zero.

\end{document}